\documentclass[lettersize,journal]{IEEEtran}
\usepackage{amsmath,amsfonts}
\usepackage{mathrsfs}
\usepackage{algorithmic}
\usepackage{algorithm}
\usepackage{array}
\usepackage[caption=false,font=normalsize,labelfont=sf,textfont=sf]{subfig}
\usepackage{textcomp}
\usepackage{stfloats}
\usepackage{url}
\usepackage{verbatim}
\usepackage{graphicx}
\usepackage{cite}
\usepackage{amsmath}
\usepackage{graphicx}
\usepackage{mdwtab}
\usepackage{booktabs}
\usepackage{multirow}
\usepackage{multicol}
\usepackage{xcolor}
\usepackage{amssymb}
\usepackage{hyperref}
\usepackage{fancyvrb}
\usepackage{tablefootnote}
\usepackage{adjustbox}
\def\L{{\cal L}}

\newcommand\Y{\mathbf Y}
\newcommand\y{\mathbf y}
\newcommand\E{\mathbf E}
\newcommand\A{\mathbf A}

\newcommand\D{\mathbf D}
\newcommand\X{\mathbf X}

\newcommand\B{\mathbf B}

\newcommand\V{\mathbf V}
\newcommand\W{\mathbf W}

\newcommand\1{\bf 1}

\begin{document}

\title{Image Processing and Machine Learning for Hyperspectral Unmixing: An Overview and the HySUPP Python Package}

\author{Behnood~Rasti,~\IEEEmembership{Senior~Member,~IEEE,}
Alexandre~Zouaoui,~\IEEEmembership{Student Member,~IEEE,} Julien~Mairal,~\IEEEmembership{Senior Member,~IEEE} and  Jocelyn Chanussot, ~\IEEEmembership{Fellow,~IEEE}
        % <-this % stops a space
\thanks{Behnood Rasti (corresponding author) is with the Faculty of Electrical Engineering and Computer Science, Technische Universität Berlin, 10623 Berlin, Germany; behnood.rasti@gmail.com}
\thanks{Alexandre Zouaoui, Jocelyn Chanussot, and Julien Mairal are with Univ. Grenoble Alpes, Inria, CNRS, Grenoble INP, LJK, 38000 Grenoble, France.}% <-this % stops a space
\thanks{Alexandre Zouaoui is currently with Data Science Experts, Grenoble, France.}
\thanks{Manuscript received April 19, 2023; revised August 16, 2023.}}

\markboth{Journal of \LaTeX\ Class Files,~Vol.~14, No.~8, August~2023}%
{Shell \MakeLowercase{\textit{et al.}}: A Sample Article Using IEEEtran.cls for IEEE Journals}

%\IEEEpubid{0000/00\~\copyright~2023 IEEE}

\maketitle

\begin{abstract}
Spectral pixels are often a mixture of the pure spectra of the materials, called endmembers, due to the low spatial resolution of hyperspectral sensors, double scattering, and intimate mixtures of materials in the scenes. Unmixing estimates the fractional abundances of the endmembers within the pixel. Depending on the prior knowledge of endmembers, linear unmixing can be divided into three main groups:  supervised, semi-supervised, and unsupervised (blind) linear unmixing. Advances in image processing and machine learning substantially affected unmixing. 
This paper provides an overview of advanced and conventional unmixing approaches. Additionally, we draw a critical comparison between advanced and conventional techniques from the three categories. We compare the performance of the unmixing techniques on three simulated and one real dataset. The experimental results reveal the advantages of different unmixing categories for different unmixing scenarios. Moreover, we provide an open-source Python-based package available at https://github.com/BehnoodRasti/HySUPP to reproduce the results.
\end{abstract}


\section{Introduction}
\begin{figure} [htbp]
\centering
\includegraphics[width=1\linewidth]{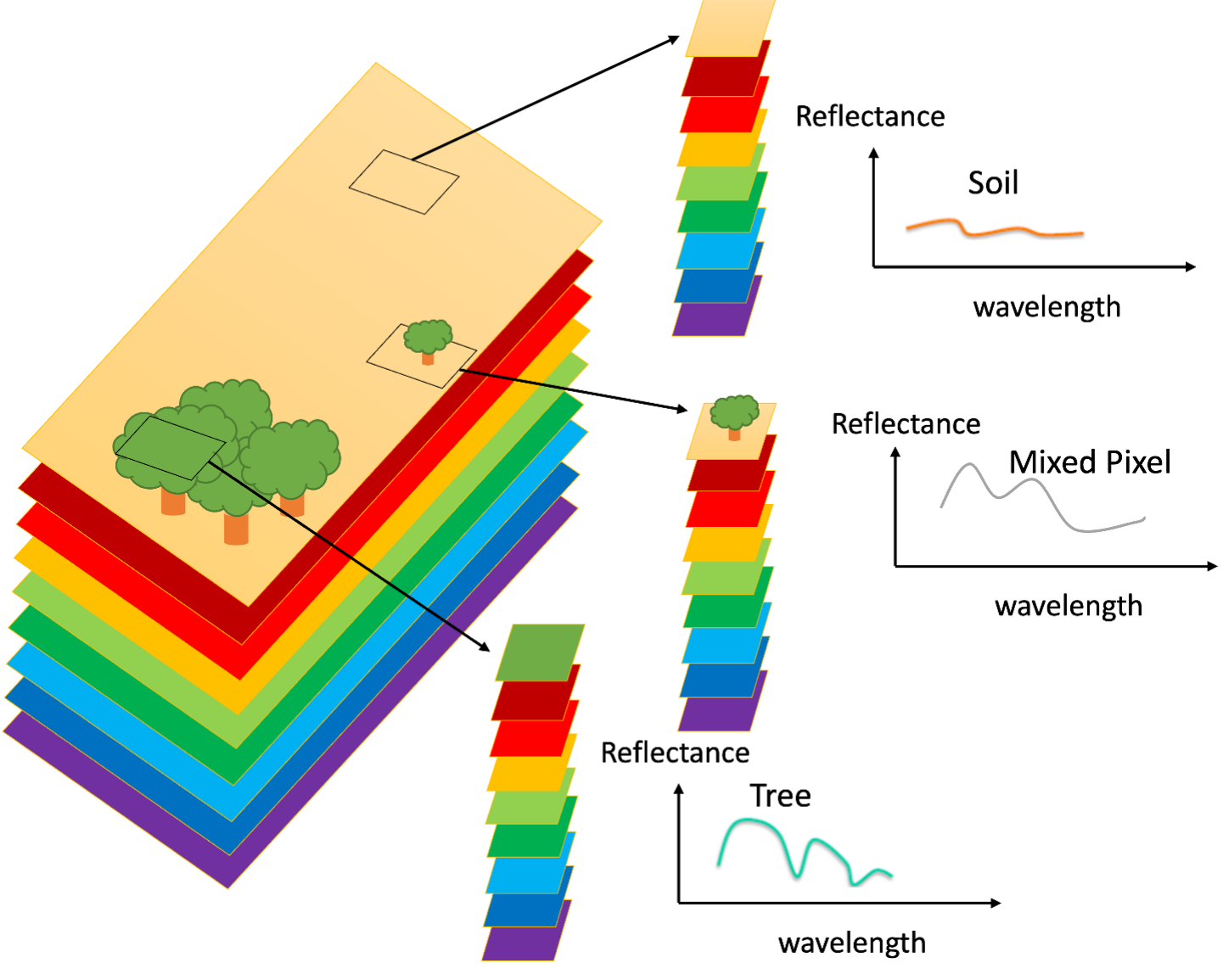}
\caption{Comparisons of mixed and pure pixels in hyperspectral data. }
\label{fig: MixedPixel}
\end{figure}

\IEEEPARstart{S}{pectral}  %unmixing is one of the most important applications in hyperspectral remote sensing. The contiguous spectra allow distinguishing the unique signature of materials called endmember. Hyperspectral sensors capture the spectral signature of materials in a range of wavelengths. Due to the low-spatial resolution, multiple scattering, and intimate mixing, the measured spectrum is generally a mixture of the pure spectra of the materials within a pixel called endmembers. 
%unmixing is a vital application in the field of hyperspectral remote sensing, where contiguous spectra are analyzed to distinguish the unique signature of materials, also known as "endmembers." Hyperspectral sensors capture the spectral signature of materials across a wide range of wavelengths. However, due to the low spatial resolution, multiple scattering, and intimate mixing, the measured spectrum is typically a combination of the pure spectra of different materials within a single pixel, resulting in mixed spectra that require unmixing. By isolating the endmembers, the unmixing process enables researchers to identify and map the spatial distribution of various materials, which is critical for many applications, including environmental monitoring, mineral exploration, and land-use classification. 
unmixing is a crucial processing technique in hyperspectral remote sensing that can play a vital role in various fields such as mineral exploration, agriculture and crop monitoring, environmental monitoring, urban planning, remote sensing of planetary surfaces, pollution monitoring, medical imaging, water quality assessment, etc. The ability to separate and identify different materials in an image is made possible by the contiguous spectra captured by hyperspectral sensors. Through the use of endmembers, which are the unique spectral signatures of materials, unmixing algorithms can decompose the mixed spectral data into its constituent parts. However, due to low spatial resolution, multiple scattering, and intimate mixing, the measured spectrum within a pixel is generally a complex mixture of the pure spectra of the constituent materials, making unmixing a challenging task.
Fig. \ref{fig: MixedPixel} demonstrates how the reflectance of a mixed pixel captured by an optical hyperspectral camera is composed of two endmembers within that pixel. In hyperspectral remote sensing, a mixing model represents the observed spectral pixel as a function of the endmembers and their corresponding fractional abundances within the pixel's area. Unmixing is the process of estimating the fractional abundances, either by estimating or extracting the endmembers or by relying on a library of endmembers. It may also involve determining the number of endmembers present. The mixing model is either linear or nonlinear, depending on the interaction of the incident light and the materials in the scene or sample.

\begin{figure*} [htbp]
\centering
\begin{tabular}{ccc} 
\includegraphics[width=.3\linewidth]{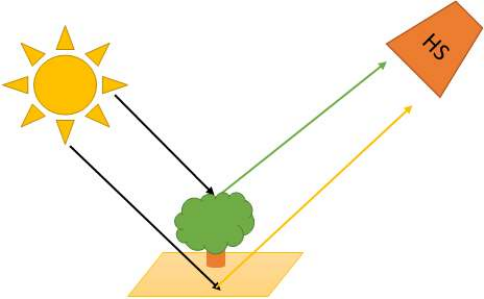}&
\includegraphics[width=.3\linewidth]{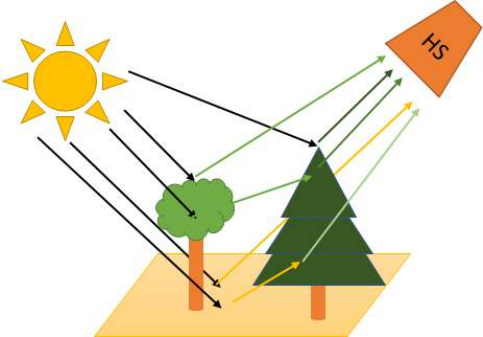}&\includegraphics[width=.3\linewidth]{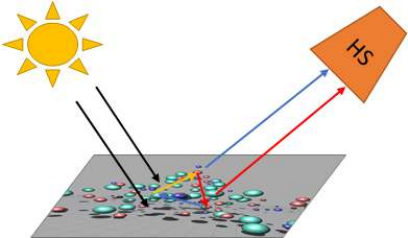}\\  (a) Linear mixture &(b) Bilinear mixture  &(c) Intimate mixture \\
 \end{tabular} %\end{center} 
\caption{Macroscopic versus microscopic assumption. This figure illustrates three major assumptions in hyperspectral imagery. (a) The linear assumption is when the light interacts with the materials only once. (b) The bilinear assumption when the light interacts with a maximum of two materials  (c) intimate mixture when the light interacts with more than two materials.}
\label{fig: Lin_Non}
\end{figure*}

In linear unmixing, the endmembers are assumed to be linearly mixed, which is valid when each light ray interacts with only one material before reaching the sensor, as shown in Figure \ref{fig: Lin_Non} (a). This assumption is common in Earth observation applications where macroscopic problems exist. The spatial resolution of the sensor plays a crucial role in macroscopic scenarios, where a pixel may contain multiple materials. In such cases, the pixel is a mixture of different materials, such as tree and soil, as shown in Figure \ref{fig: Lin_Non} (a).

Another assumption is bilinear mixing, which assumes double scattering or that the light ray interacts with two materials before reaching the sensor, as shown in Figure \ref{fig: Lin_Non} (b). However, in microscopic scenarios where the pure materials are intimately mixed within a pixel, and the light undergoes multiple scattering and reflections by several materials, the linear approximation fails, and nonlinear models that are more complex than bilinear should be used \cite{unmixing-review,Dobigeon_2014_NU}. %A mixing model models the observed spectral pixel as a function of the endmembers, corresponding to the pure spectra of the materials contained in that pixel and each of the endmembers fractional abundances within that pixel's area. Unmixing estimates the fractional abundances, often by estimating or extracting the endmembers or relying on a library of endmemebrs. Unmixing may also involve the estimation of the number of endmembers. Considering the interaction of the incident light and materials in a  scene or a sample, the mixing model is either linear or nonlinear. In linear unmixing, the endmembers are assumed to be linearly mixed. The linear model is valid when each light ray only interacts with one material before reaching the sensor Fig. \ref{fig: Lin_Non} (a). This is a common assumption in Earth observation applications due to macroscopic problems. Spatial resolution often plays a key role in macroscopic scenarios, i.e.,  the lower the resolution, the more material might be sensed within a pixel. For instance, in Fig. \ref{fig: Lin_Non} (a), assuming the scene is a pixel, then the pixel is a mixture of tree and soil. Another assumption is bilinear which assumes double scattering or the light ray interacts with two materials before reaching the sensor shown in Fig. \ref{fig: Lin_Non} (b). In microscopic scenarios, the pure materials are intimately mixed within the pixel, and the light undergoes multiple scattering and reflections by several materials. In these situations, the linear approximation often fails, and nonlinear models that are more complex than bilinear should be utilized \cite{unmixing-review,Dobigeon_2014_NU}. 

\begin{figure} [htbp]
\centering
\includegraphics[width=.7\linewidth]{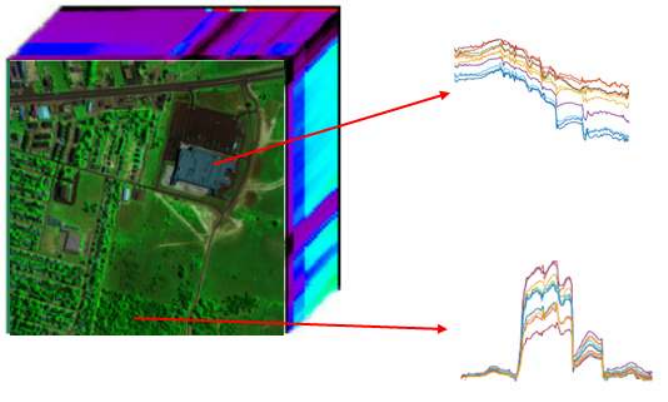}
\caption{Noise, atmospheric effects, illumination variations (caused by terrain topography, occlusion of the light), and
intrinsic variations of materials (e.g., soil signature might change dramatically by variations in its composition and moisture content) cause spectral variability. }
\label{fig: Spec_var}
\end{figure}

In macroscopic scenarios, nonlinearity is also induced by noise, atmospheric effects, temporal effects, illumination variations, and materials' intrinsic variation, which may cause spectral variability shown in Fig. \ref{fig: Spec_var} \cite{Spec_var}. Illumination variations are mainly due to two effects: varying terrain topography, which affects the angles of the incident
radiation, and the occlusion of the light source by other objects (leading to shaded areas). The intrinsic variation of spectral signatures of the materials is mainly due to physicochemical changes. For instance, soil's signature might change dramatically by variations in its composition and moisture content. Another example is the leaf's signature changes throughout the temporal seasons. Scale differences can cause spectral variability as well. For instance, a tree contains leaves, branches, fruits, bark, bark, fruits or flowers, etc., leading to the intrinsic variability of the tree spectra at different scales. The atmospheric effect should be corrected, and noise can be removed \cite{HyDe,BR_Rev1,ImRes2021}.

The unmixing process is subjective and depends on how the endmembers are defined. For example, when detecting buildings, the reflectance of a brick could be considered an endmember. However, bricks are composed of several materials, such as clay, sand, and concrete, making it challenging to define the endmembers of interest for detecting specific materials or ratios of different materials inside bricks.

In the former, linear unmixing may be sufficient to solve the problem. However, for intimate mixtures and the latter, nonlinear models must be used. For example, the Hapke model \cite{hapke_2012, BHapke1981} suggests that mixing occurs at the albedo level rather than the reflectance level. Therefore, more complex nonlinear models are necessary for such cases to accurately unmix the spectral data.
%The unmixing problem is associated with defining the endmembers, and therefore is subjective. For instance, for building detection, bricks reflectance can be considered an endmember, while bricks may include several materials, such as clay, sand, concrete, etc. In detecting clay in bricks or the ratio of different materials inside bricks, the endmembers of materials of interest define the unmixing problem. The former, maybe tackled by linear unmixing, while the latter is an intimate mixture scenario, and nonlinear models must be used. For instance, according to the Hapke model \cite{hapke_2012, BHapke1981}, the mixing occurs at the albedo level instead of the reflectance level. 
%Despite simplifications,  %it's important to note that when using unmixing to solve a specific problem, it's crucial to understand the limitations and assumptions of the models we use and to carefully consider whether they are appropriate for the problem at hand.
It is worth mentioning that, unlike the common assumption, the proportion of an endmember within a pixel or in a scene is not the percentage of the endmember material within that pixel or in that scene \cite{unmixing-review}. According to Hapke \cite{hapke_2012, BHapke1981}, the proportion of an endmember in a linear mixture indicates the relative area of that endmember. Therefore, reflectance is not a linear mixture of the mass or cross-sectional area of the endmember materials. Despite this, linear unmixing has shown significant value in remote sensing applications in the past decades \cite{unmixing-review}. However, when associating an unmixing problem with an application, we should be aware of the simplifications of the model we use. %linear unmixing in remote sensing applications, such simplifications have shown values in the past decades \cite{unmixing-review}.
%When  

%Fig. \ref{fig: LinNon} compares the difference between linear and nonlinear data. 
Assuming three materials in a scene correspond to three endmembers ($e_1$, $e_2$, and $e_3$), the observed data can be projected into a 2D subspace where all the data points are enclosed in the data simplex formed by the convex hull of endmembers (see Fig. \ref{fig: Pure_No}). In contrast, in a nonlinear scenario, the simplex does not contain all the data points. This paper mainly considers linear EO applications and macroscopic problems and focuses on linear mixing models. However, we briefly discuss the bilinear and nonlinear unmixing approaches. 

% \begin{figure} [htbp]
% \centering
% \begin{tabular}{cc} 
% \includegraphics[width=.45\linewidth]{LU_PC.png}&\includegraphics[width=.45\linewidth]{NLPC.png}\\   (a) &(b) 
%  \end{tabular} %\end{center} 
% \caption{Linear versus nonlinear unmixing. (a)  subspace projection for linear interaction and (b) nonlinear interaction.}
% \label{fig: LinNon}
% \end{figure}

% \subsection{Defining an Unmixing Problem for Real Application}
% Microscopic vs microscopic
% Endmember selection (dependent on the application)

\subsection{Key trends in the hyperspectral unmixing realm}

% As a research topic, hyperspectral unmixing has known several development stages as illustrated in figure \ref{fig: publications}. From 1992 to 2008, it was barely studied as the number of publications per year did not exceed 18. It then rose steadily from 2009 to 2016, even surpassing the 200 publications mark in 2016. Finally, it remains a hot topic since 2017 with the advent of deep learning techniques for unmixing, 2022 being the fourth most prolific year on record with 164 publications.

\begin{figure}[htbp]
    \centering
    \includegraphics[width=\linewidth]{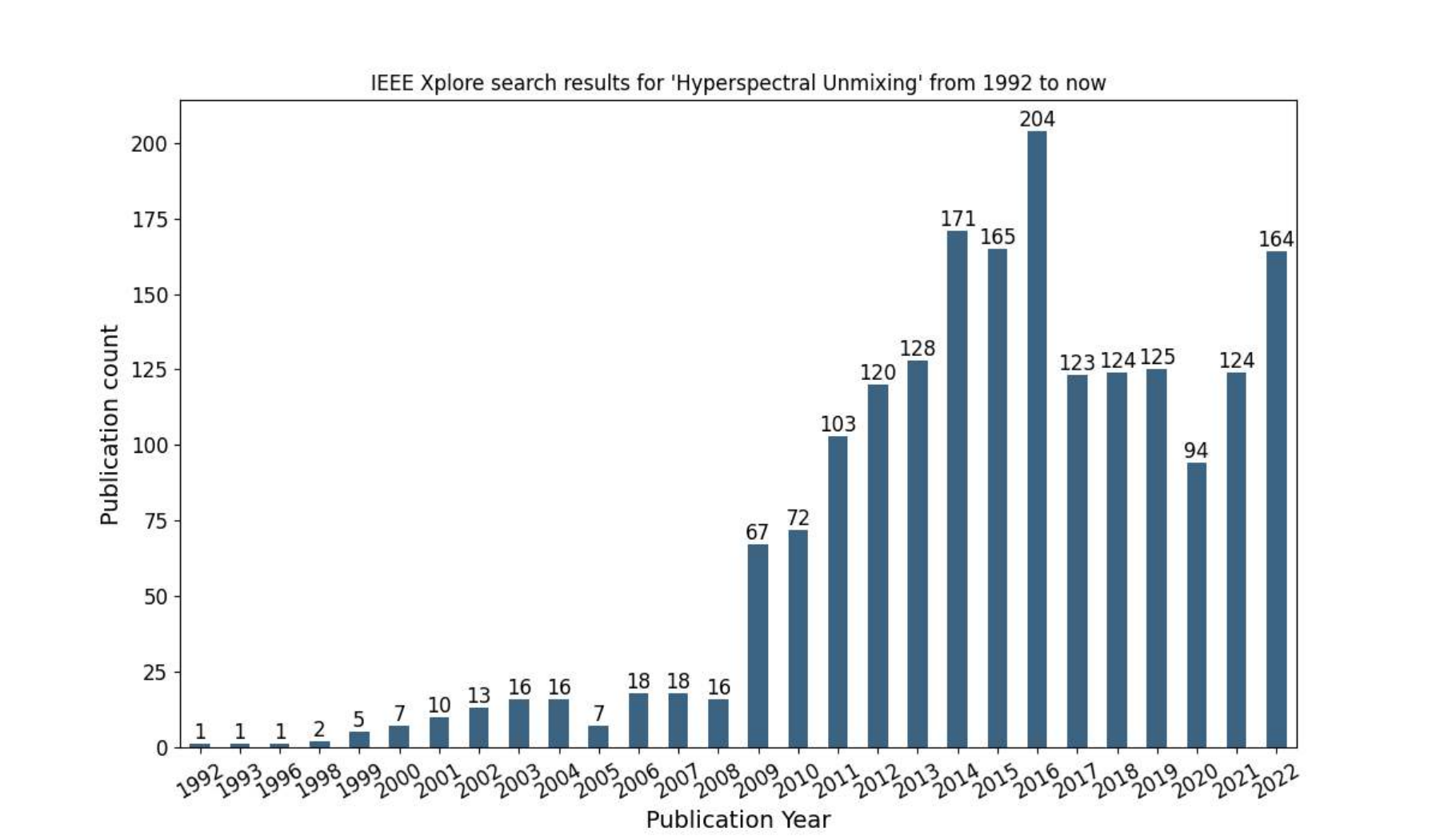}
    \caption{Publications over time based on IEEE Xplore keyword search tool using ``Hyperspectral Unmixing" as input.}
    \label{fig:publications}
\end{figure}

Hyperspectral unmixing has undergone various developmental stages, as illustrated in figure \ref{fig:publications}. 
In its early years, from 1992 to 2008, this research topic received minimal attention, with an annual publication count of no more than 18. 
However, starting from 2009 and continuing until 2016, there was a steady increase in interest, with the number of publications even surpassing 200 in 2016. 
Since 2017, hyperspectral unmixing has remained a prominent and dynamic area of research, driven by the emergence of deep learning techniques for unmixing. 
In 2022, this field reached its fourth most prolific year on record, with 164 publications.

%At the time where hyperspectral unmixing gained popularity, the main scientific programming language was MATLAB as illustrated in figure \ref{fig: matlab}. However, in recent years, interest in Python rose steadily while MATLAB declined. This can be partly explained by the rise of open science whereby open sourcing scientific code became popular. Our point is that nowadays releasing an open source unmixing package in Python is more sensible than in MATLAB, even though this may require translating existing unmixing MATLAB code into Python.

\begin{figure}[htbp]
    \centering
    \includegraphics[width=\linewidth]{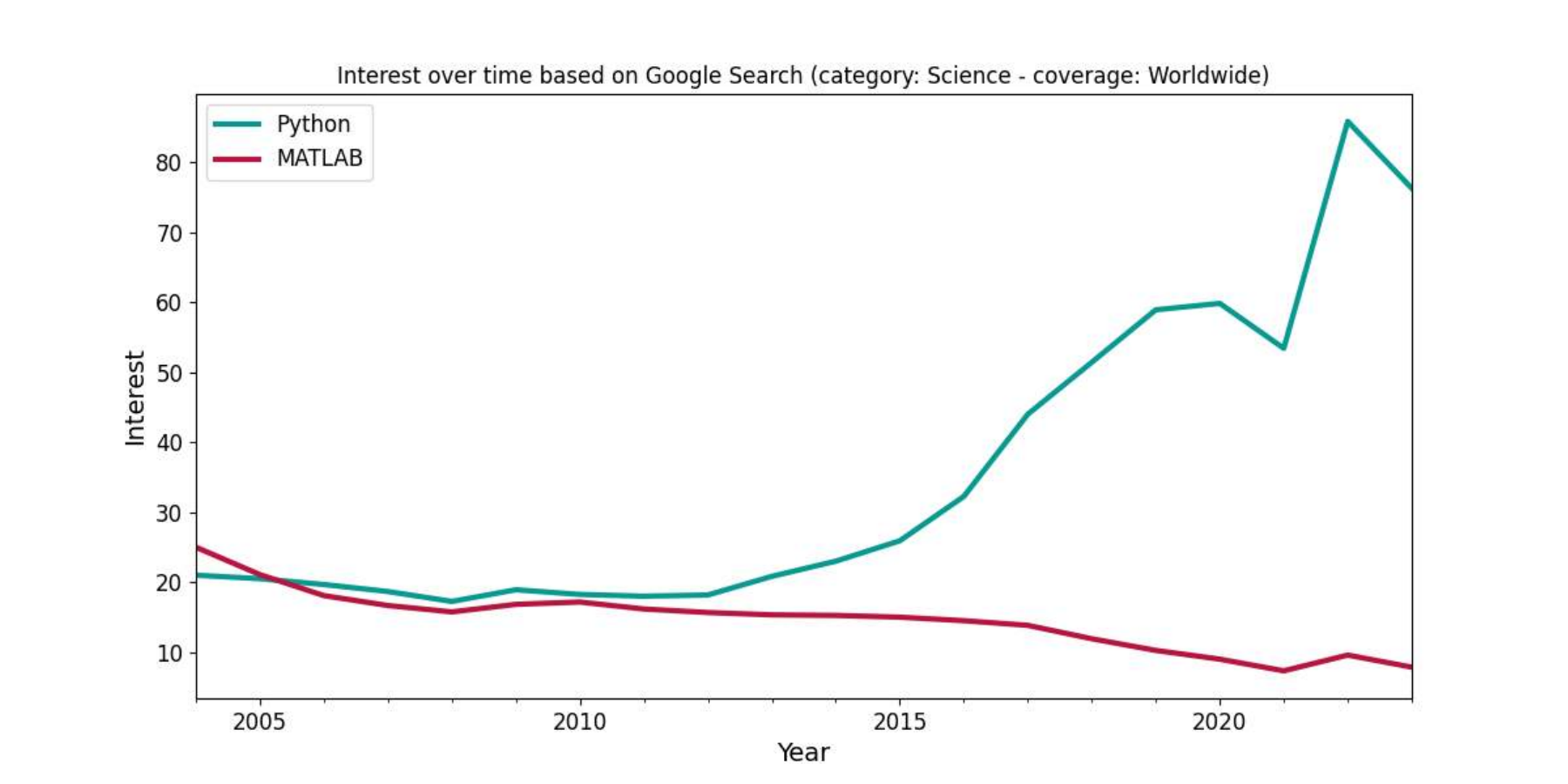}
    \caption{Worldwide interest in scientific programming languages over time according to Google Trends in the ``Science" category.}
    \label{fig:matlab}
\end{figure}

During the period when hyperspectral unmixing gained popularity, the primary scientific programming language of choice was MATLAB, as depicted in figure \ref{fig:matlab}. 
However, in recent times, there has been a steady increase in interest and adoption of Python, while MATLAB's popularity has waned. 
This shift can be attributed, in part, to the growing trend of open science, where open-sourcing scientific code has become widely embraced. 
Consequently, the present scenario suggests that developing and releasing an open-source unmixing package in Python is more advantageous than doing so in MATLAB, even though it may involve the translation of existing unmixing code from MATLAB to Python.

%\section{Contribution }
\subsection{Existing Overview Publications and Unmixing Packages}
An early survey on hyperspectral unmixing was given in \cite{Keshava2003A}, which discusses basic geometrical and statistical methods. In \cite{unmixing-review}, linear model-based unmixing techniques were divided into three categories: geometrical, statistical, and sparse regression-based approaches. A Matlab toolbox is available at https://openremotesensing.net/.  However, the toolbox is incomplete, and some methods, such as dependent component analysis (DECA), are missing. In recent years, deep learning and neural networks have become state-of-the-art for many tasks in machine learning and image processing. Consequently, many unmixing approaches were proposed based on shallow and deep neural networks. A comparison of autoencoder-based networks was drawn in \cite{Palsson_rev}. The authors discussed autoencoder-based architectures divided into five categories, i.e.,  Sparse Nonnegative Autoencoders,  Variational Autoencoders, Adversarial Autoencoders, Denoising Autoencoders, and Convolutional Autoencoders. They further discuss the choice of different modules, such as different activation functions or loss functions, and they compared shallow networks to deep ones. They also provide a TensorFlow-based Python package that is available on GitHub. However, the package is limited to blind unmixing approaches based on autoencoders. It does not discuss or compare the supervised, semisupervised, and more conventional blind unmixing methods. 

In \cite{PB_DD_Un}, some model-based and neural network-based unmixing approaches were explained but without experimental comparisons. A list of resources for the approaches was given.

%In \cite{SPM_rev}, 

A survey on endmember variability in Spectral Mixture Analysis (SMA) was given in \cite{End_Var_Rev}. In \cite{End_Var_Zare}, an overview of unmixing methods that address endmember variability was given. A comprehensive overview of the unmixing methods that address spectral variability was recently provided in \cite{Spec_var}, and a list of Matlab codes was also given. In \cite{End_extr_Plaza}, an overview of endmember extraction approaches was given. Review papers on hyperspectral remote sensing data analysis briefly discussed the unmixing methods \cite{Dias_HS_Rev2013, Ghamisi-review-2017}. In \cite{Rev_Tensor_HS}, an overview of tensor-based unmixing models was given. In \cite{Rob_NonRev}, a survey of nonlinear unmixing methods is given.

%\subsection{Existing Unmixing Tools and Packages}
%openremotesensing
There are other open-source tools such as HyperMix \cite{HyperMix}, Spectral Python (SPy) (https://www.spectralpython.net/), Spectral Library Tool (https://spectral-libraries.readthedocs.io/en/latest/), PySptools (https://pypi.org/project/pysptools/), that include basic algorithms for estimation of the number of endmembers, endmember extraction, abundance estimation, and some library tools, and library-based methods. Thus, there is a need for a comprehensive package that covers the methodologies across different unmixing categories and contains state-of-the-art image processing and machine learning techniques. %In contrast, we provide an open source python-based package which covers different unmixing categories and contains  both ML/DL and image processing state-of-the-art unmixing techniques. The package has unique features including completeness,  reproducibility, extensibility, and homogeneity which are discussed in Section \ref{sec: HySUPP} \textcolor{blue}{Alex, maybe you can add ath here}

 \subsection{Significance of Contribution}
This paper aims at the followings: 
\begin{itemize}
     \item Providing researchers with a comprehensive yet technical overview of all the essential topics regarding linear unmixing techniques. %Therefore, there is a trade-off between the technicality and the coverage of different techniques. 
     \item  Categorizing the unmixing approaches considering  prior knowledge available about endmembers.  Linear unmixing can be divided into three main categories:  supervised, semi-supervised (library-based), and unsupervised (blind) unmixing. 
     \item Compare the unmixing methods in terms of prior knowledge of the endmembers and draw conclusions that can help researchers to select an appropriate unmixing method to tackle real-world challenges. We compare conventional and deep learning-based unmixing approaches in those categories for three simulated and two real-world datasets.  For the simulated datasets, we consider three scenarios: a simple, pure pixel dataset, a dataset with spectral variations, and a challenging dataset with no pure pixel. Such comparisons provide insight to researchers into which category to use for their application. Additionally, the comparisons reveal the drawbacks of the categories, which motivate the developers to investigate new ideas to address them.
     \item  We provide an open-source HyperSpectral Unmixing Python Package (HySUPP). HySUPP is the first open-source python-based hyperspectral unmixing package to include supervised, semi-supervised, and blind unmixing methods.  The package will benefit the geoscience and remote sensing community, including researchers, developers, lecturers, and students. The package installation is straightforward since HySUPP relies on a few dependencies. In addition, all the methods can be run using a single command line instruction. 
 \end{itemize}

\section{HySUPP: HyperSpectral Unmixing Python Package}
\label{sec: HySUPP} 

%\textcolor{blue}{We need to explain the features of the package including an example. The current version of HySUPP contains 25 methods of which 20 methods are in Python and 5 can be run in a Python environment using a Matlab engine.} A table that shows which method we implemented and how much we contributed.

HySUPP exhibits a list of highly desirable properties summarized as follows: i) completeness, ii) reproducibility, iii) extensibility and iv) homogeneity. It implements common best practices and enables simple benchmarking of unmixing techniques thanks to user-friendly command line instructions.

\subsection{Features}

\paragraph{Completeness}

% As a practitioner, the ability to experiment with different unmixing techniques is crucial as one cannot expect one approach to consistently outperform the others in all unmixing scenarios. Therefore completeness is a sound requirement. Our toolbox covers all three types of unmixing (supervised, semi-supervised and unsupervised) and strives to be as representative of the various unmixing approaches as possible without aiming for exhaustiveness as this would not be feasible. As such, HySUPP currently gives access to 20 different unmixing methods (6 supervised, 6 semi-supervised and 8 unsupervised).

As a practitioner, the ability to explore and experiment with different unmixing techniques is crucial since no single approach can consistently outperform others in all unmixing scenarios. 
Thus, ensuring the completeness of our toolbox becomes essential. 
Our toolbox is designed to cover all three types of unmixing - supervised, semi-supervised, and unsupervised - while striving to be as representative as possible of the various unmixing approaches, although aiming for exhaustive inclusion would be impractical. 
Currently, HySUPP provides access to a diverse set of 20 different unmixing methods, including 6 supervised, 6 semi-supervised, and 8 unsupervised techniques.

\paragraph{Reproducibility}

%When it comes to experimenting with various unmixing techniques, experimental reproducibility is key to ensure that the conclusions drawn by users are robust. As a result, our toolbox enables experiments to be seeded, by making noise generation reproducible as well as automatically saving estimates output.

Ensuring experimental reproducibility is crucial when exploring various unmixing techniques, as it guarantees the robustness of the conclusions drawn by users. 
In line with this principle, our toolbox offers the ability to seed experiments, facilitating reproducibility through repeatable noise generation.
Additionally, HySUPP automatically saves estimates outputs, providing users with a convenient way to review and compare results, thus enhancing the reliability of their research findings.

\paragraph{Extensibility}

%HySUPP's architecture allows for seamless integration of new methods, thus facilitating future advancements in the field. We leverage configuration files to carry out experiments, while enabling users to easily plug their own model.

HySUPP's architecture is designed to support the effortless integration of new methods, providing a platform for future advancements in hyperspectral unmixing. 
Leveraging configuration files, we empower researchers to conduct experiments with ease and flexibility, enabling them to effortlessly incorporate their own models into the toolbox.

\paragraph{Homogeneity}

The uniformity of HySUPP's codebase ensures consistency across different components, enhancing its overall usability and reliability. 
More specifically, we use common methods for models and common attributes for datasets.

\paragraph{Best practices}

%We provide a simple yet powerful way to monitor each approach objective function using Python's \texttt{tqdm}. Moreover, our pipeline embeds proper endmembers auto-alignment using \texttt{munkres} algorithm as long as ground truth abundance maps are available to effectively compute unmixing performance. By establishing these best practices, we not only streamline the implementation process but also foster a collaborative environment where researchers can easily build upon existing work and share their contributions effectively.

We offer a straightforward yet potent method to monitor the objective function of each approach using Python's \texttt{tqdm} library. 
Furthermore, our pipeline incorporates precise endmember auto-alignment through the utilization of the \texttt{munkres} algorithm, provided ground truth abundance maps are accessible. 
This enhancement ensures the accurate computation of unmixing performance. 
By establishing these best practices, we streamline the implementation process and foster a collaborative environment where researchers can easily build upon existing work and share their contributions effectively.

\paragraph{Benchmarking}

%Owing to its rich set of features, HySUPP enables users to easily benchmark all implemented methods on their dataset of choice. Our toolbox currently provides 6 synthetic datasets corresponding to different unmixing scenarios, as well as 4 different metrics to comprehensively assess the various methods unmixing accuracy. Finally, leveraging \texttt{mlxp} results query tool, we empower users to analyze and visualize their results in a nice-looking manner.

Owing to its rich set of features, HySUPP enables users to easily benchmark all implemented methods on their dataset of choice. 
Our toolbox currently provides 3 synthetic datasets corresponding to different unmixing scenarios.
Moreover, we incorporate 4 distinct metrics to thoroughly evaluate unmixing accuracy across methods.
Finally, leveraging \texttt{mlxp}~\cite{Arbel2023MLXP} results query tool, we empower users to analyze and visualize their results in an appealing and informative manner.

\subsection{Example}

The following command line instruction provides an example on how to run a semi-supervised unmixing technique, \texttt{SUnCNN}~\cite{SUnCNN}, on the \texttt{DC1} dataset using an optional custom value for the signal-to-noise ratio (SNR):

\texttt{\$ python unmixing.py mode=semi data=DC1 model=SUnCNN noise.SNR=30}

%We summarize in table~\ref{tab:toolbox_desc} our implementation contributions of currently available unmixing methods. 
Table \ref{tab:toolbox_desc} lists unmixing methods included in HySUPP with their corresponding dependencies. We highlighted our main contributions and the link to the original implementations.
\begin{table*}[ht]\addtolength{\tabcolsep}{-3pt}
    \centering
    \caption{The list of unmixing methods included in HySUPP with their corresponding dependencies. We highlighted our main contributions and the link to the original implementations. }
    \begin{tabular}{c|c c c c c}
    \toprule
         Method & Original implementation & Python & Dependencies & GPU & Contributions \\
    \midrule
          FCLSU~\cite{FCLSU} & \href{https://pysptools.sourceforge.io/_modules/pysptools/abundance_maps/amaps.html#FCLS}{pysptools} & \checkmark & \texttt{numpy}, \texttt{cvxopt} & & Refactor into a \texttt{SupervisedUnmixingModel} sub-class\\
          SiVM~\cite{RHeylen_2011} & \href{https://github.com/BehnoodRasti/MiSiCNet/blob/main/UtilityMine.py}{github}& \checkmark & \texttt{numpy} & & Refactor into a \texttt{BaseExtractor} sub-class \\
          SISAL~\cite{SISAL} & \href{https://github.com/etienne-monier/lib-unmixing}{github} & \checkmark & \texttt{numpy} & & Refactor into a \texttt{BaseExtractor} sub-class \\
          UnDIP~\cite{UnDIP} & \href{https://github.com/BehnoodRasti/UnDIP}{github} & \checkmark & \texttt{torch} & \checkmark & Refactor separate scripts into a single model\\
          VCA~\cite{VCA} & \href{https://github.com/Laadr/VCA}{github} & \checkmark & \texttt{numpy} & & Replace \texttt{scipy} dependency by \texttt{numpy}. Add random seed\\
    \midrule
          CLSUnSAL~\cite{Collaborative}& \href{https://github.com/etienne-monier/lib-unmixing}{github}& \checkmark & \texttt{numpy} & & Implementation based on SUnSAL\\
          MUA\_SLIC~\cite{RABorsoi2019}& \href{https://github.com/ricardoborsoi/MUA\_SparseUnmixing}{github} & \checkmark & \texttt{numpy}, \texttt{skimage} & & MATLAB code translation using \texttt{skimage}'s SLIC\\
          S$^2$WSU~\cite{SZhang2018} & \href{https://github.com/ricardoborsoi/MUA\_SparseUnmixing}{github} & \checkmark & \texttt{numpy}, \texttt{scipy} & & MATLAB code translation\\
          SUnAA~\cite{SUnAA} & \href{https://github.com/inria-thoth/SUnAA}{github} & \checkmark & \texttt{numpy}, \texttt{spams} & & Refactor to match \texttt{SemiUnmixingModel} sub-class\\
          SUnCNN~\cite{SUnCNN} & \href{https://github.com/BehnoodRasti/SUnCNN}{github} & \checkmark & \texttt{torch} & \checkmark & Refactor separate scripts into a single model \\
          SUnSAL~\cite{SUnSAL} & \href{https://github.com/etienne-monier/lib-unmixing}{github} & \checkmark & \texttt{numpy} & & Refactor into \texttt{SemiUnmixingModel} sub-class \\
    \midrule
          ADMMNet~\cite{ADMMNet} & - & \checkmark & \texttt{torch} & \checkmark & Implemented from scratch\\
          BayesianSMA~\cite{BayesianSMA} & \href{https://ndobigeon.github.io/applications/app_hyper_SMA.html}{webpage} & \texttt{matlab.engine} & \texttt{numpy} & & Python wrapper around existing MATLAB code\\
          CNNAEU~\cite{CNNAEU} & \href{https://github.com/burknipalsson/hu\_autoencoders}{github}& \checkmark & \texttt{torch} & \checkmark & Convert existing \texttt{keras} implementation into \texttt{torch}\\
          EDAA~\cite{EDAA} & \href{https://github.com/inria-thoth/EDAA}{github} & \checkmark & \texttt{torch} & \checkmark & Refactor to match \texttt{BlindUnmixingModel} sub-class\\
          MiSiCNet~\cite{MiSiCNet} & \href{https://github.com/BehnoodRasti/MiSiCNet}{github}& \checkmark & \texttt{torch} & \checkmark & Refactor separate scripts into a single model \\
          MSNet~\cite{MSNet} & \href{https://github.com/yuyang95/JAG-MSNet}{github}& \checkmark & \texttt{torch} & \checkmark & Refactor to match \texttt{BlindUnmixingModel} sub-class \\
          NMFQMV~\cite{NMF_QMV} & \href{https://github.com/LinaZhuang/NMF-QMV\_demo}{github}& \texttt{matlab.engine} & \texttt{numpy} & & Python wrapper around existing MATLAB code \\
          PGMSU~\cite{PGMSU} & \href{https://github.com/shuaikaishi/PGMSU}{github}& \checkmark & \texttt{torch} & \checkmark & Refactor to match \texttt{BlindUnmixingModel} sub-class\\
    \bottomrule
    \end{tabular}
    \label{tab:toolbox_desc}
\end{table*}

\section{Linear Unmixing}
For frequent Earth observation, hyperspectral sensors are installed on satellites. The light passes through the atmosphere before reaching the target and then passes through the atmosphere to reach the sensor. The atmosphere absorbs and scatters light, and therefore, atmospheric corrections should be applied in addition to converting radiance to reflectance. Arguably, atmospheric corrections are unnecessary for airborne, drone-borne, and close-range imaging.

\begin{figure*} [htbp]
\centering
\includegraphics[width=.98\linewidth]{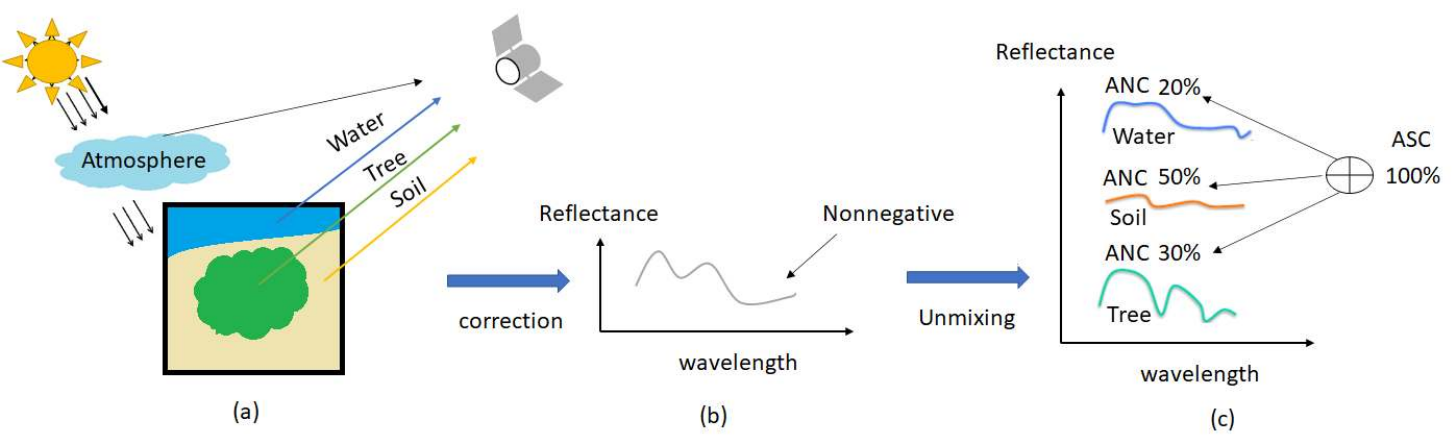}
\caption{Schematic of sensing and linear unmixing. (a) Sensing a mixed pixel (b) Reflectance of the mixed pixel and (c)  Schematic of the linear unmixing. }
\label{fig: Linabs}
\end{figure*}

Fig. \ref{fig: Linabs} (a) simplifies the concept of sensing using a satellite hyperspectral sensor. As can be seen, the sensor captures a pixel that contains three materials, i.e., Water, Tree, and Soil. At the sensor, irradiance is corrected to the reflectance in Fig. \ref{fig: Linabs} (b). Note that this reflectance is between zero and one.  Fig. \ref{fig: Linabs} (c) shows the concept of linear unmixing, which demonstrates that the measured pixel contains 20\% water (blue endmember), 50\% soil (yellow endmember), and 30\% tree (green endmember). Arguably, these portions cannot be negative, known as abundance nonnegativity constraint (ANC), and the summation of those portions is 100\%, known as abundance sum to one constraint (ASC). Additionally, the endmembers are reflectance and cannot be negative (indeed, the endmembers have to be bounded between 0 and 1), which led to the endmember nonnegativity constraint. Here, we discuss two widely used linear mixing models, i.e., the low-rank linear mixing model (so-called LMM) and the sparse and redundant linear model. The former was mainly used for supervised and blind unmixing and the latter for sparse unmixing. Some other linear models have been used in the literature, which will be discussed throughout the paper. We also discuss the components of a basic neural network model.

\begin{figure*} [htbp]
\centering
\includegraphics[width=.8\linewidth]{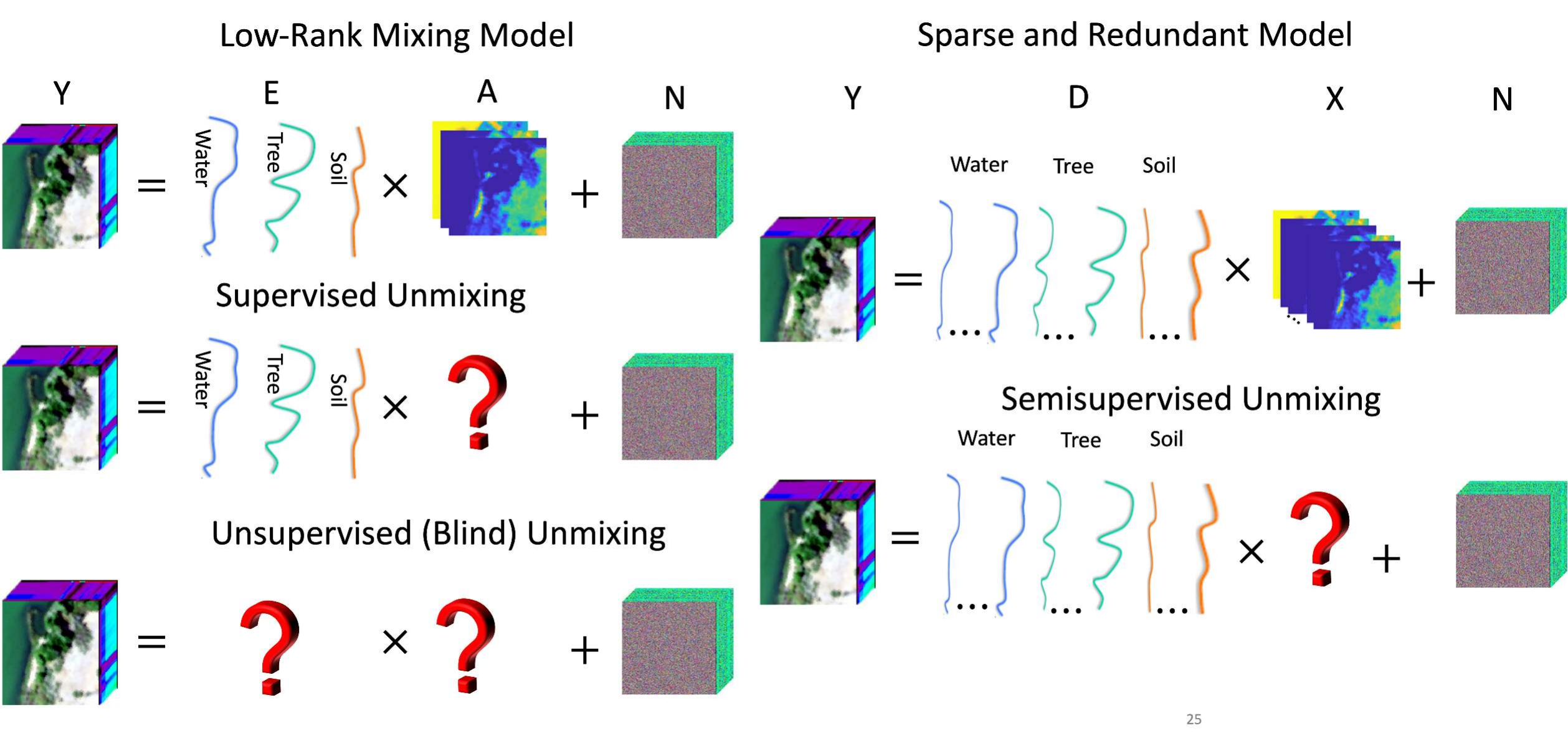}
\caption{Graphical representation of different types of linear unmixing. Supervised and blind unmixing use the same low-rank mixing model, while sparse unmixing uses the sparse and redundant linear model. Blind unmixing estimates both endmembers and abundances simultaneously.}
\label{fig: Graphabs2}
\end{figure*}

Unmixing techniques can be categorized into three main groups considering the prior knowledge of endmembers. Supervised and unsupervised (blind) unmixing use the same low-rank mixing model, while semisupervised unmixing uses the sparse and redundant linear model. The endmember library is overcomplete; consequently, the abundances are desired to be sparse. Blind unmixing estimates both endmembers and abundances simultaneously.  Fig. \ref{fig: Graphabs2} graphically compares those three groups.     
It is worth mentioning that unsupervised (blind) unmixing  can be applied to at sensor irradiance, however, the data cannot be interpreted and the endmembers extracted/estimated cannot be associated with the corresponding materials. One can apply supervised unmixing to irradiance if the pure pixels are also irradiance. Semisupervised  unmixing cannot be applied to sensor radiance since they rely on a spectral library.

\subsection{Low-rank Linear Mixture Model}
Every $p$-dimensional pixel ${\bf y}$ (i.e., the sensor has $p$ bands) captured by a hyperspectral sensor can be represented as a linear combination of the endmembers within the pixel. Let us assume that matrix ${\bf E}\in \mathbb{R}^{p\times r}$ contains $r$ endmembers within the pixel. Then,  ${\bf y} \in  \mathbb{R}^{p}$ is represented as
 \begin{equation}\label{eq: M0}
{\bf y} = {\bf E}{\bf a} + {\bf n}, ~~{\rm s.t.~~}\sum_{i=1}^ra_i=1, a_i\geq 0, i=1,2,..,r,
\end{equation} 
where ${\bf n}$ denotes the $p$-dimensional random vector denoting the additive random Gaussian noise. Using matrix notations, we can represent all the pixels ${\bf Y}$ as
 \begin{equation}\label{eq: LR}
{\bf Y} = {\bf E}{\bf A} + {\bf N}, ~~{\rm s.t.~~}{\bf A}\geq 0,{\bf 1}_{r}^{T}{\bf A}={\bf 1}_{n}^{T},
\end{equation} 

where {\bf Y}$ \in  \mathbb{R}^{p\times n}$ is the observed HSI, with $n$ pixels and $p$ bands, {\bf N}  $\in \mathbb{R}^{p\times n}$ is noise, and {\bf A} $\in \mathbb{R}^{r\times n}$  contain the $r$ endmembers and their fractional abundances, respectively. ${\bf 1}_n$ indicates an $n$-component column vector of ones. Model \ref{eq: LR} is known as the linear mixture model (LMM). %In blind unmixing scenarios, the task is to estimate both ${\bf E}$ and ${\bf A}$ simultaneously. 

\subsection{Low-rank LMM and Simplex Volume}

Suppose that the endmembers are affinely independent i.e., ${\bf e}_2-{\bf e}_1, ..., {\bf e}_r-{\bf e}_1$  are linearly independent, then 
\begin{equation}\label{eq: Data_S}
    \mathcal{S}\triangleq\{{\bf E}{\bf a}\in \mathbb{R}^{p}|\sum_{i=1}^ra_i=1, a_i\geq 0, i=1,2,..,r\}
\end{equation}
is $(r-1)$-Simplex.
Indeed, this is the convex hull of the vertices, i.e., $ {\bf e}_i$'s assuming no noise. Therefore, if we ignore the noise, any point of the dataset belongs to $ \mathcal{S}$ (see Fig. \ref{fig: Pure_No} (a)). We should note that columns of ${\bf A}$ belong to $(r-1)$-probability simplex 
\begin{equation}\label{eq: Ab_S}
    \Delta_r\triangleq\{{\bf a}\in \mathbb{R}^r|\sum_{i=1}^r a_i=1, a_i\geq 0, i=1,2,..,r\}.
\end{equation}
Therefore, ${\bf a}_{i}\in \Delta_r$ means ASC and ANC. Indeed, the vertices of $\Delta_r$ are $r$ unit vectors in $\mathbb{R}^r$. Hereafter, we refer to (\ref{eq: Data_S}) as data simplex and to (\ref{eq: Ab_S}) as abundance simplex. It is worth mentioning that, unlike data simplex, the abundance simplex is also valid for nonlinear models.

Let us assume $\mathcal{S}_r$ is an $r$-simplex in $\mathbb{R}^r$ i.e., ${\bf e}_i \in \mathbb{R}^r$ for $i=0,1,2,..,r$. Then the volume of this simplex is given by %(note that ${\bf E} \in \mathbb{R}^{r \times r+1}$)
\begin{equation}
    Vol( \mathcal{S}_r)=\frac{1}{r!}|det[{\bf e}_1- {\bf e}_0, ... , {\bf e}_r- {\bf e}_0]|=\frac{1}{r!}|det\begin{bmatrix}
    1 & ... & 1\\
 {\bf e}_0 & ... &  {\bf e}_r 
\end{bmatrix}|,
\end{equation}
where $det$ is the determinant of a matrix. In the linear mixture model (LMM), we have an $(r-1)$-Simplex in $\mathbb{R}^r$ (after the projection into an $r$ dimensional subspace, see \ref{subsec:SubPro}) i.e., ${\bf e}_i \in \mathbb{R}^r$ for $i=1,2,..,r$  (${\bf E}_r \in \mathbb{R}^{r \times r}$). The volume of an $(r-1)$-simplex is zero in $\mathbb{R}^r$. Therefore, the extended simplex, which contains the origin (i.e., ${\bf E}_0=[{\bf 0},{\bf E}]$), can be used to calculate the volume. Hence,
\begin{equation}
    Vol({\bf E}_0)=\frac{1}{r!}|det[{\bf e}_1, ... , {\bf e}_r]|=\frac{1}{r!}|det({\bf E})|.
\end{equation}

Alternatively, the data points can be shifted to the origin, and the volume of the  $(r-1)$-simplex can be computed in $\mathbb{R}^{r-1}$ given by
\begin{align}\nonumber
    Vol({\bf E}_r)=&\frac{1}{(r-1)!}|det[{\bf e}_2- {\bf e}_1, ... , {\bf e}_r- {\bf e}_1]|=\\& \frac{1}{(r-1)!}|det\begin{bmatrix}
    1 & ... & 1\\
 {\bf e}_1 & ... &  {\bf e}_r 
\end{bmatrix}|.
\end{align}
It is worth mentioning that the projection into a subspace (\ref{subsec:SubPro}) is necessary to form a squared matrix and calculate the determinant. 
%This is geometrical approaches which were extensively discussed in \cite{unmixing-review}, rely on estimation or exploits 

\subsection{Sparse and Redundant Linear Mixture Model}

The observed spectra can be represented using a sparse and redundant linear mixture model given by 
 \begin{align}\label{eq: SR}\nonumber
&{\bf Y} = {\bf D}{\bf X} + {\bf N}, \\&
~~~{\rm s.t.}~~~{\bf X}\geq 0,{\bf 1}_{m}^{T}{\bf X}={\bf 1}_{n}^{T},  
\end{align} 
where $ {\bf D}  \in \mathbb{R}^{p\times m} $  ($p\ll m$) denotes the spectral library containing $m$ endmembers and $ {\bf X}\in \mathbb{R}^{m\times n}$ is the unknown  fractional abundances to estimate. Note that ${\bf D}$ is an overcomplete dictionary and therefore, should be a well-designed dictionary. A well-designed dictionary contains endmembers of the materials in the scene and can sparsify the redundant ${\bf X}$. Hence, a spectral library can be pruned based on the spectral angles of the spectra (i.e., spectra with small degree differences will be removed). However, this comes with a risk of losing material endmembers if they are scaled versions of each other.  In the case of a well-designed dictionary, the pixels are a mixture of a few atoms of the dictionary, and therefore ${\bf X}$ is a sparse matrix. Note that we get a zero row if an endmember material does not exist in the observed spectra. The rest of the rows will contain zeros since the abundances are often sparse.  This model is often used in sparse unmixing. The fractional abundances ${\bf X}$  are estimated using sparsity-enforcing penalties/constraints in a sparse regression formulation. 
\subsection{Shallow/Deep Neural Network Model}
\label{SD_model}
\begin{figure} [htbp]
\centering
\includegraphics[width=.9\linewidth]{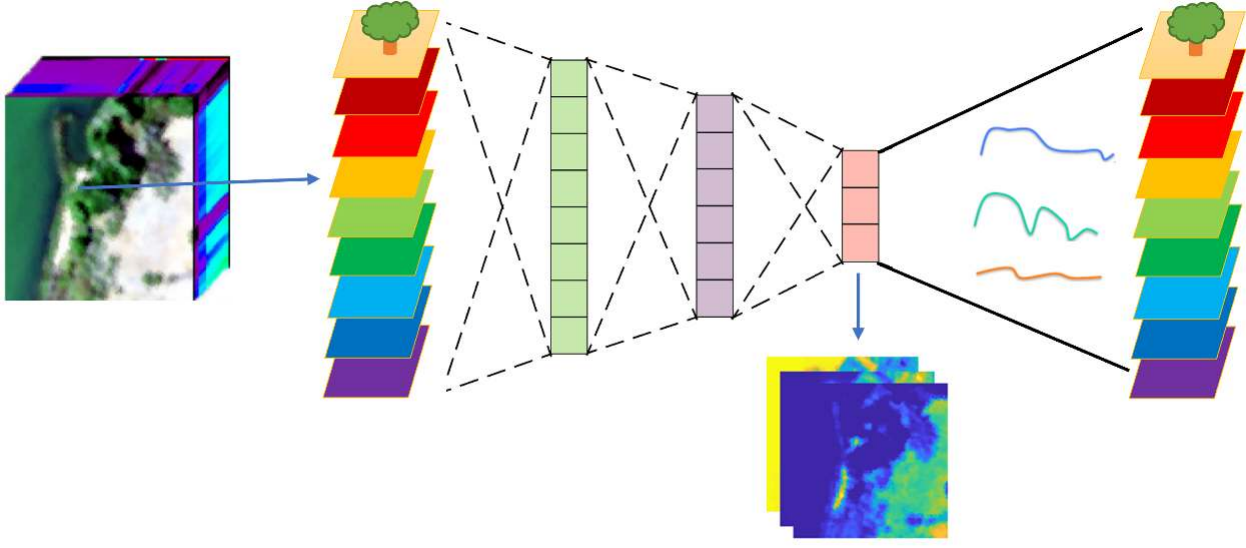}
\caption{A basic encoder-decode architecture used for unmixing. The (shallow or deep) encoder encodes the pixels to abundance. ASC and ANC can be enforced on the bottleneck. The shallow decoder reconstructs the pixels, and the decoder's weights are the endmembers.}
\label{fig: EDU}
\end{figure}

Shallow or deep neural networks can be used for supervised, semi-supervised, or blind unmixing. The network is often based on autoencoder-based architecture, as shown in Fig. \ref{fig: EDU}. The encoder $\mathscr{E}$, which could be deep or shallow, encodes the spectral pixels into the abundances given by
  \begin{equation}
      {\bf a} = \mathscr{E}({\bf y}),
  \end{equation}
and the decoder $\mathscr{D}$ reconstruct the pixel given by
\begin{equation}
      \hat{\bf y} = \mathscr{D}({\bf a}).
  \end{equation}
  In every layer, an activation function is used to apply nonlinearity. The most common choices are Rectified Linear Unit ReLU (ReLU$(x)=max(0,x)$) and Leaky ReLU (LReLU$(x)=max(ax,x)$, where $a$ is very small, and a common choice is $a=0.1$).  Batch normalization (BN) is commonly used in each layer to speed up the learning process. The BN function is given by 
\begin{equation}
    BN(x)=\alpha_2 \frac{x-\mu}{\sqrt{\sigma^2+\epsilon}} + \alpha_1,
\end{equation}
  where $\alpha_2$ and $\alpha_1$ are learnable parameters, $\mu$ and $\sigma$ are the mean and standard deviation of the batch, and $\epsilon$ is a minimal value. 
  
 A linear layer can be used for the decoder to reconstruct the pixel. The decoder weights are the endmembers fixed for supervised unmixing and can be learned through the learning process in the case of unsupervised (blind) unmixing. ASC and ANC can be realized differently; Applying nonnegative activation functions such as ReLU and normalization for ASC given by \cite{Palsson_rev}
  \begin{equation}
      {\bf a} = \frac{{\bf a}}{\sum_{i=1}^ra_i},
  \end{equation}
alternatively, one can use
  \begin{equation}
      {\bf a} = \frac{|{\bf a}|}{\sum_{i=1}^r|a_i|},
  \end{equation}
A common way is to use a softmax function
\begin{equation}
    softmax({\bf a})=\frac{e^{{\bf a}}}{\sum^r_{i=1}e^{a_i}}.
\end{equation}
The scaled version of the softmax can be used as $ softmax(\gamma{\bf a}_i)$ where $\gamma>1$ is constant. ASC and ANC can also be implemented by adding a penalty term to the loss function. However, it degrades the stability of the training process and decreases the convergence time, and therefore, it is not recommended \cite{Palsson_rev}. We should note that those functions are realized as the last hidden layer of the decoder. The loss function is often chosen as a reconstruction term plus a regularization term given by 
\begin{equation}
    \L_{reg}(\y,\hat\y)=\L(\y,\hat\y)+ \lambda J({\bf a},\hat\W_{en},\hat\W_{de}),
\end{equation}
where $\L$ is often selected as a reconstruction term such as $\ell_2$ norm (or MSE)
\begin{equation}
\L_{\ell_2}(\y,\hat\y)=||\y-\hat\y||_2^2,
\end{equation}
or 
\begin{equation}
\L_{SAD}(\y,\hat\y)=arccos\frac{\y^T\hat\y}{||\y||_2||\hat\y||_2},
\end{equation}
or
\begin{equation}
\L_{SID}(\y,\hat\y)=\sum_{i=1}^pp_ilog(\frac{p_i}{q_i})+\sum_{i=1}^pq_ilog(\frac{q_i}{p_i}),
\end{equation}
where 
\begin{equation}
p_i=   \frac{y_i}{\sum_{i=1}^py_i},~~q_i=   \frac{\hat y_i}{\sum_{i=1}^p\hat y_i},
\end{equation}
or a combination of them. %The regularizer $J$ can be a combination of a Frobenius norm on $\W_{en}$ with  sparsity norms such as  $\ell_1$ norm on ${\bf a}$. 
The regularizer $J$ can be a combination of several functions, such as a Frobenius norm on $\W_{en}$ with a sparsity norm such as  $\ell_1$ norm on ${\bf a}$, and a Frobenius norm or a geometrical (MV induced) norm on $\W_{de}$. Note that SAD is scale-invariant and, therefore, can capture the spectral variability of the pixel. This is a big advantage for real-world datasets. On the other hand, utilizing SAD for datasets containing materials having scaled endmembers is not recommended since the network cannot be trained to distinguish those materials. Assuming a single-layer encoder-decoder, the abundances are given by
  \begin{equation}
      {\bf a} = softmax(BN(RelU({\bf W}_{en}\y+{\bf b}))),
  \end{equation}
where ${\bf b}$ denotes the biases of the layer. The reconstructed pixel is given by 
\begin{equation}
      \hat{\bf y} = \W_{de}{\bf a}.
  \end{equation}

\subsection{The Effect of Noise in Unmixing}
Hyperspectral data are usually downgraded by different sources, including atmospheric effects, imaging artifacts, and instrumental noise. These sources can further affect hyperspectral analysis, including unmixing \cite{BR_Rev1}. Unmixing is an inverse problem that could be very sensitive to noise, mainly when endmembers are highly correlated. Therefore, unmixing techniques may fail due to noise. In \cite{unmixing-review}, so-called signal-to-noise-ratio spectral distribution (SNR-SD) was suggested to determine if the unmixing inverse problem gives acceptable results. However, the hyperspectral denoising field has considerably evolved during the past decade \cite{BR_Rev1, HyDe, BR_rev_2021} and noise reduction as a preprocessing step could improve the unmixing performance \cite{BR_UnDN}.  We should note that unmixing techniques usually consider Gaussian noise and even can perform as a denoiser but are not as efficient as denoising techniques \cite{BR_UnDN}. Additionally, other types of noise, such as strip and sparse, can be removed by applying denoising before unmixing \cite{BR_rev_2021}. Alternatively, some methods were proposed for performing denoising and unmixing in a unified framework for boosting the performance of each other \cite{TInce_2019, JYang_2016}. As a result, we suggest applying a well-established noise reduction technique before estimating the abundance or characterizing the unmixing inverse problem using SNR-SD.  

\subsection{Unmixing Approaches w.r.t. Prior Knowledge of Endmembers}
\begin{figure*} [htbp]
\centering
\includegraphics[width=1\linewidth]{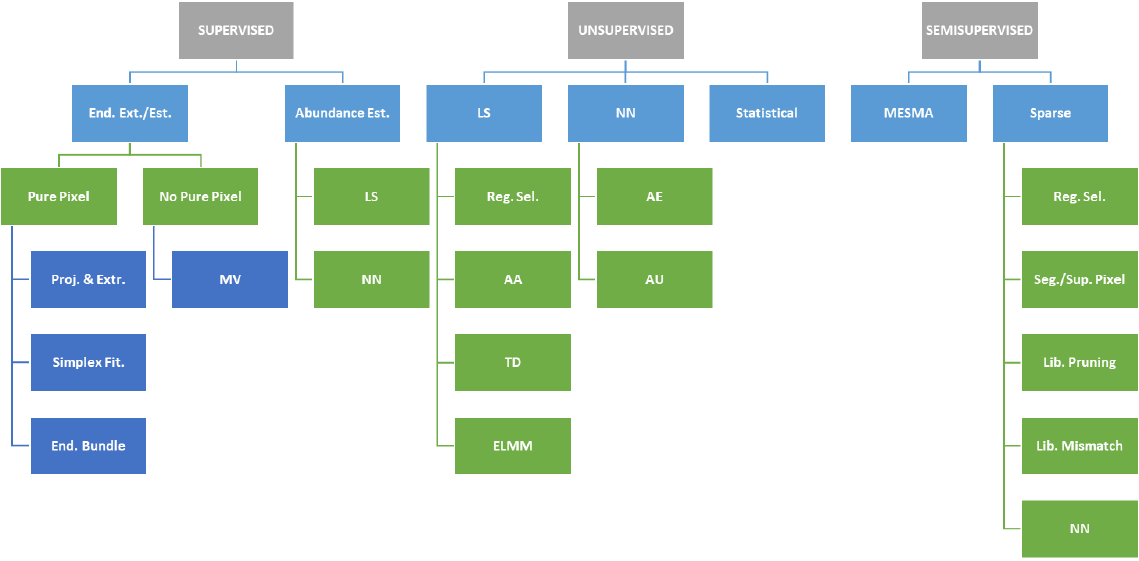}
\caption{\textcolor{black}{Taxonomy of the linear unmixing methods described in this paper. The top categorization is based on the prior knowledge of endmembers.}}
\label{fig: Tax}
\end{figure*}

Unmixing techniques can be categorized in terms of prior knowledge of the endmembers. Depending on the prior knowledge available about endmembers, the unmixing problem can be divided into three main categories: (1) supervised unmixing, (2)  unsupervised (blind) unmixing, and (3) Semi-supervised unmixing. In supervised unmixing, abundances are estimated by relying on known endmembers, whereas blind unmixing estimates both the endmembers and the abundances simultaneously. Semi-supervised unmixing relies on a library of endmembers that ideally contains the endmembers in the scene. We discuss each category separately.  Fig. \ref{fig: Graphabs2} demonstrates the graphical abstracts of these three linear unmixing categories. The taxonomy of the linear unmixing methods discussed in this paper is given in Fig. \ref{fig: Tax}.

\section{Supervised Unmixing}
\begin{figure*} [htbp]
\centering
\includegraphics[width=.78\linewidth]{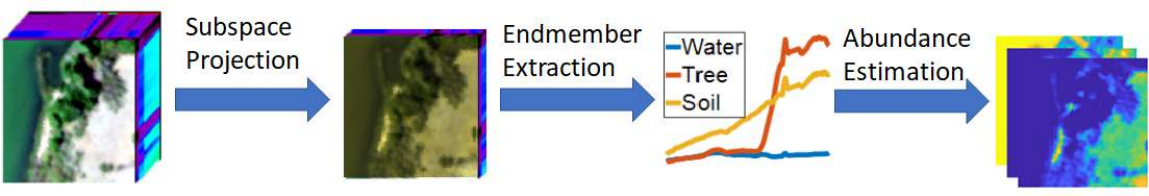}
\caption{The processing chain of supervised unmixing. First, the data is projected into a subspace. Then, endmembers are extracted using a geometrical approach. If the endmembers are available, then the endmember extraction can be skipped. In the final step, abundances can be estimated using an ML/DL-based approach.}
\label{fig: SUabs}
\end{figure*}

In supervised unmixing, we assume that endmembers are known, and the abundance matrix needs to be estimated. In practice, the endmember can be measured in the field or laboratory. One can select them from a spectral library, however, due to the variations in the imaging setup, such a selection will not lead to a desirable abundance estimation. Alternatively, they can be extracted/estimated from data points. Geometrical approaches are often used at this step. Endmember extraction is not often easy since the captured data might not contain pure pixels to represent all the endmembers in the scene. Most endmember extraction methods rely on pure pixels or some pixels on the facet of the simplex. Geometrical approaches are less effective in no pure pixel scenarios. The processing chain of supervised unmixing is shown in Fig. \ref{fig: SUabs}. Note that % this processing chain is known as a hyperspectral unmixing processing chain in \cite{unmixing-review}, however, 
we classify methods based on such sequential process (as opposed to simultaneous estimation of endmembers and abundances) as supervised unmixing since in the step of abundance estimation the endmembers are assumed to be known. The processing chain often includes three main steps; 1) Subspace projection: the data is projected into a subspace, 2) endmember extraction/estimation: Often a geometrical approach is used to extract the endmembers, and 3) Abundance estimation: abundances can be estimated using a least square (LS) or a neural network (NN) -based approach. 

\subsection{Subspace Projection and Estimation of number of Endmembers}
\label{subsec:SubPro}
Hyperspectral data often live in subspace having dimension much lower than the dimension of spectral bands defined by the sensor. Assuming $r$ endmembers in the scene, the intrinsic/subspace dimension is $r-1$, i.e., the data points can be represented by $r-1$ linearly independent vectors or bases (in the case of orthogonal projections). Therefore, identifying such a subspace and projecting the data into that reduces the computational cost, memory consumption, and removes the noise and outliers.  

The subspace projection is given by 
 \begin{equation}\label{eq: sub}
{\bf Y} = {\bf V}^T{\bf F} + {\bf N},% ~~{\rm s.t.~~}{\bf V}^T{\bf V}={\bf I}_{r-1}
\end{equation} 
where columns $ {\bf V} \in  \mathbb{R}^{p\times r}$ spans the subspace and columns of ${\bf F}$ are the projected spectral pixels. When $ {\bf V}$ is semi orthogonal, i.e., ${\bf V}^T{\bf V}={\bf I}_{r}$, the project is orthogonal which is a common choice in HS subspace projection. In (\ref{eq: sub}), both ${\bf V}$ and ${\bf F}$ should be estimated \cite{WSRRR, OTVCA}. Alternately, we can assume \cite{unmixing-review}
 \begin{equation}\label{eq: sub2}
{\bf Y}_r = {\bf V}^T{\bf EA} + {\bf V}^T{\bf N},% ~~{\rm s.t.~~}{\bf V}^T{\bf V}={\bf I}_{r-1}
\end{equation} 
where ${\bf Y}_r = {\bf V}^T{\bf Y}$. Reduction methods such as minimum noise fraction (MNF), principal component analysis (PCA), and noise adjusted principal component (NAPC) estimate ${\bf V}$ \cite{FE_Rev_BR20}. Not that ${\bf Y}$ is usually not a square matrix; hence, singular value decomposition (SVD) can be used to estimate ${\bf V}$ \cite{HySime, HySURE}. Please see \cite{FE_Rev_BR20, Ghamisi-review-2017} for more details on hyperspectral feature reduction.  

Estimating the number of endmembers is not a trivial task. Unmixing approaches are vulnerable to this parameter and under or overestimating $r$ may considerably affect the error of the models. We should note that sparse regression-based methods which do not use endmember bundles do not rely on the estimation of $r$. In the literature, this problem was addressed with 
 alternative names such as hyperspectral subspace identification, intrinsic
order selection, virtual dimension, and estimation of the number of spectrally distinct
signal sources \cite{HySime,HySURE, Sub_id_Chang_Du}. This problem can be addressed using eigenvalue-based detection techniques \cite{HFC,Sub_id_Chang_Du,HFC_2} or estimating the mean square errors \cite{HySime, HySURE}. Geometrical-based approaches were also proposed for endmember estimation\cite{ICE,Zare_ICE,GENE}. 
%  \begin{equation}\label{eq: SVD}
% {\bf Y} = {\bf V}{\bf S}{\bf U}^T%+ {\bf V}^T{\bf N},% ~~{\rm s.t.~~}{\bf V}^T{\bf V}={\bf I}_{r-1}
% \end{equation}
% and 
\subsection{Endmember Extraction for Linear Unmixing}
\begin{figure} [htbp]
\centering
\begin{tabular}{cc} 
\includegraphics[width=.4\linewidth]{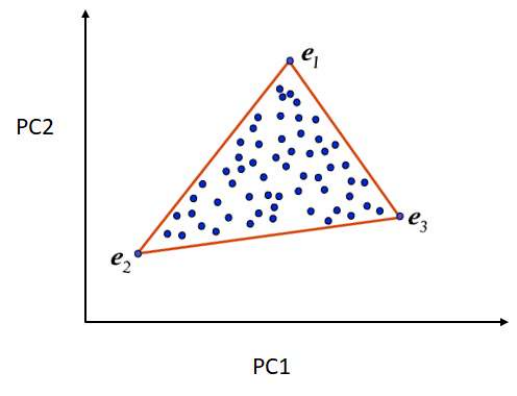}&\includegraphics[width=.4\linewidth]{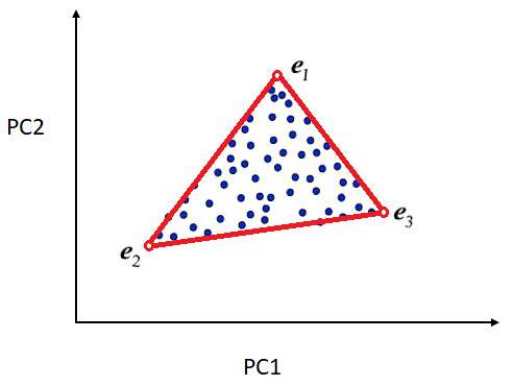}\\  (a) &(b) \\\includegraphics[width=.4\linewidth]{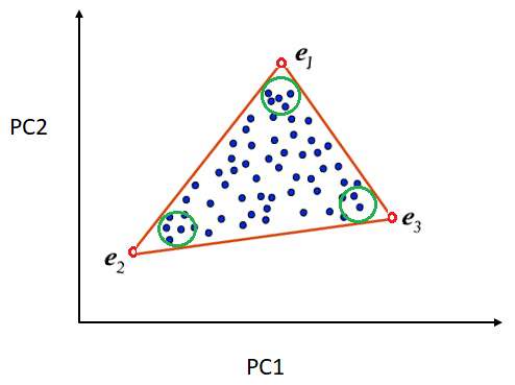}&\includegraphics[width=.4\linewidth]{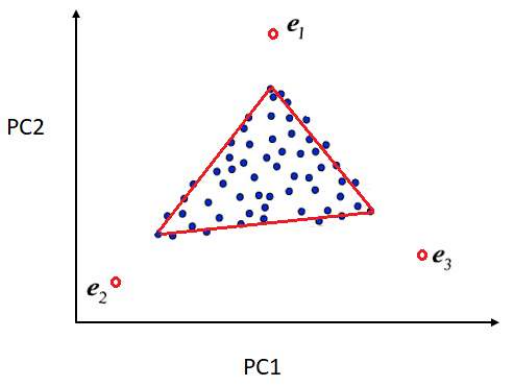}\\ (c)& (d)
 \end{tabular} %\end{center} 
\caption{Pure pixel versus no pure pixel assumption. (a) Pure pixels scenario (b) No pure pixels scenario with enough pixels on the facets of the simplex (c) endmember bundle, and (d) No pure pixels scenario without pixels on the facets of the simplex (highly mixed scenario).}
\label{fig: Pure_No}
\end{figure}

As can be seen in Fig. \ref{fig: Pure_No} (a), the endmembers are the vertices of the simplex enclosing the data points. Therefore, the geometrical concept inspires the mainstream of endmember extraction techniques. Geometrical approaches can be divided into two main groups; pure pixel-based approaches and approaches without pure pixel assumption. Fig. \ref{fig: Pure_No} compares the pure pixel with no pure pixel assumption. In Fig. \ref{fig: Pure_No} (b), the red circles demonstrate the positions of the missing pure pixels. In this case, the endmembers  cannot be extracted and they should be estimated. 
This is feasible using geometrical approaches if enough pixels are located on the facets of the simplex. This becomes much more challenging for higher dimensional cases when noise and other nonlinearities are involved. Alternatively, a group of endmembers (endmember bundle) can be extracted based on the assumption of endmember variability caused by spectral variability as shown in Fig. \ref{fig: Pure_No} (c).  Fig. \ref{fig: Pure_No} (d) shows the highly mixed scenario that geometrical approaches cannot cope with.

\subsubsection{Pure Pixel Assumption}
Many techniques consider pure pixel assumption for simplicity, while in real applications, pure pixels of some endmembers are often missing. The methods that rely on pure pixels for endmember extraction can be divided into three main groups: projections and extremes, simplex fitting methods, and multiple endmember extraction methods (endmember bundles).
\begin{itemize}
    \item Projections and Extremes: This group often searches for extremes by iteratively projecting data points. The vertices can be selected as the extreme points after iteratively projecting the data in a particular direction. For instance, Pixel Purity Index (PPI) \cite{PPI} scores the spectral vectors by projecting them onto a large set of random vectors (called skewers) and counting the number of times that each vector is an extreme point. Orthogonal subspace projection (OSP) \cite{OSP} and Vertex Component Analysis (VCA) \cite{VCA} selects endmembers iteratively by projecting the data into an orthogonal direction to the subspace spanned by the already selected endmember.
    \item Simplex Fitting: The endmembers are assumed to be located at the vertices of the simplex enclosing the data points. Therefore, they can be extracted by maximizing the data simplex. N-FINDR \cite{N-FINDR} searches for pure pixels that form the largest simplex by gradually inflating a simplex inside the data. Simplex volume maximization (SiVM) \cite{RHeylen_2011} extracts the endmembers by iteratively maximizing the simplex volume using 
\begin{equation}\label{eq: SiVM}
 \arg\max_{{\bf E}_r}V(\bf{E}_r) = \arg\max_{{\bf E}_r}\sqrt{\frac{(-1)^r \cdot \textup{ cmd }(\mathbf{E}_r)}{2^{r -1}(r -1)!}}\textup{, }
 \end{equation}
 where $\textup{cmd}$ is the Cayley–Menger determinant:
 \begin{equation}\nonumber
 \textup{ cmd }(\mathbf{E}_r)= \det {\tiny 
 \begin{bmatrix} 0 &{}\quad 1 &{}\quad 1 &{}\quad 1 &{}\quad \ldots &{}\quad 1\\ 1 &{}\quad 0 &{}\quad d_{1,2}^2 &{}\quad d_{1,3}^2 &{}\quad \ldots &{}\quad d_{1,r }^2\\ 1 &{}\quad d_{2,1}^2 &{}\quad 0 &{}\quad d_{2,3}^2 &{}\quad \ldots &{}\quad d_{2,r }^2\\ 1 &{}\quad d_{3,1}^2 &{}\quad d_{3,2}^2 &{}\quad 0 &{}\quad \ldots &{}\quad d_{3,r }^2\\ \vdots &{}\quad \vdots &{}\quad \vdots &{}\quad \vdots &{}\quad \ddots &{}\quad \vdots \\ 1 &{}\quad d_{r ,1}^2 &{}\quad d_{r ,2}^2 &{}\quad d_{r ,3}^2 &{}\quad \ldots &{}\quad 0 
 \end{bmatrix}},
\end{equation} 
and $d_{i,j}$ is the Euclidean distance between endmembers $\mathbf{e}_{i}$ and $\mathbf{e}_{j}$. 

\item Multiple Pure Pixels and Endmember Bundles: 
The multiple endmembers were also referred to as the endmember bundles.  The endmember bundle and bundle unmixing were proposed in \cite{MESM, End_Bund1, End_Bund2}. PCA was used to identify the (projected) endmembers using the extremes. More than one extreme was selected for materials. The bundle unmixing was formed using linear programming to determine the minimum, average, and maximum of the fractional abundances induced by endmember variability. In \cite{End_bundle}, endmember bundles were extracted from the observed data. Endmember variability is modeled using the endmember bundle, which will be discussed in more detail in semisupervised unmixing.

Some techniques incorporate spatial information for extracting the endmembers \cite{AMEE, SSEE, SSEE_2,SSEE_3}. These methods often utilize information from neighboring pixels in the endmember extraction. For instance, in \cite{SPP}, the similarity of the pixel and its neighbors was used to weigh the pixel and consequently adjust the simplex so that its vertices better represent the homogeneous regions in natural images. The primary assumption of these methods is that the neighboring pixels of pure pixels are also pure or have some purity level that can boost the endmember extraction.

\end{itemize}

\subsubsection{No Pure Pixel Assumption}
If there are no pure pixels, virtual endmembers can be successfully located at the vertices if enough data points are available on the facets of the data simplex. Here, the main idea is to minimize the simplex volume subject to a convex combination of the endmembers containing the data points. For instance, the minimum volume enclosing simplex (MVES) \cite{MVES,MVES2} seeks the minimum simplex, which encloses the data points. Maximum Volume Inscribed Ellipsoid (MVIE)  \cite{MVIE} seeks the maximum volume ellipsoid contained within the convex hull of the data points. This volume  is maximal when touching the facets of the data simplex. The contact points  aid in estimating the virtual endmembers.

    Alternatively, the minimum volume simplex analysis (MVSA) \cite{MVSA} and the simplex identification via variable splitting and augmented Lagrangian (SISAL) \cite{MVSA,SISAL} form a non-convex minimization (or a non-concave maximization) problem which estimates endmembers by minimizing volume data simplex, subject to the ASC and ANC \cite{MVSA,SISAL}. Assuming ${\bf Q}={\bf E}_r^{-1}$, SISAL solves
   \begin{align}\label{eq: SISAL}\nonumber
  &\hat{\bf Q}=\arg\min_{{\bf Q}} -log|det(Q)|+\lambda{\1}_r^Thinge({\bf Q}{\bf Y}_r){\1}_n\\&
{\rm s.t.}~~{\bf 1}_{r}^{T}{\bf Q}={\bf 1}_{n}^{T}{\bf Y}^{-1}_r. 
\end{align}
    where the hinge function ($hinge(x)=max\{-x,0\}$) is the regularization allowing violations to the non-negativity constraint and makes SISAL robust to noise. SISAL and MVSA solve the same problem however, SISAL is more efficient since it uses an alternating direction method of multipliers (ADMM) \cite{ADMM}.
    
%    We should mention that the unmixing review \cite{unmixing-review} extensively discusses the endmember extraction/estimation under the category of geometrical approaches. Therefore, we do not repeat those methods and discussions. Here, we aim to discuss a few well-known and widely-used approaches, including those which will be used in the experimental section of this paper. For more detail on endmember extraction/estimation using geometrical approaches, we refer to \cite{unmixing-review}.

%    Some approach also incorporates spatial information so-called spatial-spectral endmember extraction \cite{End_extr_Plaza}.  They
%    \textcolor{blue}{a few more endmember extraction, from Zare, recent method see \cite{End_extr_Plaza}}

\subsection{Abundance Estimation}

\subsubsection{Least Squares-based Approaches}

In supervised unmixing, when the endmember ${\bf E}$ are known, the abundances should be estimated. This step is also known as inversion, referring to the inverse problem of estimating abundances when the endmembers are known. The presence of noise, inevitable errors in endmember estimation/extraction, and the physical constraints make such an inverse problem very challenging. 

Unconstrained least squares unmixing (UCLSU) via the orthogonal subspace projection was proposed for abundance estimation \cite{ULS_CI}. Non-negative constrained least squares unmixing (NCLSU) \cite{NCLS, NCLSU} was proposed to estimate the abundances subjected to ANC.  A weighted least square is suggested in \cite{WLS} for estimating the abundances of multispectral remotely sensed data. There are several attempts to solve the least squares problem subjected to both ANC and ASC \cite{CLS, WLS}. In \cite{WLS}, quadratic programming was suggested for the constrained least squares. The first efficient algorithm was proposed in \cite{FCLSU} and called fully constrained least squares unmixing (FCLSU). Later, fully constrained least squares unmixing by simplex projection was proposed \cite{FCLSU_Simplex} to derive a more efficient algorithm. A recursive algorithm minimizes the least square by performing orthogonal projections while holding the ASC and ANC. However, we should note that with the advances in graphical processing units (GPU), FCLSU can be efficiently solved using convex optimization techniques. Therefore, FCLSU is the most widely used method for abundance estimation given by: 
\begin{align}\label{eq: FCLSU}\nonumber
  &\hat{\bf A}=\arg\min_{{\bf A}} \frac{1}{2} || {\bf Y}-{\bf EA}||_{F}^{2} \\&
{\rm s.t.}{\bf A}\geq 0,{\bf 1}_{r}^{T}{\bf A}={\bf 1}_{n}^{T}. 
\end{align}
where the $||.||_F$ denotes the Frobenius norm. Problem (\ref{eq: FCLSU}) is convex and can be solved using any convex optimization solver. %Here, we provide a fast Python-based solver ..... cite,   which was accelerated using GPU. 
To promote spatial regularization, one can solve a penalized least square for abundance estimation given by
\begin{align}\label{eq: ab_est}\nonumber
  &\hat{\bf A}=\arg\min_{{\bf A}} \frac{1}{2} || {\bf Y}-{\bf EA}||_{F}^{2} +\lambda\phi({\bf A})\\&
{\rm s.t.}{\bf A}\geq 0,{\bf 1}_{r}^{T}{\bf A}={\bf 1}_{n}^{T}, 
\end{align}
where the first term is the fidelity term. The penalty term is defined based on function $\phi$ applied to the unknown abundances. $\phi$ can be a total variation or sparsity-based function that usually incorporates spatial information. $\lambda$ controls the trade-off between the fidelity and penalty terms. Problem (\ref{eq: ab_est}) is convex (${\bf E}$ is known) as long as the penalty term is convex and, therefore, it can be solved using any convex optimization, least squares, or quadratic programming solver. Problem (\ref{eq: ab_est}) will lead to better estimation in terms of RMSE compared to (\ref{eq: FCLSU}) for some values of $\lambda$. However, the selection of optimum $\lambda$ is not trivial.

\subsubsection{Shallow/Deep Neural Networks}
In supervised unmixing networks, the endmembers or the decoder weights are fixed. Therefore, a regularized loss function can be given by
\begin{equation}
    \L_{reg}(\y,\hat\y)=\L(\y,\hat\y)+ \lambda J({\bf a},\W_{en}).
\end{equation}
The regularizer $J$ can be a combination of a Frobenius norm on $\W_{en}$ with  sparsity norms such as  $\ell_1$ norm on ${\bf a}$. Autoencoders  only utilize spectral information. Therefore, convolutional autoencoders were suggested for spectral unmixing to use both spatial and spectral information. Deep convolutional autoencoders were used for supervised hyperspectral unmixing in \cite{Ghassemian2020}, exploiting 3D convolutional filters.

In \cite{UnDIP}, unmixing using deep image prior was proposed (UnDIP). Selection of the prior $\phi$ can be data dependent. Inspired by deep image prior \cite{DIP, DIP_CVPR},  UnDIP shows that the selection of the image prior $\phi$ in the minimization problem (\ref{eq: ab_est}) can be shifted into optimizing a deep network's parameters. Then we have, 
\begin{equation}\label{eq: UnDIP}
\hat{\theta}=\arg\min_{{\bf \theta}} \frac{1}{2} || {\bf Y}-{\bf E}f_{\theta} ({\bf Z})||_{F}^{2}%+\lambda R({\bf A})
~~~{\rm s.t.}~~ \hat{\bf A}=f_{\hat\theta}({\bf Z}).
%{\bf A}\geq 0,1_{r}^{T}{\bf A}=1_{n}^{T},
\end{equation}
 ${\bf Z}$ is the network input fixed throughout the learning. $f_\theta$ is a deep network with parameters $\theta$ representing the weights and biases. UnDIP uses SiVM to extract ${\bf E}$. Therefore, it is suitable for pure pixel-based scenarios. UnDIP uses an encoder-decoder network for $f_\theta$ and (\ref{eq: UnDIP}) as the loss function and a softmax layer to enforce ASC and ANC. In \cite{EGU_Net}, a two-stream network was proposed. The network contains an encoder in parallel with an autoencoder. First, bundle endmembers were extracted using VCA, and the corresponding abundances were estimated using the linear mixing model. Then, the encoder was trained based on the bundle endmembers and the abundances. The encoder shares the weights with the encoder of the autoencoder, whose decoder is nonlinear. A similar idea was also used in \cite{EGU_Net2}. 

% \textcolor{blue}{ a few more methods with deep encoder should be added \cite{Palsson_rev}}

\section{Unsupervised (Blind) Unmixing}
In supervised unmixing, the processing chain is sequential. Usually, the estimation of abundance does not affect the estimation of endmembers due to the order in the processing chain. In unsupervised unmixing, we assume that both  endmember and abundances are unknown and  blind unmixing endmember and abundances are estimated simultaneously. We consider three major paradigms; 1) Least Squares-based Approaches 2) Shallow/Deep learning-based Approaches, and 3) Statistical-Based Approaches. Due to the inherent nonconvexity  of blind unmixing methods, they are often vulnerable to initialization, and therefore they are always initialized using a geometrical endmember extraction approach.  %Most recent advanced techniques developed for blind unmixing fall into these two categories. There are, however, other statistical-based approaches, such as the Bayesian-based models (e.g., joint maximum a posteriori (MAP) estimator and the minimum mean squared error (MMSE) estimator was used in \cite{Un_MAP} and \cite{Un_MMSE}, respectively), which are closely related to least squares problems \cite{Elad_AvS}. 

\subsection{Least Squares-based Approaches}
Least squares-based approaches usually solve a minimization problem with respect to both ${\bf E}$ and ${\bf A}$ in the form of
\begin{align}\label{eq: BU}\nonumber
  &(\hat{\bf A},\hat{\bf E})=\arg\min_{{\bf A},{\bf E}} \frac{1}{2} || {\bf Y}-{\bf EA}||_{F}^{2} +\lambda\phi({\bf A})+\beta\psi({\bf E})\\&
{\rm s.t.}{\bf A}\geq 0,{\bf 1}_{r}^{T}{\bf A}={\bf 1}_{n}^{T}, 0\leq {\bf E} \leq 1
\end{align}
The main difference between the unmixing approaches in this group is the selection of the penalty functions $\phi$ and $\psi$. $\phi$ is often selected for incorporating spatial information and capturing the spatial correlation. The common choices for $\phi$ are TV and sparsity-promoting norms. $\psi$ is usually a geometrical regularizer for incorporating geometrical information. 
\subsubsection{Selection of a Regularizer}
The common choices for $\psi$ are simplex volume terms and norms with simplex volume flavor, such as TV \cite{TV_End}. Total variation penalty enforces the data simplex to have minimum volume \cite{TV_End}
\begin{align}\label{eq: TV}\nonumber
  &(\hat{\bf A},\hat{\bf E})=\arg\min_{{\bf A},{\bf E}} \frac{1}{2} || {\bf Y}-{\bf E}{\bf A}||_{F}^{2} +\beta TV_e({\bf E})\\&
{\rm s.t.}{\bf A}\geq 0,{\bf 1}_{r}^{T}{\bf A}={\bf 1}_{n}^{T}
\end{align}
where  $ TV_e({\bf E})$ is a TV function given by
\begin{equation}\label{eq: TV21}
 TV_e({\bf E})=\sum_{i,j=1}^r|| {\bf e}_i-{\bf e}_j||_2^{2}=||{\bf E}({\bf I}_r-\frac{1}{r}{\bf 1}_r{\bf 1}_r^T)||_F^2.  
\end{equation}
See Appendix \ref{app: TV} for more details. We should mention that the method proposed in \cite{TV_End} solves the problem after subspace projection using MNF.

Minimum volume-constrained nonnegative matrix factorization (MVC-NMF) was proposed in \cite{MVC}. MVC-NMF suggests a minimum volume penalty for the constraint nonnegative least squares given by
\begin{align}\label{eq: MVC}\nonumber
  &(\hat{\bf A},\hat{\bf E})=\arg\min_{{\bf A},{\bf E}} \frac{1}{2} || {\bf Y}-{\bf EA}||_{F}^{2} +\beta Vol^2({\bf E}_r)\\&
{\rm s.t.}{\bf A}\geq 0,{\bf 1}_{r}^{T}{\bf A}={\bf 1}_{n}^{T}, {\bf E} \geq 0.
\end{align}

% \begin{equation}
% V(\bf{E}) =  \frac{1}{(r-1)!} \left | \det 
%  \begin{bmatrix} 
%  1 & \hdots & 1\\  \mathbf{e}_{(1)}&  \hdots & \bf{e}_{(r)}
%  \end{bmatrix}\right |,
%  \label{Simplex_volume}
% \end{equation}
Collaborative nonnegative matrix factorization (CoMNF) \cite{CoNMF} suggests a sparsity-promoting term on abundances and MV on the simplex given by 
\begin{align}\label{eq: CoNMF}\nonumber
  &(\hat{\bf A},\hat{\bf E})=\arg\min_{{\bf A},{\bf E}} \frac{1}{2} || {\bf Y}-{\bf EA}||_{F}^{2} +\beta\left\| \bf E - {\bf m}{\bf 1}_{r}^{T} \right\|_F^2\\&+\lambda\sum_{i=1}^r\left\|{\bf a}_{(i)}\right\|_2^q~~
{\rm s.t.}~~{\bf A}\geq 0,{\bf 1}_{r}^{T}{\bf A}={\bf 1}_{n}^{T},
\end{align}
where, $0\leq q \leq 1$, ${\bf m}$ contains the mean values of the spectral pixels, i.e., ${\bf m}=\frac{1}{n}{\bf Y}{\bf 1}_n$, and ${\bf a}^T_{(i)}$ is the $i$th row of the matrix ${\bf A}$. This term pulls the endmembers toward the center of mass. CoNMF uses both spatial and MV (geometrical) regularizers and solves the problem by projecting the data into a subspace. In \cite{R-CoNMF
}, Robust CoMNF (RCoNMF) was proposed, which utilizes a geometrical penalty that minimizes the distances between the endmembers to be estimated and the boundary pixels (${\bf P} \in \mathbb{R}^{p\times r}$). The main assumption is that the endmembers are close to the extremes of the data simplex (so-called boundary pixels). Hence, RCoNMF solves 
\begin{align}\label{eq: RCoNMF}\nonumber
  &(\hat{\bf A},\hat{\bf E})=\arg\min_{{\bf A},{\bf E}} \frac{1}{2} || {\bf Y}-{\bf EA}||_{F}^{2} +\beta\left\| \bf E - {\bf P} \right\|_F^2\\&+\lambda\sum_{i=1}^r\left\|{\bf a}_{(i)}\right\|_2~~
{\rm s.t.}~~{\bf A}\geq 0,{\bf 1}_{r}^{T}{\bf A}={\bf 1}_{n}^{T},
\end{align}
where the boundary pixels, ${\bf P}$, can be obtained using pure pixel-based endmember extraction. In \cite{R-CoNMF}, VCA was used to estimate ${\bf P}$. In \cite{L_0.5}, $\ell_{0.5}$-NMF was proposed which enforces $\ell_q=\sum_{i=1}^r\sum_{j=1}^n |a_{ij}|^{0.5}$ sparsity penalty ($q=1/2$) on the abundances. A multiplicative update rule \cite{NMF_multiplic, NMF_L1_multiplic} was proposed to solve the $\ell_{0.5}$-NMF problem. 

In such minimization problems, selecting the regulation parameters is not trivial. In \cite{NMF_QMV}, Nonnegative Matrix Factorization-Quadratic Minimum Volume (NMF-QMV) was proposed. NMF-QMV unifies the three geometrical regularizers mentioned above (i.e., TV, center, and boundary) in a single framework. NMF-QMV proposes the unified cost function using the MV-regularizers given by 
\begin{align}\label{eq: NMF_QMV}\nonumber
  &(\hat{\bf A},\hat{\bf E})=\arg\min_{{\bf A},{\bf E}} \frac{1}{2} || {\bf Y}-{\bf EA}||_{F}^{2} +\beta\left\| {\bf E}{\bf G}  - {\bf O} \right\|_F^2\\&~~
{\rm s.t.}~~{\bf A}\geq 0,{\bf 1}_{r}^{T}{\bf A}={\bf 1}_{n}^{T},
\end{align}
where the selection of ${\bf G}$ and ${\bf O}$ are given in Table \ref{tab:mv}. We emphasize that those terms do not compute the volume of a simplex. They are convex penalties with volume flavor. A parameter selection technique was proposed in \ref{eq: NMF_QMV} by searching for an optimum $\beta$ in a large interval. $\beta$ optimum is when the estimated endmember simplex boundary is close to the data points boundary. Note that the spatial regularizer was ignored in (\ref{eq: NMF_QMV}), simplifying the selection technique. 
\begin{table}[htbp]\scriptsize
\centering
 	\caption{Selection of ${\bf G}$ and ${\bf O}$ for the unified geometrical regularizer.}
\begin{tabular}{c|ccc}
		\toprule
		    &\textbf{TV}   &\textbf{Center}             &\textbf{Boundry} \\		
		\midrule
		${\bf G}$  &  ${\bf I}_r-\frac{1}{r}{\bf 1}_r{\bf 1}_r^T$ & ${\bf I}_r$ & ${\bf I}_r$\\ \midrule
 		{\bf O}    &  ${\bf 0}$ & ${\bf m}{\bf 1}_r$ & Extracted Extremes  \\
		\bottomrule
	\end{tabular}   
	\label{tab:mv}
\end{table}
\begin{figure} [htbp]
\centering
\begin{tabular}{ccc} 
\includegraphics[width=.3\linewidth]{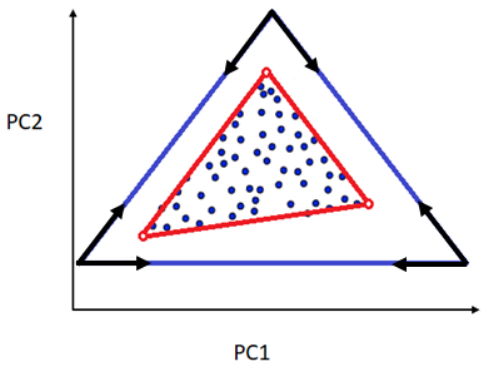}&\includegraphics[width=.3\linewidth]{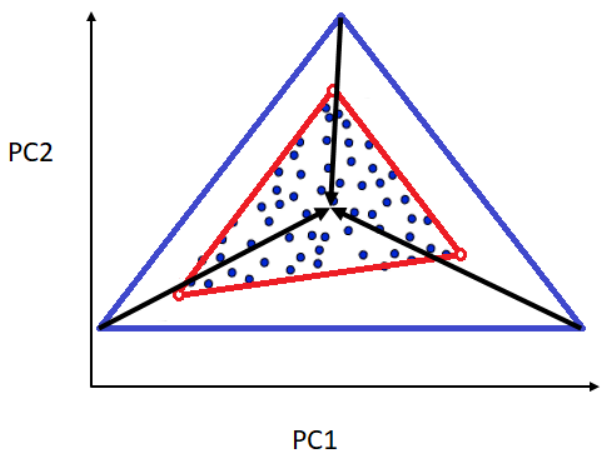}&\includegraphics[width=.3\linewidth]{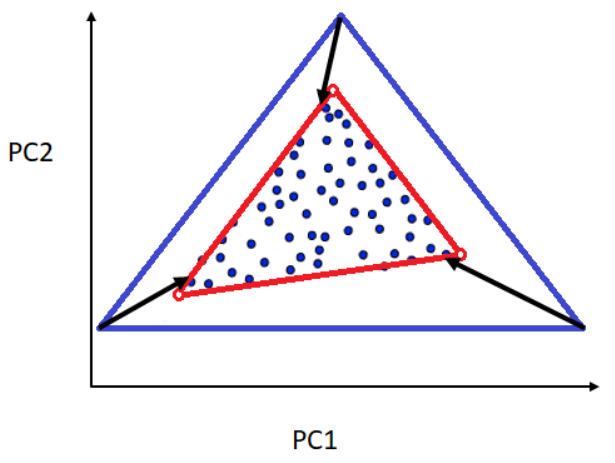}\\   (a) &(b) & (c)
 \end{tabular} %\end{center} 
\caption{The intuition of how the three geometrical regularizers enforce the endmembers using (a) TV, (b) center, and (c) boundary terms to change an initial simplex (blue) towards the endmember simplex (red). The black arrows show the direction of the force.}
\label{fig: mv}
\end{figure}
We should note that some blind unmixing techniques\cite{TV_End, CoNMF, R-CoNMF, NMF_QMV}, solve the problems in a subspace. This has two main advantages: 1) reducing the computations, and 2) reducing noise and removing outliers. Hence, the endmembers are assumed to belong to an $r-1$ dimensional affine set, which best represents the data using an orthogonal projection. Note that the abundances do not change and can be estimated in a subspace. However, in real-world scenarios, the estimated endmembers in a subspace after reprojection  into the original space may have negative values due to noise and nonlinearity since they cannot be bounded between zero and one in a subspace.

In \cite{JakubLq}, the spectral roughness penalty was proposed for imposing the smoothness to the endmembers, and the $\ell_q$ norm was applied to the abundances to enforce sparsity on the abundances. The proposed cost function is given by
\begin{align}\label{eq: lq}\nonumber
  &(\hat{\bf A},\hat{\bf E})=\arg\min_{{\bf A},{\bf E}} \frac{1}{2} || {\bf Y}-{\bf EA}||_{F}^{2} +\frac{\beta}{2} \left\|{\bf R}{\bf E}\right\|_F^2+\lambda\sum_{i=1}^n\left\|{\bf a}_i\right\|_q\\&+\frac{\alpha}{2} \left\| {\bf 1}_{r}^{T}{\bf A}-{\bf 1}_{n}^{T} \right\|_2^2~~
{\rm s.t.}~~{\bf A}\geq 0, 
\end{align}
where $0\leq q \leq 1$ and ${\bf R}\in \mathbb{R}^{(p-1)\times p}$ is the first-order difference matrix given by
\begin{equation}
{\bf R}=
\left[
\begin{array}[pos]{cccccc}
 -1 & 1 & 0  & \ldots & 0  & 0 \\
  0 & -1 & 1  & 0 & \cdots  & 0 \\
  \vdots & \vdots & \ddots & \ddots & \vdots  & \vdots \\
  0 & 0 & 0  & \ldots & -1  & 1 \\
\end{array}
\right].
\end{equation}
We should note that the roughness penalty does not incorporate geometrical information, and ASC constraint was relaxed and added to the cost function as a penalty. This makes the algorithm robust to the noise however, it adds one extra regularization parameter, $\alpha$, which together with $\beta$ and $\lambda$ become very challenging to select, particularly in real-world applications. In \cite{JackobTV}, the (2 dimensional) isotropic total variation was applied to abundances  to promote piece-wise smoothness spatially. The cost function is given by 
\begin{align}\label{eq: lqTV}\nonumber
  &(\hat{\bf A},\hat{\bf E})=\arg\min_{{\bf A},{\bf E}} \frac{1}{2} || {\bf Y}-{\bf EA}||_{F}^{2} +\lambda_1\sum_{i=1}^rTV_{i}({\bf a}_{(i)})+\\&\lambda_2\sum_{i=1}^r\left\|{\bf a}_{(i)}\right\|_q~~
{\rm s.t.}~~{\bf A}\geq 0, {\bf E}\geq 0, \left\|{\bf e}_{(i)}\right\|_2^2=1,i=1, 2, ...,r,
\end{align}
where $0\leq q \leq 1$ and $TV_{i}$ is the nonisotropic TV function given by
\begin{equation}
    TV_{i}({\bf x})=\left\|\sqrt{({\bf H}_h({\bf x}))^2+({\bf H}_v({\bf x}))^2}\right\|_1.
\end{equation}
${\bf H}_v={\bf I}\otimes {\bf R}$ and ${\bf H}_h={\bf R}\otimes {\bf I}$ are the matrix operators to calculate the first order vertical and horizontal differences, respectively, for the vectorized image ${\bf x}$. We should note that ASC was ignored in (\ref{eq: lqTV}), and the endmember norm constraint (ENC) was applied to enforce a unit norm on the endmembers. ENC captures the spectral variability but has a big disadvantage, which does not let the algorithm distinguish the materials with similar endmembers with some scaling factors, such as tree and grass. In \cite{NMF_TV_RS}, NMF subject to ASC was suggested using a reweighted $\ell_1$ norm and the total variation penalty applied to the abundances.  Many NMF-based blind unmixing approaches were proposed using the form of minimization (\ref{eq: BU}) \cite{NMF_1,NMF_2,NMF_3,NMF_4}.  

A plug-and-play (PnP) priors framework was proposed in \cite{PnP} for blind unmixing. Problem (\ref{eq: BU}), omitting the penalty and constraint on $\E$, was solved using an ADMM. The penalized least squares step in the ADMM can be solved using a denoiser. Several denoisers, including a deep network, were suggested to solve that step. Additionally, the penalty term was considered not only applied to the abundances but also to the reconstructed data, i.e., $\phi(\E\A)$.

\subsubsection{Archetypal Analysis Models}
Several archetypal analysis (AA)-based \cite{cutler1994archetypal} algorithms were proposed for blind unmixing. Let's assume that the endmembers are convex combinations of the spectral pixels, then using AA we have
\begin{align}\label{eq: AA}\nonumber
  &(\hat{\bf A},\hat{\bf B})=\arg\min_{{\bf A,B}} \frac{1}{2} || {\bf Y}-{\bf YBA}||_{F}^{2} \\&
{\rm s.t.}~~{\bf B}\geq 0,{\bf 1}_{n}^{T}{\bf B}={\bf 1}_{r}^{T}, {\bf A}\geq 0,{\bf 1}_{r}^{T}{\bf A}={\bf 1}_{n}^{T},
\end{align}
where ${\bf B} \in \mathbb{R}^{n\times r}$ and the columns of  ${\bf B}$ belong to the simplex $\Delta_n$. In \cite{EDAA}, Entropic Descent Archetypal Analysis (EDAA) was proposed. EDAA uses an entropic descent algorithm to solve the AA problem (\ref{eq: AA}). A model selection technique suggested where the coherence $\mu=\max_{i\neq j}{\bf e}^T_i{\bf e}_j$ 
 and $\ell_1$ residual $|| {\bf Y}-\hat{\bf E}\hat{\bf A}||_1$ are minimized. This makes the solution robust to the initialization of ${\bf B}$. The convex combination constraint on the endmembers is very strict. Therefore, in \cite{ME_AA}, a variation of AA suggested which uses a relaxation factor $\delta$ and an $r$ dimensional scaling vector ${\bf s} \in \mathbb{R}^{r}$
\begin{align}\label{eq: KAA}\nonumber
  &(\hat{\bf A},\hat{\bf B},\hat{\bf S})=\arg\min_{{\bf A,B, S}} \frac{1}{2} || {\bf Y}-{\bf YB}({\bf S}\odot{\bf I}){\bf A}||_{F}^{2}~~
{\rm s.t.}~~ \\&{\bf B}\geq 0,{\bf 1}_{n}^{T}{\bf B}={\bf 1}_{r}^{T}, {\bf A}\geq 0,{\bf 1}_{r}^{T}{\bf A}={\bf 1}_{n}^{T}, 1-\delta\leq{\bf s}\leq1+\delta
\end{align}
where ${\bf S}={\bf s1}^T$ and $\odot$ denotes the Hadamard product. Additionally, Kernel AA (KAA) was suggested to take into account the endmember variability by multiple endmember extraction. The proposed KAA was computationally expensive, and therefore, in \cite{FKAA}, the Nystr\"{o}m method is used to construct a low-rank approximation of the high-dimensional kernel matrix using K-means. 

An $\ell_1$ penalized variation of AA was proposed in \cite{L1AA} given by 
\begin{align}\label{eq: L1AA}\nonumber
  &(\hat{\bf A},\hat{\bf B},\hat{\bf S})=\arg\min_{{\bf A,B, S}} \frac{1}{2} || {\bf Y}-{\bf PB}({\bf S}\odot{\bf I}){\bf A}||_{F}^{2}+||\A||_1~~
{\rm s.t.}~~ \\&{\bf B}\geq 0,{\bf 1}_{n}^{T}{\bf B}={\bf 1}_{r}^{T}, {\bf A}\geq 0,{\bf 1}_{r}^{T}{\bf A}={\bf 1}_{n}^{T}, 1-\delta\leq{\bf s}\leq1+\delta
\end{align}
where the sparsity was enforced on the abundance and ${\bf P}\in \mathbb{R}^{n\times c}, c>>r$ contains boundary pixels extracted by using PPI to cope with the endmember variability.

\subsubsection{Tensor Models}
\textcolor{black}{Tensor decomposition (TD) was also suggested for blind unmixing motivated by the tensor structure of hyperspectral data, which can be represented as a three-dimensional array (tensor) with dimensions corresponding to spatial coordinates (row and columns) and spectral bands (channels) \cite{NTF}. Hyperspectral data are often decomposed into nonnegative tensors. The interpretation of endmembers and abundances in a tensor decomposition makes the unmixing modeling challenging. The canonical polyadic decomposition (CPD) \cite{NTF} and Tucker decomposition (or higher order SVD) \cite{TD} have been proposed for unmixing. However, they lack such an interpretation. Among tensor decompositions, matrix-vector third-order tensor decomposition (also known as rank-($L_i$, $L_i$, 1) block-term decomposition) is the most popular model for unmixing due to the interpretation of endmembers and abundances in the decomposition. It is a specific case of block
term decompositions (BTDs) \cite{BTD} which is generally defined as a sum of r rank-($L_i$, $M_i$, $N_i$) terms ($i$ = 1, 2, ..., r). BTD can be assumed as a generalization of CPD and Tucker decomposition where it decomposes
a tensor into a sum of component tensors (same as CPD), while
each component tensor is factorized using a Tucker decomposition. 
Considering the hyperspectral data as a 3rd-order tensor ${\mathcal Y}\in \mathbb{R}^{n_1 \times n_2 \times p}$ (where $n_1 \times n_2= n$), rank-($L_i$, $L_i$, 1) decomposition is written as
\begin{equation}
    {\mathcal Y}=\sum_{i=1}^r{\bf A}_i\odot{\bf e}_i =\sum_{i=1}^r{\bf K}_i{\B}_i^T\odot{\bf e}_i+ {\mathcal N},{\bf K}_i\geq 0, {\bf B}_i\geq 0, {\bf e}_i\geq 0
\end{equation}
where ${\bf A}_i$ denotes $i$th $n_1 \times n_2$ abundance (i.e., $vec^{-1}({\bf a}_i)$) of rank $L_i$. Therefore, ${\bf A}_i$ can be written as a matrix product ${\bf A}_i={\bf K}_i{\B}_i^T$ where ${\bf K}_i \in \mathbb{R}^{n_1 \times L_i}$ and ${\B}_i \in \mathbb{R}^{n_2 \times L_i}$ are full column rank matrices of rank $L_i$. The rank-($L_i$, $L_i$, 1) decomposition shown in Fig \ref{fig: TensorDec} is unique and a sufficient uniqueness condition is
that the partitioned matrices ${\bf K}=[{\bf K}_1, {\bf K}_2, ..., {\bf K}_r]$ and ${\bf B}=[{\bf B}_1, {\bf B}_2, ..., {\bf B}_r]$ are full column rank and ${\bf E}$ does not have any collinear columns \cite{BTD}. It was shown that full column rank of ${\bf A}$ and ${\bf B}$  can be met if $\min(n1,n2)=\sum_{i=1}^r{L}_i$. Those conditions may be met in unmixing applications however they are not ensured. We should note that the abovementioned uniqueness conditions were discussed for BTD without nonnegativity constraints. Nevertheless, the final solution always depends on the initialization due to the nonconvexity in blind unmixing.
\begin{figure}
    \centering
    \includegraphics[width=.5\textwidth]{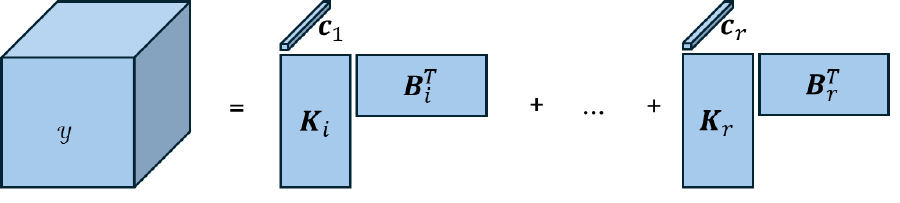}
    \caption{The  rank-($L_i$, $L_i$, 1) decomposition which is widely used for blind linear unmixing.}
    \label{fig: TensorDec}
\end{figure}
Matrix-Vector Nonnegative Tensor Factorization (MVNTF) was proposed in \cite{MVNTF} which solves 
\begin{align}\label{eq: MVNTF}\nonumber
     &(\hat{\bf K},\hat{\bf B},\hat{\bf E})=\arg\min_{{\bf K,B,E}}   \frac{1}{2} \|{\mathcal Y}-\sum_{i=1}^r{\bf K}_i{\B}_i^T\odot{\bf e}_i\|^2_F+\\& \frac{\lambda}{2} \|\sum_{i=1}^r{\bf K}_i{\B}_i^T-{\bf 1}_{n_1}{\bf 1}_{n_2}^T\|^2_F
~~
{\rm s.t.}~~  {\bf K}\geq 0, {\bf B}\geq 0, {\bf E}\geq 0 
\end{align}
where the Frobenius norm for a third-order tensor is given by $\mathcal{X}=\sqrt{\sum_i\sum_j\sum_k x^2_{ijk}}$ and the second term ensures the ASC. Several variations of (\ref{eq: MVNTF}) were proposed in the literature. In \cite{SLNTF}, sparse and low-rank NTF was proposed which enforces both $\ell_1$ sparsity and nuclear norm on abundances. This method was further improved in \cite{WSLNTF} by exploiting weighted norms for the abundances.  In \cite{TV_NTF}, the TV regularized MVNTF was proposed which utilizes the TV regularization on abundances to promote piecewise smoothness on the abundances and to capture spatial information. Tensor decomposition was also proposed for multi-feature hyperspectral unmixing to incorporate spatial features extracted by mathematical morphology \cite{MultiHU_TD}.}

\subsubsection{Extended Linear Mixing Models and Spectral variability}
An Extended Linear Mixing Model (ELMM) was proposed in \cite{ELMM1} to address the  endmember variability given by 
\begin{equation}
     {\bf Y}={\bf E}{\bf S}{\bf A}+{\bf N},{\bf A}\geq 0, {\bf S}\geq 0, {\bf E}\geq 0
\end{equation}
where  ${\bf S}={\bf s}\1^T$ and ${\bf s}$ contains the scaling factor of the endmembers. Nonnegative constraint least square (NCLS) was used to estimate the scaled abundances. Assuming the scaling factor is the same for all the endmembers and knowing that sum of the abundances is the scaling factor, therefore, the scaling factor was removed from abundances with a final normalization step. ELMM considers the spectral variation caused by illumination and topography. We should note that ELMM only considers uniform variations. In \cite{ELMM2}, a pixel-dependent ELMM was given by
\begin{equation}
    {\bf y}_i = {\bf E}_i{\bf a}_i + {\bf n}_i, {\bf A}\geq 0,{\bf E}_i={\bf E} \odot {\bf c}_{i}{\1}^T , {\bf a}_{i}\geq 0,{\bf 1}_{r}^{T}{\bf a}_{i}=1,
\end{equation}
where ${\bf C}\in \mathbb{R}^{r \times n}$ contains the scaling factors in its column. A minimization problem was proposed as
\begin{align}\label{eq: ELMM2}\nonumber
  &(\hat{\bf A},\hat{\mathcal E},\hat{{\bf C}})=
 \arg\min_{{\bf A},\mathcal{E, C}} \frac{1}{2}\sum_{i=1}^n || {\bf y}_i-{\bf E}_i{\bf a}_i||_{2}^{2}+\\ &\lambda_1  \sum_{i=1}^n||{\bf E}_i-{\bf E}\odot{\bf c}_i\1^T||_F^2 +\lambda_2 \psi(\bf{C})  + \lambda_3\phi({\bf A})\\ \nonumber& ~~
{\rm s.t.}~~{\bf A}\geq 0,{\bf 1}_{r}^{T}{\bf A}={\bf 1}_{n}^{T}
\end{align}
where $\mathcal{E}\in \mathbb{R}^{p\times r\times n}$ is a ${p\times r\times n}$ tensor with ${\bf E}_i$ located in band $i$ and 
\begin{align}
&\psi(\bf{C})=\frac{1}{2}(||{\bf R}{\bf C}||_F^2+||{\bf C}{\bf R}^T||_F^2)\\&
 \phi({\bf A})=  \sum_{j=1}^r||{\bf H}_h({\bf a}_{(j)})||_{2}+||{\bf H}_v({\bf a}_{(j)})||_{2}.
\end{align}
Generalized LMM (GLMM) was proposed in \cite{GLMM} to address the nonuniform variation. GLMM is given by
 \begin{equation}\label{eq: GLMM}
{\bf y}_i = {\bf E}_i{\bf a}_i + {\bf n}_i, ~~{\rm s.t.~~} {\bf E}_i={\bf E}\odot{\bf C}_i, {\bf a}_i\geq 0,{\bf 1}_{r}^{T}{\bf a}_i=1,
\end{equation} 
where ${\bf C}_i\in \mathbb{R}^{p\times r}, i =1,..., n$ are scaling matrices dedicated to every pixel. The proposed minimization is given by
\begin{align}\label{eq: GLMM1}\nonumber
  &(\hat{\bf A},\hat{\mathcal{E}},\hat{\mathcal{C}})=\arg\min_{{\bf A},\mathcal{E, C}} \frac{1}{2}\sum_{i=1}^n || {\bf y}_i-{\bf E}_i{\bf a}_i||_{2}^{2}+\\ &\lambda_1  \sum_{i=1}^n||{\bf E}_i-{\bf E}\odot{\bf C}_i||_F^2 +\lambda_2    \sum_{i=1}^n\psi({\bf C}_i)   + \lambda_3\phi({\bf A}) \\ \nonumber& ~~
{\rm s.t.}~~{\bf A}\geq 0,{\bf 1}_{r}^{T}{\bf A}={\bf 1}_{n}^{T}, 
\end{align}
where $\mathcal{C}\in \mathbb{R}^{p\times r\times n}$ is ${p\times r\times n}$ tensor with ${\bf C}_i$ located in band $i$. In \cite{GLMM2}, a framework was suggested to estimate the scaling matrices ${\bf C}_i$ using pure pixel endmember extraction techniques. 
A perturbed linear mixing model (PLMM) was proposed in \cite{PLMM}. This simplifies the problem, however, the algorithm considerably relies on the presence of pure pixels and the extraction technique. The proposed model is given by
 \begin{equation}\label{eq: PLMM}
{\bf y}_i = {\bf E}_i{\bf a}_i + {\bf n}_i, {\rm s.t.} {\bf E}_i={\bf E}+{\bf C}_i\geq 0, {\bf E}\geq 0, {\bf a}_i\geq 0,{\bf 1}_{r}^{T}{\bf a}_i=1,
\end{equation} 
where ${\bf C}_i$ additive perturbation to model the spectral variability for every pixel. The proposed minimization is given by 
\begin{align}\label{eq: PLMM1}\nonumber
  &(\hat{\bf A},\hat{\mathcal{E}},\hat{\mathcal{C}})=\arg\min_{{\bf A},\mathcal{E, C}} \frac{1}{2}\sum_{i=1}^n || {\bf y}_i-{\bf E}_i{\bf a}_i||_{2}^{2}+\frac{\lambda_1}{2} ||{\bf A}{\bf H}_d||_F^2\\&+\lambda_2 \psi({\bf E})+\lambda_3  \sum_{i=1}^n||{\bf C}_i||_F^2 \nonumber    \\& ~~
{\rm s.t.}~~{\bf A}\geq 0,{\bf 1}_{r}^{T}{\bf A}={\bf 1}_{n}^{T}, {\bf E}\geq 0, {\bf E}_i\geq 0, i=1,...,n
\end{align}
where $\psi$ can be $Vol^2$, $TV_e$, or the Euclidean distance between ${\bf E}$ and some reference spectral signatures and ${\bf H}_d$ is a matrix to calculate the deference of an abundance pixel with 4 nearest neighboring pixels (see \cite{PLMM} for more details). PLMM was used in \cite{PLMM_MT}, and GLMM was used in \cite{GLMM_MT}  for multitemporal unmixing. %PLMM was also used in \cite{PLMM_DMT} for distributed unmixing.
There exist other approaches that tackle spectral variability such as Sparsity-Enhanced Convolutional Decomposition (SeCoDe) \cite{yao_sparsity-enhanced_2021} (a tensor-based model),  Augmented Linear Mixing Models (ALMM) \cite{hong2018augmented}, and Subspace Unmixing with Low-Rank Attribute Embedding (SULoRA) \cite{SULoRA}.

\subsection{Shallow/Deep Neural Networks}

Shallow or deep neural networks can be used for blind unmixing, as shown in Fig. \ref{fig: EDU}. The decoder is often shallow, and a linear layer is often used for linear unmixing. In blind unmixing networks, the decoder weights are the network's parameters representing the endmembers to be learned. Therefore, the regularized loss function is given by  
\begin{equation}
    \L_{reg}(\y,\hat\y)=\L(\y,\hat\y)+ \lambda J({\bf a},\W_{en},\W_{de})
\end{equation}
where $J$ can be a combination of several functions, such as a Frobenius norm on $\W_{en}$ with a sparsity norm such as  $\ell_1$ norm on ${\bf a}$, and  a Frobenius norm or a geometrical (MV induced) norm on $\W_{de}$. The loss function $\L$ might also be selected as a combination of several functions. For instance, EndNet \cite{EndNet} proposes a loss function with several terms, including a Kullback-Leibler divergence term, a SAD similarity, and a sparsity term, making the parameter selection very challenging.

The blind unmixing networks can be divided into two main groups: 1) autoencoder-based networks and 2) algorithm unrolling-based networks.  

\subsubsection{Autoencoder-based networks}
The majority of the blind unmixing networks are variations of autoencoder-based networks. The main components of such networks were explained in Section \ref{SD_model}. However, the encoder can be much deeper and more complicated. Five main types of AEs were considered for blind unmixing \cite{Palsson_rev}; 1) Sparse Nonnegative Autoencoders (SNAEs): SNAEs enforce  sparsity on the bottleneck layer often via a regularization term in the loss, 2) Variational Autoencoders (VAEs): For VAEs, the decoder samples from a probability distribution model from the data and the encoded (hidden) features.  The loss function uses a reconstructed loss plus a (KL-divergence) regularization term, ensuring that the encoded data stay close to a particular distribution (often a unit normal distribution), 3) Adversarial Autoencoders (AAEs): AAEs are based on generative adversarial networks (GANs). The network also uses a discriminator, and the AE's decoder often acts as a generator for the adversarial network. AAEs use a reconstructed plus an adversarial loss. 
4) Denoising Autoencoders (DAEs): DAEs are trained based on partially corrupted input while minimizing the loss between the reconstructed and original (uncorrupted) inputs. Therefore, the network learns to remove the noise.  5) Convolutional Autoencoders (CAEs): CAEs use convolutional layers instead of fully connected layers. The weights are the convolutional filters that be learned throughout the training process. The convolutional filters capture the spatial correlation, which benefits the unmixing process. A comparative study was provided on the performance of AEs for hyperspectral unmixing in \cite{Palsson_rev}. The probabilistic generative model for spectral unmixing (PGMSU) was proposed in \cite{PGMSU}. The network uses a VAE to generate the endmembers and a parallel encoder to generate the abundances.

SNSA \cite{YSU2018} utilizes a stack of nonnegative sparse autoencoders. The last one performs the task of unmixing, and the others are exploited to improve the robustness concerning the outliers. Deep AutoEncoder Network (DAEN) \cite{DAEN} exploits a stacked autoencoder to initialize a variational autoencoder which performs the unmixing task.  An untied Denoising Autoencoder with Sparsity (uDAS) \cite{uDAS} exploits an additional denoising constraint on the decoder and a $\ell_{2,1}$ sparsity promoting constraint on the decoder. Note that the autoencoder receives a spectral pixel at a time for training the network. Training a network based on a single spectrum at a time ignores the spatial information. The advantage of incorporating spatial information for spectral unmixing has been confirmed in the literature. Therefore, patch-wise or cube-wise CNN was proposed to utilize the spatial information. In \cite{Hua2021}, an unmixing technique based on an autoencoder network is proposed that incorporates the spatial correlation between pixels by utilizing an adaptive abundance smoothing method.  In \cite{CNN_unmixing}, patch-wise or cube-wise convolutional autoencoders were proposed to incorporate the spatial information. In \cite{Burkni20SSAE}, parallel autoencoders were applied to spectral patches to exploit the spatial information. A CAE was proposed in \cite{Burkni20CAE}. The proposed network uses two convolutional layers as the decoder and a convolutional layer with linear activation as the decoder. ASC and ANC were enforced using a softmax, and SAD was used as the loss function. The cycle-consistency unmixing network \cite{CyCUNet} utilizes two convolutional autoencoders, which are cascaded and performed cyclically. The proposed loss function contains two terms for spectral reconstruction and one for abundance reconstruction to incorporate high-level semantic information. In \cite{Trans_SU},  a combination of a CAE and a transformer was proposed for unmixing. The attention module of the transformer captures the nonlocal contextual feature dependencies between the image patches. In \cite{AAE}, VCA and FCLSU were first used to drive prior abundances from superpixels. Then, the adversarial training guides the encoder of an AAE to transfer the spatial information into the network by matching the abundances. A VAE was used in \cite{DeepGUn} to address the spectral variability. First, a deep generative endmember model was learned using the pure pixel extracted from the data. Then, an optimization was suggested and solved using the endmember model as a prior. A deep generative model was proposed in \cite{GAN_VAE}, which uses three types of encoders, i.e., CNN, convolutional graph networks, and self-attention-based networks. A discriminator was used to enforce the generated endmembers to have a distribution similar to the endmember models either extracted from the data or selected from a library.

In AEs, a deeper encoder does not necessarily improve the performance of AE-based unmixing, however, it helps to enlarge the receptive field \cite{Palsson_rev}. Therefore, in \cite{MSNet}, a Multi-Stage CAE Network (MSNet) was suggested, which uses a multi-stage framework. Every stage uses a shallow encoder-decoder architecture. The input data are spatially downsampled at every stage. The features at the next stage were concatenated with the previous stage's input to preserve the abundance consistency of different stages.

A drawback of the many DL-based approaches is the absence of geometrical information. Most of them rely on SAD and/or MSE for training. As we already discussed and it has been proven \cite{MV_proof}, geometrical information could be very beneficial without pure pixels. To incorporate geometrical and spatial information, in \cite{MiSiCNet}, a minimum simplex convolutional network (MiSiCNet) was proposed. In \cite{MiSiCNet}, it was shown that the minimization problem 
 \begin{align}\label{eq: PLSU}\nonumber
(\hat{\bf A},\hat{\bf E})=&\arg\min_{{\bf A},{\bf E}} \frac{1}{2} || {\bf Y}-{\bf EA}||_{F}^{2}+\beta \left\| \bf E - {\bf m}{\bf 1}_{r}^{T} \right\|_F^2\\&+\lambda \phi({\bf A})
{~~\rm s.t.~~}{\bf A}\geq 0,{\bf 1}_{r}^{T}{\bf A}={\bf 1}_{n}^{T},0\leq {\bf E} \leq 1
\end{align}
can be turned into 
 \begin{align}\label{eq: PLSU2}\nonumber
(\hat\theta_1,\hat\theta_2)=&\arg\min_{\theta_1,\theta_2} \frac{1}{2} || {\bf Y}-\theta_2f_{\theta_1}({\bf Z})||_{F}^{2}+\lambda \left\|  \theta_2 - {\bf m}{\bf 1}_{r}^{T} \right\|_F^2\\& ~~~{\rm s.t.}~~\hat{\bf E} \hat{\bf A}=\hat\theta_2f_{\hat\theta_1}({\bf Z})
% \hat{\bf Y}=\hat\theta_2f_{\hat\theta_1}({\bf Z}),
\end{align}
where $\hat{\bf Y}=\hat{\bf E}\hat{\bf A}$, $\theta_1$ and $\theta_2$ are the parameters of the encoder and the decoder of the network, respectively. Therefore, the minimization problem (\ref{eq: PLSU}) can be shifted to the minimization of the parameters of a deep network $f$ with a loss function given by:
 \begin{equation}\label{eq: loss}
\mathcal L({\bf Y},\hat{\bf Y},\hat\E,{\bf m})=\frac{1}{2} || {\bf Y}-\hat{\bf Y}||_{F}^{2}+\lambda \left\| \hat \E - {\bf m}{\bf 1}_{r}^{T} \right\|_F^2,
\end{equation}
where $f$ is an encoder-decoder network and $\hat\E$ is the weight of the decoder. MiSiCNet uses a softmax layer to enforce both the ASC and the ANC. 

\subsubsection{Algorithm Unrolling-based Networks}
The main idea behind algorithm unrolling-based networks is to incorporate model-based with learning-based methods. In this group, an iterative algorithm (such as ADMM) is considered to solve a mixture model. Then, the iterations of the algorithm will be unrolled and implemented using neural networks. The updating equations will be substituted using NNs. Sequential repetitions of such networks will be trained using a loss function such as MSE. An advantage of such networks is that the tuning parameters can be trainable. SUnSAL with ADMM implementation was unrolled in \cite{SUNSAL_Unrol} with an extra linear layer augmented to the network to implement blind unmixing (ADMMNet). The sequential networks in \cite{SUNSAL_Unrol} were tested with and without parameter sharing. In \cite{ISTA_Unrol}, $\ell_1$-NMF was considered for blind unmixing. The iterative soft-thresholding algorithm \cite{ISTA}, was unrolled to estimate the abundances. An extra linear layer was used for the endmember estimation and added to the sequential networks. The scenarios of the networks  with and without sharing parameters were also considered in \cite{ISTA_Unrol}. A solution to the $\ell_p$-NMF was unrolled for blind unmixing  in \cite{NMF_Lp_Unrol}.  Instead of using a linear layer, the endmember estimation step was also unrolled. Therefore, the sequential network was constructed of endmember networks followed by abundance networks. For the abundance and endmember networks, the generalized shrinkage thresholding solution given in \cite{GST} and the projected gradient descent were unrolled, respectively.

Finally, we should note that often blind NN-based unmixing can be turned into supervised unmixing by fixing the weights of the decoders using extracted endmembers.

\subsection{Statistical-based Approaches}
Most recent advanced techniques developed for blind unmixing fall into the first two categories. There are, however, other statistical-based approaches that
%Bayesian-based models (e.g., joint maximum a posteriori (MAP) estimator and the minimum mean squared error (MMSE) estimator was used in \cite{Un_MAP} and \cite{Un_MMSE}, respectively), which are closely related to least squares problems \cite{Elad_AvS}. 
%Not all the statistical approaches
cannot be reformulated as the constraint penalized least squares. The Bayesian approach is widely used for blind unmixing. The maximum a posteriori (MAP) estimator and the minimum mean squared error (MMSE) estimator were used in \cite{Un_MAP} and \cite{Un_MMSE}, respectively. The Bayesian approaches using the MAP estimator can be reformulated as the penalized least square in the form of (\ref{eq: BU}) when the random noise is assumed to be Gaussian and uncorrelated, i.e., the noise covariance matrix is diagonal \cite{Elad_AvS, unmixing-review}. This assumption simplifies the inverse problem and its solution. Independent Component Analysis (ICA) \cite{ICA}, as a solution to blind source separation, was proposed for blind unmixing \cite{ICA_BU, ICA_BU2}. The abundances are dependent (ASC), therefore, ICA is not a suitable solution to blind unmixing \cite{ICA_No}. Dependent component analysis (DECA) was proposed for blind unmixing of highly mixed datasets \cite{DCA_1, DECA}. DECA assumes a mixture of Dirichlet densities as the prior for abundances. Dirichlet density satisfies ANC and ASC. The Bayesian estimation problems are often hard to solve, and the estimation of unknown parameters is computationally expensive. In \cite{Bayes_Gibbs}, a Gibbs sampling strategy was proposed to decrease the complexity of the proposed hierarchical Bayesian model. A similar Bayesian approach was proposed in \cite{BayesianSMA} for spectral mixture analysis (BayesianSMA) from Raman spectroscopy.

\section{Semisupervised (Library-based) Unmixing}

When neither the pure pixels nor the sufficient spectra on the facets of the data simplex are available, the endmembers can not be successfully extracted/estimated, which results in poor abundance estimations. Alternatively, blind unmixing techniques can be used. However, blind unmixing is a non-convex problem; therefore, a good initialization of endmembers using endmember extraction techniques will often benefit in finding a better solution. Additionally, in highly mixed scenarios, blind unmixing methods often fail due to the large sets of solutions that fit the data. Fig. \ref{fig: Pure_No} (d) demonstrates that in highly mixed scenarios, finding the real simplex is not trivial. Therefore, semi-supervised unmixing was suggested, relying on an endmember library. Here, we should note that we categorize all techniques that entirely or partially rely on a spectral library as semi-supervised unmixing techniques. Therefore, before discussing the semisupervised techniques, we discuss the library selection and construction. 

%\subsection{Library Selection and/or Construction} 
Library selection or construction is a crucial step to the success of a semisupervised unmixing method. Blind selection of a library without extra attention and some processing steps will lead to poor results for semisupervised unmixing. There are two major paradigms in semisupervised unmixing i) Multiple Endmember Spectral and Mixture Analysis and ii) Sparse unmixing. The former was designed to address the spectral variability using the endmember variability. Therefore, the spectral library is designed to represent the variability of the endmembers. The latter seeks a sparse solution relying on the library, and therefore, the high correlation of the library endmembers avoids the sparse solution. In both paradigms, the library must well represent the materials in a sense, i.e., it must contain all the endmembers of the materials in the scene. Generally, a library can be obtained using either of the following ways \cite{Lib_Chall,Spec_var}.
\begin{enumerate}
    \item In situ field or/and laboratory measurements: endmember libraries can be built using field or/and laboratory measurements \cite{MESMA}. Creating such libraries has several major drawbacks. It is hard, time-consuming, expensive, and might be sensor-dependent. Due to the different measuring conditions, systems, or instruments, there are often mismatches and scaling discrepancies between the endmembers from the library and the observed data \cite{MisMat_Lib, SUn}. A list of available spectral libraries was given on https://specchio.ch/. %List of endmember library should be given here (https://specchio.ch/) and DIRSIG spectral library.
    \item Construction using observed data: Extracting multiple pure pixels from the observed datasets representing every material in the observed dataset. These spectra can be further clustered to represent the endmember bundles. These libraries are data-dependent and often fail in highly mixed scenarios since there are not enough pure pixels in the dataset \cite{MESMA_Lib_EE, GSIMN, End_bundle}.
    \item Construction using physical models: Radiative transfer models can be used to generate endmember libraries \cite{Rad_Tra_Lib}. A radiative transfer model is defined as a function of physicochemical parameters. Therefore, different instances of endmembers can be obtained by varying physicochemical parameters. For instance, PROSPECT model \cite{PROSPECT} for vegetation and Hapke model \cite{BHapke1981} for reflectance modeling of intimate mixtures and densely packed grains or particles can be used. These libraries are sensor- and data-independent. However, they depend on the availability and accuracy of a model for the target materials. Additionally, selecting representative samples from such complex models is a big challenge.
\end{enumerate}
A combination of the libraries mentioned above also may be used. Alternatively, generative machine learning models can be used to augment a library. In \cite{MESMA_VAE}, a variational autoencoder was used for library augmentation.

\subsection{Multiple Endmember Spectral and Mixture Analysis} 
As we discussed, endmember variability is a challenge in hyperspectral unmixing. A solution is to use a library containing endmembers' variations for materials. Then, the unmixing problem can be formulated as a semisupervised problem relying on such a dictionary.  
The pioneer method was proposed in \cite{MESMA} called multiple endmember spectral and mixture analysis (MESMA). MESMA was proposed to address the endmember variability. Assuming a structured library, $\D=[\D_1,\D_2, ..., \D_r]$ contains the endmember bundles, ${\bf D}_i$'s, for all the materials, MESMA allows  different and scaled endmembers for every pixel solving the problem 
\begin{align}\nonumber
      &\hat{\bf a}_i,\hat{\bf E}_i=\arg\min_{{\bf a}_i, {\bf E}_i} \frac{1}{2} || {\bf y}_i-{\bf E}_i{\bf a}_i||_{2}^{2}
~~~{\rm s.t.}~~~\\ &{\bf E}_i\in{\bf D}, {\bf a}_i\geq 0,{\bf 1}_{r}^{T}{\bf a}_i=1.
\end{align}
This is a combinatorial problem, and MESMA searches for all combinations of endmembers (from every endmember bundle) and selects the one with the lowest reconstruction error. Endmember candidates are sometimes called models. This is solved for all the pixels. MESMA is a highly computationally demanding algorithm without any stopping criteria $ \Pi_im_i$ ($m_i$'s are dimensions of endmember bundles) times FCLSU must be computed for every pixel. However, this can be performed in parallel for all the pixels, it is computationally costly, and therefore, several algorithms have been proposed to reduce the computational burden. As a result, instead of FCLSU, the sum to one constrained least square unmixing (SCLSU) was used with a threshold for the error (RE). Note that ignoring the nonnegativity constraint leads to SCLSU, which has a closed-form solution and can be implemented efficiently. Then, solutions that are positive and less than one were selected \cite{MESMA,End_Rob}. Additionally, the number of iterations and the threshold value can be changed to decrease the computations, but it can highly affect the performance \cite{Comp_MESMA}. In \cite{MELSUM}, Unconstrained Least Squares (UCLS) were used to speed up the search for the solutions to the combinatorial problem. 

Alternatively, the size of the library can be decreased \cite{MESMA_LibR1,MESMA_LibR2,MESMA_LibR3}. In \cite{MESMA_Rob}, the angle between a pixel and the projections into hyperplanes of the fixed point (selected endmembers) and the endmember candidates were used to find the optimal endmember. The minimum angle is equivalent to the minimum error, and therefore, the SCLSU problem (note that projection into an affine set or a hyperplane only satisfies ASC) was shifted to more efficient angle computations. FCLSU was used to estimate the abundances using the selected endmembers (the optimal model), and the result with minimum reconstruction error was selected. On the other hand, in \cite{MESMA_VAE}, a deep generative model using variational autoencoders was suggested for augmenting the library endmembers to address the library mismatch. First, VAE was used to augment the library, and then MESMA was used for unmixing. However, this makes the algorithm computationally very expensive. A similar but more efficient approach was proposed in \cite{MESMA_Rob2}, which only uses one unmixing step. %\textcolor{blue}{BSMA,AutoMCU}

%    \subsubsection{Endmember Bundle, and Bundle Unmixing} 
% Endmember variability is a major challenge in spectral unmixing. Endmember variability and spectral variability were sometimes exchangeably used in the literature. We use endmember variability to model spectral variability. However, as discussed, corrections and denoising are also necessary to remove spectral variability. Fig. \ref{} (a) shows how a group of endmembers (endmember bundle) may play roles in determining the data simplex. As shown in Fig. \ref{} (b), the endmember bundles for the highly mixed scenario may be inside the simplex. \textcolor{blue}{reasons of End var}

%\textcolor{blue}{The multiple endmembers were also referred to as the endmember bundles.  The endmember bundle and bundle unmixing were proposed in \cite{MESM, End_Bund1, End_Bund2}. PCA was used to identify the (projected) endmembers using the extremes. More than one extreme was selected for materials. The bundle unmixing was formed using linear programming to determine the minimum, average, and maximum of the fractional abundances induced by endmember variability. In \cite{End_bundle}, endmember bundles were extracted from the observed data.} 

%In \cite{End_var_Rob}

\subsection{Sparse Unmixing}
The first efficient idea was based on sparse and redundant modeling \cite{SUnSAL} and sparse regression, and therefore, it is often referred to as sparse unmixing. We should note that semisupervised unmixing refers to all the unmixing methods relying on a library and is not limited to sparse unmixing methods. %Some semisupervised approaches are not based on sparse regression, e.g., \cite{MESMA, Semi_SU_bays}. However, in this paper, we pay more attention to the sparse unmixing due to the numerous publications in recent years on this topic.  

In sparse unmixing, the fractional abundances are estimated using sparse regression techniques. These methods describe each spectrum as a sparse linear combination of the elements of a rich library of pure spectra.

The main idea of sparse unmixing was first suggested in \cite{SUnSAL}. If a well-designed dictionary is available, the sparse regression can be used to estimate the abundances given by 
\begin{align}\label{eq: SUnSAL}\nonumber
  &\hat{\bf X}=\arg\min_{{\bf X}} \frac{1}{2} || {\bf Y}-{\bf DX}||_{F}^{2}+\lambda ||{\bf X}||_1\\&
~~~{\rm s.t.}~~~{\bf X}\geq 0,{\bf 1}_{m}^{T}{\bf X}={\bf 1}_{n}^{T}.
\end{align}
An ADMM-based algorithm was proposed to solve the problem (\ref{eq: SUnSAL}), and therefore it was called sparse unmixing by variable splitting and augmented Lagrangian (SUnSAL).
%where $\ell_q$ norm is often selected to be a (weighted) sparsity-promoting norm. .   when $q=1$ was proposed in \cite{SUnSAL}   
In  \cite{SUnSAL}, also the constraint SUnSAL (CSUnSAL) proposed given by 
\begin{align}\label{eq: CSUnSAL}\nonumber
  &\hat{\bf X}=\arg\min_{{\bf X}}||{\bf X}||_1
~~~{\rm s.t.}~~~|| {\bf Y}-{\bf DX}||_{F}<\delta,\\& {\bf X}\geq 0,{\bf 1}_{m}^{T}{\bf X}={\bf 1}_{n}^{T},  
\end{align}
and was solved using an ADMM-based algorithm. However, it is suggested to use SUnSAL without ASC due to the conflict with $\ell_1$ \cite{JakubLq}. Additionally, ASC was found to be a rigorous constraint that often does not occur in the real world due to noise and signature variability \cite{SUn}. Therefore, SUnSAL often refers to the problem (\ref{eq: SUnSAL}) without ASC, and problem (\ref{eq: SUnSAL}) refers to FCLS version of SUnSAL which can be seen in the arXiv version of SUnSAL \cite{SUNSAL_arX}. In \cite{SU_3Jose}, problem (\ref{eq: SUnSAL}) is suggested for implementation on a GPU when $\lambda=0$. In \cite{SUn}, the nonconvex $\ell_0$ sparse regression  
\begin{align}\label{eq: L0}\nonumber
  &\hat{\bf X}=\arg\min_{{\bf X}}||{\bf X}||_0
~~~{\rm s.t.}~~~|| {\bf Y}-{\bf DX}||_{F}<\delta,\\& {\bf X}\geq 0,  
\end{align}
were compared with SUnSAL and CSUnSAL. The nonnegative $\ell_0$ sparse regression problem (\ref{eq: L0}) was solved using a greedy algorithm called orthogonal matching pursuit \cite{OMP}. It was shown that SUnSAL outperforms the other two sparse regressions. Additionally, the experiments in \cite{SUn} confirmed the advantage of ANC.  

The success of sparse unmixing depends on the following assumptions; i) the library must well represent the materials in a sense, i.e., it must contain all the endmembers of the materials in the scene; ii) sparse solutions of the underdetermined linear systems exist. The latter depends on the sparseness degree of the observed spectra over the library (i.e., the number of library members representing the spectral pixels) and the coherence of library members. The coherence of a library can be measured using mutual coherence \cite{MC} (or restricted isometric constants (RICs) \cite{RICs}). A higher mutual coherence requires a lower sparseness degree to achieve a sparse solution. In sparse unmixing, the mutual coherence of the spectral library is very high \cite{SUn}, which is the main drawback, however,  the sparseness degree is low (i.e., pixels are mixtures of a few pure spectra). To decrease the mutual coherence, the library must be pruned before applying the sparse regression or throughout the estimation process. Alternatively, the sparsity-inducing prior can be better defined by incorporating spatial information and/or applying segmentation, localization, and group sparsity, which will be discussed in detail in this section.

\subsubsection{Selection of a Regularizer}
In sparse unmixing, many attempts have been made to improve the sparse representation of the abundances over a spectral library by selecting different regularizers. In \cite{SUnSAL-TV}, SunSAL was improved by incorporating spatial information and applying a TV penalty on abundances (SUnSAL-TV) given by
\begin{align}\label{eq: SUnSALTV}\nonumber
  &\hat{\bf X}=\arg\min_{{\bf X}} \frac{1}{2} || {\bf Y}-{\bf DX}||_{F}^{2}+\lambda_1||{\bf X}||_1+\\& \lambda_2TV_{ni}(X)
~~~{\rm s.t.}~~~{\bf X}\geq 0.
\end{align}
where $TV_{ni}(X)$ is a nonisotropic TV function given by
\begin{equation}
    TV_{ni(X)}=||{\bf X}{\bf H}_h^T||_1+||{\bf X}{\bf H}_v^T||_1
\end{equation}
One of the main challenges of sparse regression is the high mutual coherence of the spectral library, which might lead to poor abundance estimation \cite{SUn}. Therefore, in \cite{GSUNSAL}, group SUNSAL (GSUNSAL) was suggested by dividing the library into $L$ groups (structured dictionary) and adding a group LASSO (Least Absolute Shrinkage and Selection Operator) \cite{GLASSO}
    \begin{align}\label{eq: GSUNSAL}\nonumber
  &\hat{\bf X}=\arg\min_{{\bf X}} \frac{1}{2} || {\bf Y}-{\bf D}{\bf X}||_{F}^{2}+\\&\sum_{i=1}^n\sum_{g=1}^L\lambda^g_1||{\bf x}^{g}_i||_2+\lambda_2||{\bf X}||_1
~~~{\rm s.t.}~~~{\bf X}\geq 0.
\end{align}
where ${\bf x}^{g}$ is the $g$th subvector of abundances.
As we mentioned, $\ell_1$ norm conflicts with ASC and therefore $\ell_q$ norm also investigated for sparse unmixing \cite{SU_lq} %when $0 <q< 1$
 \begin{align}\label{eq: SU_lq}\nonumber
  &\hat{\bf X}=\arg\min_{{\bf X}} \frac{1}{2} || {\bf Y}-{\bf DX}||_{F}^{2}+\lambda \sum_{i=1}^n||{\bf x_{i}}||^q_q\\&
~~~{\rm s.t.}~~~{\bf X}\geq 0,{\bf 1}_{m}^{T}{\bf X}={\bf 1}_{n}^{T}, 0 <q< 1 
\end{align}
 where 
 \begin{equation}
          ||{\bf x}||^q_{q}\triangleq \sum_{i=1}^m| x_i|^q,
 \end{equation}
 Collaborative sparse unmixing \cite{Collaborative} applies $\ell_{2,1}$ (i.e., the sum of $\ell_2$ on the abundances) to promote the sparsity on the abundances. 
    \begin{align}\label{eq: colab}\nonumber
  &\hat{\bf X}=\arg\min_{{\bf X}} \frac{1}{2} || {\bf Y}-{\bf DX}||_{F}^{2}+\lambda \sum_{i=1}^m||{\bf x}_{(i)}||_2\\&
~~~{\rm s.t.}~~~{\bf X}\geq 0.
\end{align}
 Collaborative sparse unmixing using $\ell_0$ norm was proposed in \cite{collab_l0} given by 
     \begin{align}\label{eq: colab2}\nonumber
  &\hat{\bf X}=\arg\min_{{\bf X}} \frac{1}{2} || {\bf Y}-{\bf DX}||_{F}^{2}+\lambda ||{\bf X}||_{2,0}\\&
~~~{\rm s.t.}~~~{\bf X}\geq 0.
\end{align}
 where $||{\bf X}||_{2,0}=\sum_{i=1}^mId(||{\bf x}_{(i)}||_2>0)$ and $Id$ is the Heaviside step indicator function, i.e., it is one when the argument is positive. 
In \cite{GSIMN}, the endmember variability was considered using a dictionary of endmember bundles generated from the data. Unlike the methods mentioned  above,  in \cite{GSIMN}, the dictionary is extracted from data points. In \cite{GSIMN}, a framework of sparse regression was proposed given by 
    \begin{align}\label{eq: Gnorm}\nonumber
  &\hat{\bf X}=\arg\min_{{\bf X}} \frac{1}{2} || {\bf Y}-{\bf D}{\bf X}||_{F}^{2}+\\&\lambda ||{\bf X}||_{g,q,d}
~~~{\rm s.t.}~~~{\bf X}\geq 0,{\bf 1}_{m}^{T}{\bf X}={\bf 1}_{n}^{T},
\end{align}
where the grouped sparse norm is defined as 
\begin{equation}
    ||{\bf X}||_{g,q,d}\triangleq\sum_{i=1}^n (\sum_{g=1}^L||{\bf x}^{g}_i||^d_{q})^\frac{1}{d}.
\end{equation}
Three cases are compared, i.e., $\{q=1,d=2\}$, $\{q=1, 0<d<1\}$, and $\{q=2,d=1\}$. 
We should note that ASC was used in the endmember bundles method \cite{GSIMN}, unlike the structured dictionary methods \cite{Collaborative, GSUNSAL}.

In \cite{DRSU_TV}, a spatial-spectral weighted $\ell_1$ norm used for (\ref{eq: SUnSALTV}) given by
\begin{align}\label{eq: DRSUTV}\nonumber
  &\hat{\bf X}=\arg\min_{{\bf X}} \frac{1}{2} || {\bf Y}-{\bf DX}||_{F}^{2}+\lambda_1||{\bf W}\odot{\bf S}{\bf X}||_1+\\& \lambda_2TV_{ni}({\bf X})
~~~{\rm s.t.}~~~{\bf X}\geq 0.
\end{align}
where ${\bf S}={\bf s}\1^T$, $s_i=(||{\bf x}_{(i)}||_1+\epsilon)^{-1}$, and ${\bf W}=({\bf X}+\epsilon)^{-1}$ are the weight matrices updated in every iterations. The weights further promote the sparsity on ${\bf X}$. 

Local collaborative sparse unmixing (LCSU) \cite{SZhang2016} and a spectral-spatial weighted sparse unmixing (S$^2$WSU) \cite{SZhang2018} further incorporate spatial information by adapting the collaborative sparse unmixing and the reweighted $\ell_1$ norm locally to neighboring pixels, respectively. Indeed, S$^2$WSU uses problem (\ref{eq: DRSUTV}) locally and when $\lambda_2=0$.

In \cite{SU_LR}, the weighted nuclear norm was added to the weighted $\ell_1$ norm to utilize the low-rankness of the abundances. The proposed sparse regression is given by  
\begin{align}\label{eq: SU_LR}\nonumber
  &\hat{\bf X}=\arg\min_{{\bf X}} \frac{1}{2} || {\bf Y}-{\bf DX}||_{F}^{2}+\lambda_1||{\bf W}\odot{\bf X}||_1+\\& \lambda_2||{\bf X}||_{t,*}
~~~{\rm s.t.}~~~{\bf X}\geq 0.
\end{align}
where the weighted nuclear norm is given by
\begin{equation}
    ||{\bf X}||_{t,*}=\sum_{i=1}^rt_i\sigma_i({\bf X}),
\end{equation}
and $\sigma_i({\bf X})$ are the singular values of ${\bf X}$ and the weights are given by $t_i=(\sigma_i({\bf X})+\epsilon)^{-1}$. In \cite{SU_LR_SSW}, the spatial-spectral weighted $\ell_1$ was used with the weighted nuclear norm. In \cite{SU_SpNorm}, the weighted Schatten p-norm \cite{SpNorm} and a collaborative norm was suggested to induce the low rankness and the sparsity, respectively, given by 
\begin{align}\label{eq: SU_LR2}\nonumber
  &\hat{\bf X}=\arg\min_{{\bf X}} \frac{1}{2} || {\bf Y}-{\bf DX}||_{F}^{2}+\lambda_1\sum_{i=1}^m||{\bf x}_{(i)}||^q_{2,q}+\\& \lambda_2||{\bf X}||^q_{t,Sp}
~~~{\rm s.t.}~~~{\bf X}\geq 0, 0< q \leq 1.
\end{align}
where
\begin{align}
     &||{\bf x}_{(i)}||^q_{2,q}\triangleq \sum_{i=1}^m||{\bf x}_{(i)}||^q_{2},\\
     &
     ||{\bf X}||^q_{t,Sp}\triangleq \sum_{i=1}^rt_i\sigma^d_i({\bf X}).
\end{align}
\textcolor{black}{Tensor decomposition can be also used for sparse unmixing formulation using the mode-n tensor multiplication \cite{TD_SU1, TD_SU2}.} %Several methods were proposed using such representation and enforcing a verity of penalties.  }
%\textcolor{blue}{In \cite{XJiang2021}, an efficient two-phase multiobjective sparse unmixing approach was presented to exploit the spatial-contextual information for improving the abundance estimation.}

\subsubsection{Segmentation/Super-pixel and Spectral Variability}

Some methods apply segmentation prior to sparse unmixing for grouping. In this way, they remove spectral variation and capture spatial information. A fast Multiscale Sparse Unmixing Algorithm (MUA) was proposed in \cite{RABorsoi2019}. MUA first applies segmentation techniques such as a binary partition tree (BPT), simple linear iterative clustering (SLIC), or K-means to group pixels. Then SUnSAL was utilized for unmixing the average pixels of every segment. The coarse fractional abundance matrix estimated using SUNSAL (${\bf X}_{seg.}$) is used as a prior for the final sparse regression given by 
     \begin{align}\label{eq: MUA}\nonumber
  &\hat{\bf X}=\arg\min_{{\bf X}} \frac{1}{2} || {\bf Y}-{\bf DX}||_{F}^{2}+\lambda_1 ||{\bf X}||_1+\lambda_2 ||{\bf X}-{\bf X}_{seg.}||_{2}^2\\&
~~~{\rm s.t.}~~~{\bf X}\geq 0.
\end{align}

This idea is also called "super-pixel" in the literature, and numerous methods have been proposed using this terminology. In \cite{TanerInce2020}, SLIC was utilized for the pixel grouping and pixel-based sparse unmixing was performed using superpixel-based graph Laplacian regularization. In \cite{SUn_SV}, the spectral variability was considered using a spectral variability dictionary. The dictionary was obtained by applying PCA on an endmember library extracted from the dataset using a pure pixel-based endmember extraction \cite{SPEE}. The PCs with high variance were used to create the spectral variability library.

\subsubsection{Library pruning}
Library pruning is a necessary preprocessing step in sparse unmixing to reduce the mutual coherence of the spectral library. The pruning step considerably improves the computational time. However, it risks losing endmembers that are scaled versions of each other. Nevertheless, in real-world applications, the library pruning must be done cautiously, particularly when the SNR is low, and the number of endmembers is high \cite{SU_LibP_BS}.  A simple pruning often used before applying sparse unmixing is to compute SAD. Then, spectra with small angles are removed from the library to reduce the mutual coherence. A pruning strategy was proposed in \cite{Lib_Prun}. The library was projected into the data subspace, and spectra with high projection errors (the normalized Euclidean distance) were removed from the library. Arguably, data-dependent library pruning is more suitable for sparse unmixing. In \cite{MUSIC_CSR}, a library pruning-based sparse unmixing called multiple signal classification collaborative sparse unmixing (MUSIC-CSR) was proposed. The library was pruned (before applying the collaborative sparse unmixing) using an orthogonal projection (${P}^\perp={\bf I}-{\V\V}^T$)  \cite{MUSIC} where $\V$ is the bases of the $\Y$'s subspace obtained by HySime \cite{HySime} to reduce the noise effect. The normalized projection error of each library endmember $||{P}^\perp{\bf d}_i||_2/||{\bf d}_i||_2$ was sorted by increasing order. Hence, the first $r$ defines the reduced library. A similar idea was used in \cite{SU_LibP_MO}, which formulates the library reduction into a multiobjective optimization w.r.t. a binary vector representing the support of the abundances matrix.  

\subsubsection{Library Mismatch}
As mentioned, the endmember library is fixed in the problems and methods discussed, and the ultimate goal is abundance estimation. However, even a pruned and well-selected spectral library cannot represent all the endmembers of materials in a real-world dataset. There are several factors, such as noise, atmospheric effects, illumination variations, and the intrinsic variation of materials, which may change the endmembers and induce scaling factors into the endmembers in the scene compared to the ones from the library \cite{SUn, SUnSAL_Leg} known as library mismatches. This problem was partially addressed in \cite{SUn}. A bandwise scaling $ {\bf SY}$ was suggested to apply to the dataset where the diagonal matrix ${\bf S}$ was obtained by solving
\begin{align}\label{eq: rescale}\nonumber
  &(\hat{\bf X},\hat{\bf S})=\arg\min_{{\bf X,S}}  || {\bf SY}-{\bf DX}||_{F} \\&
{\rm s.t.}~~{\bf X}\geq 0,{\bf 1}_{m}^{T}{\bf X}={\bf 1}_{n}^{T},
\end{align}
A cyclic descent solution was suggested with the initialization of ${\bf S}={\bf I}$ turning the problem into iterative problems of least squares (w.r.t. S) and constrained least squares (w.r.t. X). The nonsparse fractional abundances of $\X$ and the corresponding spectral vectors of ${\bf Y}$ were removed after the first iteration to speed up the algorithm. Problem (\ref{eq: rescale}) is noncovex, therefore, the solution will be suboptimal. Additionally, the bandwise scaling ${\bf S}$ was estimated for a suboptimal sparse $\X$.

In \cite{SU_LibP_Mis}, the library mismatch was incorporated into the sparse unmixing problem given by
\begin{align}\label{eq: LibP_Mis}\nonumber
  &(\hat{\bf X}, \hat{\bf D}')=\arg\min_{{\bf X}} \frac{1}{2} || {\bf Y}-{\bf D}'{\bf X}||_{F}^{2}+\lambda\sum_{i=1}^m||{\bf x}_{(i)}||^q_{2,q},\\& {\rm s.t.} ~~{\bf X}\geq 0, ||d_i-d'_i||_2<\delta, i=1, 2, ...,m.
\end{align}
We should note that a surrogate function was suggested to be minimized instead of (\ref{eq: LibP_Mis}) to reduce the complexity of the algorithm (see \cite{SU_LibP_Mis}). Additionally, the MUSIC pruning method was adapted (so-called robust MUSIC) to consider the mismatches given by 
\begin{equation}
    \arg\min_{{\bf z}}\frac{({\bf d}_i-{\bf z})^TP^\perp({\bf d}_i-{\bf z})}{||{\bf d}_i-{\bf z}||_2^2}~~ {\rm s.t.} ~~||{\bf z}||_2<\delta.
\end{equation}
Recently, sparse unmixing using archetypal analysis (SUnAA) was proposed in \cite{SUnAA}, which  addresses the spectral library mismatch from a different viewpoint. Instead of scaling data, SUnAA assumes that endmembers can be a convex combination of the library endmembers. Therefore, an extra matrix defines the contributions of the endmember from the library. The proposed linear model in \cite{SUnAA} utilizes the advantages of the low-rank model  and the sparse and redundant model defined as
 \begin{align}\label{eq: SRLR}\nonumber
&{\bf Y} = {\bf D}{\bf B}{\bf A} + {\bf N}, 
~~{\rm s.t.}~~{\bf B}\geq 0,{\bf 1}_{m}^{T}{\bf B}={\bf 1}_{r}^{T}, \\& {\rm and }~~{\bf A}\geq 0,{\bf 1}_{r}^{T}{\bf A}={\bf 1}_{n}^{T}, 
\end{align} 
where ${\bf B}\in \mathbb{R}^{m\times r}$ determines the contributions of the endmembers from the the library ${\bf D}$ and ${\bf A}\in \mathbb{R}^{r\times n}$ is the (low-rank) abundance matrix. To simultaneously estimate  ${\bf B}$ and ${\bf A}$ SUnAA solves a minimization problem given by 
\begin{align}\label{eq: SUnAA}\nonumber
  &(\hat{\bf B},\hat{\bf A})=\arg\min_{{\bf B,A}} \frac{1}{2} || {\bf Y}-{\bf DBA}||_{F}^{2} \\&
{\rm s.t.}{\bf B}\geq 0,{\bf 1}_{m}^{T}{\bf B}={\bf 1}_{r}^{T},  {\rm and }{\bf A}\geq 0,{\bf 1}_{r}^{T}{\bf A}={\bf 1}_{n}^{T}, 
\end{align}
SUnAA assumes that the unknown endmembers are a convex combination of the library's endmembers which leads to enforcing non-negativity and sum to one constraint on ${\bf B}$. Unlike conventional sparse unmixing, the problem  (\ref{eq: SUnAA}) is nonconvex. A  cyclic descent algorithm based on an active set method \cite{RAA} was proposed to solve (\ref{eq: SUnAA}), iteratively. We should note that SUnAA is a parameter-free technique.  

Numerous techniques were tailored utilizing a combination of the aforementioned ideas, i.e., spatial regularization (TV), spatial/spectral weighted norm, collaborative norm, low-rank inducing, superpixel (segmentation), structured library/ group norm, and localization, library pruning \cite{SU_GM1,SU_GM2,SU_GM3,SU_GM4,SU_GM5,SU_GM6,SU_GM7,SU_Mix}. Here, we have drawn the main ideas behind the majority of sparse techniques. 

%\subsubsection{Endmember Variability}
%In \cite{SU_End_Var}
\subsubsection{Shallow/Deep Neural Networks}
There are a few NN-based sparse unmixing approaches proposed in the literature. The pioneer method is called sparse unmixing using an unsupervised convolutional neural network (SUnCNN) \cite{SUnCNN}. As mentioned, the $\ell_1$ penalties cannot be applied to the abundance while holding the ASC. This was addressed in \cite{SUnCNN} by implicitly applying an image prior and holding the ASC and ANC. It was shown that the problem of selecting a suitable prior for a sparse regression given by 
\begin{align}\label{eq: IP}\nonumber
  \hat{\bf X}=&\arg\min_{{\bf X}} \frac{1}{2} || {\bf Y}-{\bf DX}||_{F}^{2}+\lambda \phi({\bf X})\\&
{\rm s.t.}{\bf X}\geq 0,{\bf 1}_{m}^{T}{\bf X}={\bf 1}_{n}^{T},  
\end{align}
could be moved to optimizing the parameters of a deep network ($f_{\theta}$) given by
\begin{equation}\label{eq: SUnCNN}
\hat{\theta}=\arg\min_{{\bf \theta}} \frac{1}{2} || {\bf Y}-{\bf D}f_{\theta} ({\bf Z})||_{F}^{2}%+\lambda R({\bf A})
~~~{\rm s.t.}~~ \hat{\bf X}=f_{\hat\theta}({\bf Z}),
%{\bf A}\geq 0,1_{r}^{T}{\bf A}=1_{n}^{T},
\end{equation}
where $f_\theta$ is the deep CNN with parameters $\theta$. and ANC and ASC can be enforced using a softmax layer. In \cite{SMALU}, an asymmetric encoder-decoder architecture is used for sparse unmixing. Instead of softmax, a sparse variation of softmax is used to avoid the full support of softmax while enforcing ASC. 

Spectral Variability Augmented Two-Stream Network (SVANT) was proposed in \cite{SVANT}. SVATN uses the convolutional encoder-decoder used in
SUnCNN for both streams. The first stream uses a spectral library to estimate the abundances, and the second stream uses a spectral variability library to estimate spectral variation. A correlation-based variability extraction method was proposed to create a spectral variability library. A spectral–spatial feature learning encoder-decoder network was proposed in \cite{SS_Net}. The encoder utilizes two branches with spatial and spectral filters to extract the spatial and spatial features, respectively. The concatenated features go through three spectral–spatial residual blocks, which utilize MLP for spectral attention. On the other hand, the decoder is shallow and uses the endmember library to reconstruct data.

Algorithm unrolling-based approaches were also proposed for sparse unmixing. An unrolling-based shallow network was recently proposed for sparse unmixing in \cite{ISTA_Unrol_SUn,SU_Unroll}. An ADMM-based solution to the nonnegative $\ell_1$ sparse regression problem (so-called SUnSAL), i.e., problem (\ref{eq: SUnSAL}), was unrolled. An intermediate convolution was applied to the abundance to incorporate spatial information. A combination of SAD, MSE, and SID was used as the loss function to train the shallow network. In \cite{ISTA_Unrol}, the iterative soft-thresholding algorithm (ISTA) \cite{ISTA} was unrolled to solve the nonnegative $\ell_1$ sparse regression problem. SUnSAL was also unrolled in \cite{SUNSAL_Unrol} for sparse unmixing. 

\section{Nonlinear Unmixing Approaches}
The linear mixture model is a simplified model and often fails in the case of intimate mixtures or/and when the light undergoes multiple reflections before reaching the sensor \cite{Dias_HS_Rev2013}. Alternatively, nonlinear models are used \cite{unmixing-review,Dobigeon_2014_NU}. 

Bilinear approaches are often used for double scattering. They assume that the received light at the sensor interacts with two materials. This is modeled using an extra mixing term by the Hadamard product between the endmembers of the materials. The Fan model \cite{Fan}, \cite{Rob_NonRev} is a variation of this model. One disadvantage of the Fan model is the generalization; therefore, it performs poorly for linearly mixed datasets. Polynomial post nonlinear mixing model (PPNM) \cite{PPNM}, generalized bilinear model (GBM) \cite{AHalimi2011}, and linear-quadratic model (LQM) \cite{Rob_NonRev} were proposed to generalize the Fan model for the linear mixtures. However, they contain hyperparameters to describe the trade-off between the linear and nonlinear terms. In \cite{LianruGAO2022}, a nonlinear low-rank tensor unmixing algorithm was proposed to solve the GBM.  Bilinear models have physical interpretation in some specific applications, such as canopy scenarios; however, they have several disadvantages. They have many parameters, and the estimated abundances are hard to interpret. Additionally, they often do not include self-interactions or consider the reflections from objects outside the instance field of view \cite{Rob_NonRev}. Moreover, they are limited to secondary interactions. Therefore, several nonlinear mixing models have been developed,  such as the multilinear mixing model (MLM) \cite{RHeylenMLM} and the p-linear (\emph{p} $>$ 2) mixture model (pLMM) \cite{AMarinoni20159,AMarinoni20158,AMarinoni20169} were proposed for multiple interactions of the incident light. 

Kernel methods have also been employed for nonlinear unmixing \cite{Rob_NonRev}. They represent data in a higher dimensional space to linearize the problem. Hence, linear models become effective in high-dimensional space. In \cite{KFCLSU}, a kernel nonnegative matrix factorization (kernel-NMF) was proposed in \cite{KNMF} for nonlinear unmixing. The FCLSU problem can be solved using support vector machines (SVMs). Pure pixels in  data are the common support vectors that span the data simplex and allow the estimation of the abundances of all pixels enclosed within the simplex. Hence, Kernel SVMs were proposed for nonlinear spectral unmixing in \cite{SVM_Un}.

Radiative transfer models are 
mathematical physic-based mixing models that can be used to reconstruct intimately mixed materials' reflectance spectrum \cite{Rob_NonRev}. The inverse problems based on such models are very hard to solve. The Hapke model estimates the areal fractions of the materials within the mixture by transforming the reflectance spectra to their single-scattering albedos (SSA) and applying linear unmixing. The simplified version of the Hapke model was used for predicting intimate mixtures' composition \cite{BHapke1998, BHapke1981}. The proposed nonlinear unmixing methods based on Radiative transfer models often assume that spectral reflectance of the pure materials is available for estimating fractional abundances. 

Deep networks were also proposed for nonlinear unmixing. Deep AE-based architectures have been broadly used for linear unmixing. In \cite{NU_DAE}, an AE network was proposed where the encoder utilizes an extra nonlinear layer to capture the nonlinearity of the data. A deep AE with multitask learning was suggested in \cite{BU_DAE_MTL}. A long short-term memory-based autoencoder was proposed for PPNM in \cite{DL_PPNM}. The proposed method in \cite{3DCNN_PPNM} exploits a 3-D CAE-based network for PPNM. A supervised AE was used in \cite{AE_RBF_PPNM} for Fan, bilinear, and PPNM.  The Radial basis function (RBF) kernels and K-means clustering were used to estimate the number of endmembers and the endmember spectra, respectively. DL-based nonlinear unmixing techniques are often AE-based networks using PPNM, and as we discussed, they may have disadvantages of the bilinear models mentioned above \cite{MB_DAE_PPNM,AE_RBF_PPNM2,Simplex_AE_PPNM}.  GAN architectures were also explored  for nonlinear unmixing. In \cite{NU_GAN}, a cycle-consistent loss was used to ensure the reconstruction in addition to two GAN losses. A CNN was designed based on Hapke model (HapkeCNN) in \cite{HapkeCNN} to incorporate the physical model in the learning process.

\section{Experimental Results}
We use three simulated and one real datasets. The simulated datasets were designed according to different mixing scenarios briefly explained in Table \ref{tab:data_description}. We avoid using the widely used benchmark datasets such as Samson and Jasper since their abundances are generated synthetically. For real-world experiments, we use the Cuprite dataset, a well-studied site with geological reference maps. The simulated experiments were carried out for 30 dB SNR. The tuning parameters were fine-tuned for the methods up to some levels. We should note that some methods have several parameters to be tuned; therefore, searching for the optimum is cumbersome. All the results are averaged over 10 experiments, and the standard deviations are shown by error bars. In experiments, we compare 20 unmixing methods from different categories as follows. For supervised methods, we used three endmember extraction/estimation techniques, i.e., VCA, SiVM, and SISAL with FCLSU and UnDIP. All six combinations of them were considered. For blind unmixing, we use PGMSU, MSNET, CNNAEU, ADMMNet, BayesianSMA, NMFQMV, MiSiCNet, and EDAA. For sparse unmixing, we used SUnSAL, CLSUnSAL, MUA\_SLIC, S$^2$WSU, SUnCNN, and SUnAA. The codes for all those methods were provided in  HySUPP for reproducibility.  

For the quantitative evaluation, we use the SRE in dB for estimated abundances given by %given by 
\begin{equation}
    \text{SRE}(\X, \hat{\X})=20\log_{10}\frac{\|{\X}\|_F}{\|{\X}-\hat{\X}\|_F}.
\end{equation}
and the spectral angle distance (SAD) in degrees between the estimated and ground truth endmembers: 
\begin{equation}
\mbox{SAD}({\bf E},\hat{\bf E})=\frac{1}{r}\sum_{i=1}^r\arccos{\left(\frac{ \left\langle{\bf e}_{(i)}, \hat{\bf e}_{(i)}\right\rangle}{\left\|{\bf e}_{(i)}\right\|_2\left\|\hat{\bf e}_{(i)}\right\|_2}\right)}\frac{180}{\pi},
\end{equation} 
where $\langle . \rangle$ denotes the inner product. % and ${\bf e}_{(i)}$ indicates the $i$th column of ${\bf E}$. }
\subsection{Data Description}

\subsubsection{Synthetic Datasets with Spatial Structure}
We simulated three data cubes (DC1, DC2, and DC3). DC1 was simulated using a linear mixing model with 5 endmembers selected from the USGS library and 75$\times$75 pixels. The abundance maps are composed of five rows of square regions uniformly distributed over the spatial dimension. This dataset contains pure pixels for all endmembers. DC2 has 100$\times$100 pixels and was simulated using a linear mixing model with 9 endmembers. The abundance maps were sampled
from a Dirichlet distribution centered at a Gaussian
random field to have piece-wise smooth maps with steep transitions. Therefore, DC2 contains spectral variations. For DC1 and DC2, an endmember library $\D \in \mathbb{R}^{188\times 240}$, composed of 240 spectral signatures were selected from the USGS library with a minimum pair-spectra angle of 4.44\textdegree. DC3 was simulated with no pure pixels, and it has two mixed pixels on the facet of the data simplex. DC3 is composed of  105$\times$105 pixels using the linear combination of six endmembers. For DC3, we use a library $\D \in \mathbb{R}^{188\times 498}$ composed of 498 spectral pixels from the USGS library. Note that we remove the water absorption and noisy bands, and therefore, the final pixels are of dimension $p=188$.

%\subsubsection{Synthetic Datasets with different Mixing Ratio} \textcolor{blue}{Alex}

% \subsubsection{Apex Dataset}
% In experiments, we use a crop of Apex datasets \cite{SCHAEPMAN2015207} containing 111$\times$122 pixels and 285 bands covering the wavelength range from 413 to 2420 nm. Four ground truth endmembers i.e., Water, Tree, Road, and Roof were selected which are the dominant materials in the scene. This dataset was only used to compare the supervised and blind unmixing methods.  
\subsubsection{Cuprite Dataset}
The Cuprite dataset used in this paper contains 250$\times$ 191 pixels. Cuprite is a well-studied mineral site, and the dominant minerals are demonstrated using a geological ground reference therefore, the estimated abundance maps by different techniques can be compared visually. %The geological ground reference for the dominant minerals existing in the scene is shown in Fig. \ref{Real dataset} (a). 
We use the same library as explained for DC3.

\begin{table*}[ht]
    \centering
    \caption{Specifications of the synthetic datasets used in the experiments.}
    \begin{tabular}{c|c c c c }
    \toprule
         & \# endmembers ($r$) & \# atoms in ${\bf D}$ ($m$) & \# pixels ($n$) & Main features \\
    \midrule
        DC1 & 5 & 240 & 5625 (75 $\times$ 75)  & Pure pixels\\
        DC2 & 9 & 240 & 10000 (100 $\times$ 100)  & Pure pixels, Spectral variability\\
        DC3 & 6 & 498 & 11025 (105 $\times$ 105)  & No pure pixels, 2 points on the data simplex facets\\
%    \midrule
%        DC4\_100 & 6 & 498 & 1000 & 1.00 & \emph{Pure pixels scenario} \\
%        DC4\_85 & 6 & 498 & 1000 & 0.85 & \emph{Moderately mixed scenario} \\
%        DC4\_70 & 6 & 498 & 1000 & 0.70 & \emph{Highly mixed scenario} \\
        
    \bottomrule
    \end{tabular}
    
    \label{tab:data_description}
\end{table*}

%\subsubsection{Pure Pixel Scenario}
%\subsubsection{Hihghly Mixed Scenario}
%\subsection{Experimental Results: Real Datasets with simulated Ground Truth}

%\subsubsection{Apex}

\subsection{Experimental Results: Synthetic Datasets}
\begin{figure*}[htbp]
   \centering
   \begin{tabular}{c} 
   \includegraphics[width=.8\textwidth]{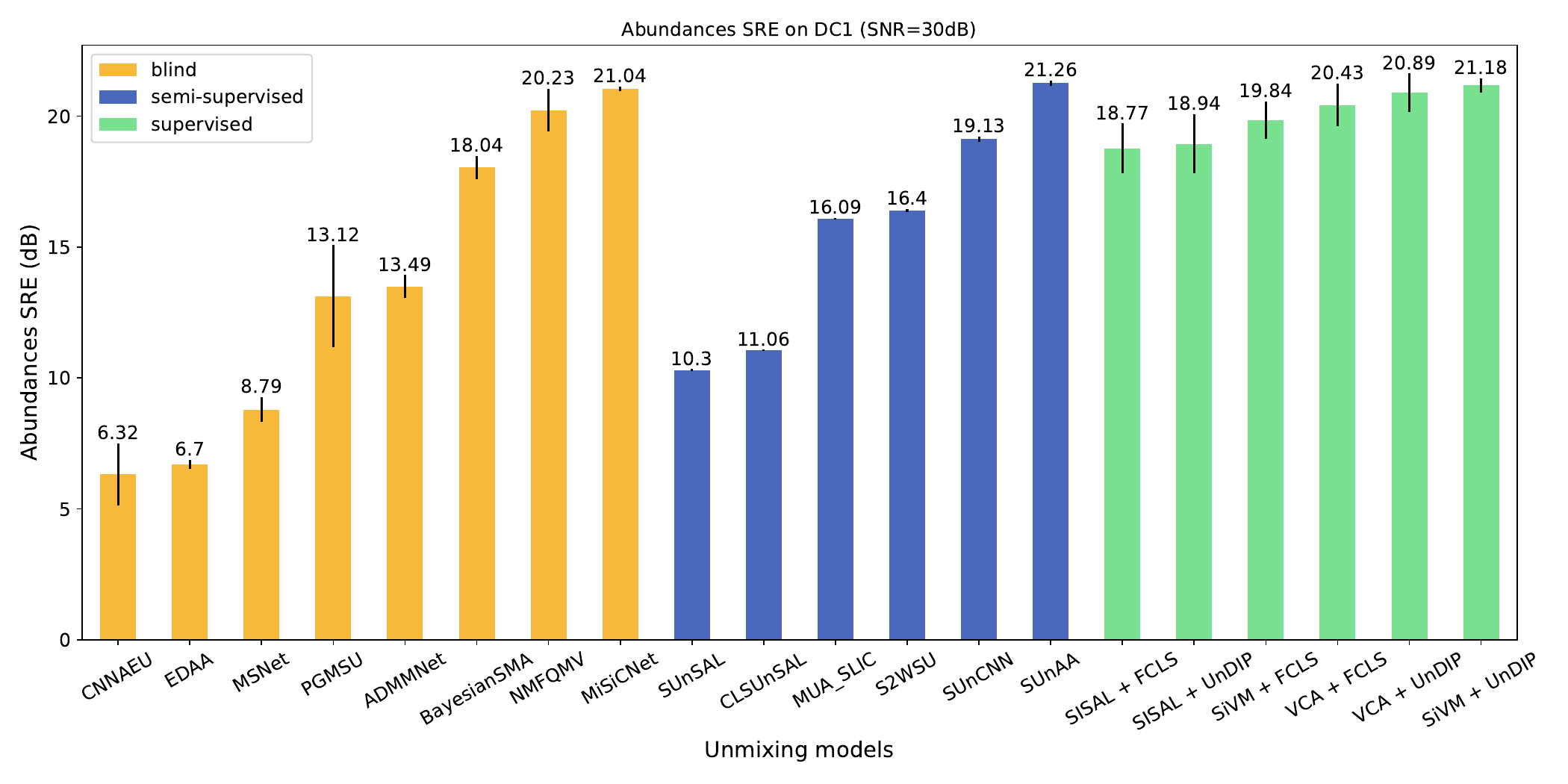}\\   
   \includegraphics[width=.8\textwidth]{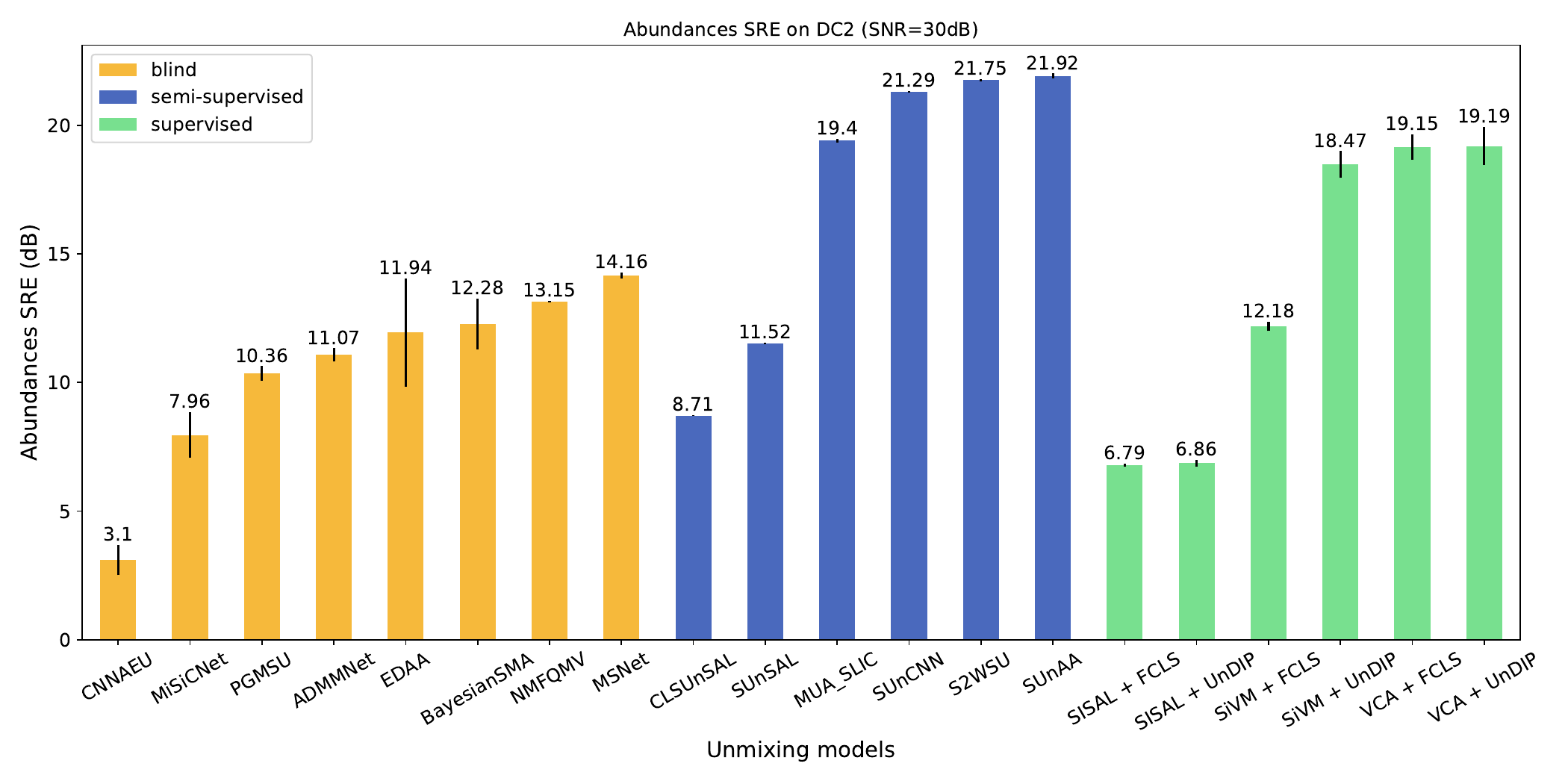}
\\ 
   \includegraphics[width=.8\textwidth]{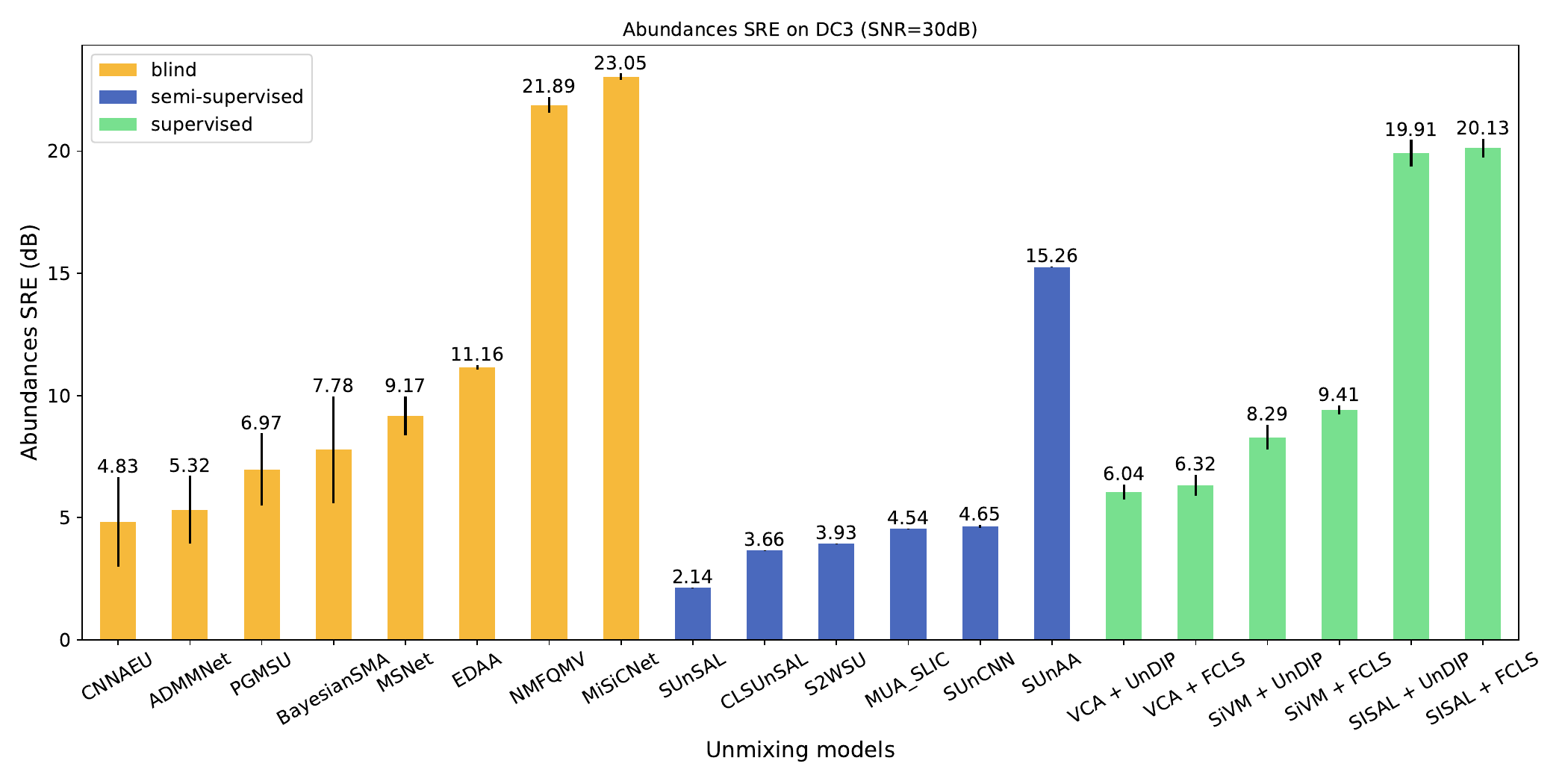}
 \end{tabular} 
   \caption{Comparing abundance SRE ($\uparrow$) in dB using different unmixing techniques applied to (from top to bottom) synthetic DC1, DC2, and DC3.}
   \label{fig:DC1_30dB_SRE}
\end{figure*}

Figure \ref{fig:DC1_30dB_SRE} demonstrates the unmixing results in terms of abundance SRE in DB for different techniques applied to DC1, DC2, and DC3. The outcomes of the results can be summarized as follows: 
\begin{itemize}
    \item For the pure pixel scenario, supervised techniques perform very well (Fig. \ref{fig:DC1_30dB_SRE}, DC1). Overall, they perform better than the semisupervised and blind techniques. This confirms the importance of geometrical information for endmember extraction/ estimation techniques when there are pure pixels in the dataset. This is further confirmed in blind methods where MiSiCNet and NMFQMV, which exploit geometrical information, outperform the other blind techniques and provide competitive results compared to supervised ones. Sparse techniques show moderate results except SUnAA, which outperforms all the other techniques. The results confirm that the sparse unmixing techniques are not suitable when there are pure pixels for the endmembers in the dataset. We should mention that SUnAA does not match the characteristics of conventional sparse techniques. Even though SUnAA relies on a library, it uses a nonconvex optimization to estimate endmembers. 
    \item On the other hand, for DC2, which contains spectral variations, as can be seen from Fig. \ref{fig:DC1_30dB_SRE}, sparse unmixing techniques outperform the supervised and blind techniques. The results confirm that sparse unmixing techniques are more suitable for capturing the spectral variability. SUnAA outperforms the other technique. Note that SUnAA is a parameter-free technique. Supervised techniques outperform blind techniques due to the presence of pure pixels. 
    \item In the case of no pure pixel scenario (Fig. \ref{fig:DC1_30dB_SRE}, DC3), blind unmixing techniques that exploit geometrical information outperform the other techniques. Sparse unmixing provides very poor results. Although SUnAA considerably significantly outperforms the performance of sparse unmixing techniques, it is very far from the best performance which is obtained by MiSiCNet. Among supervised techniques, SISAL (+ FCLSU/UnDIP) shows competitive results because it uses geometrical information to estimate the endmembers.  
    \item Table \ref{SAD} compares the SAD values for the endmember estimation/extraction techniques with the blind unmixing techniques for the simulated datasets at 30dB. In the case of pure pixel scenarios, SiVM and VCA outperform the other techniques confirming the advantage of geometrical approaches for endmember extraction when there are pure pixels for all the endmembers. However, in the no-pure pixel scenario, they cannot estimate the endmembers. On the other hand, the blind unmixing methods that incorporate geometrical information i.e., NMFQMV and MiSiCNet outperform the other techniques. Overall, the results reveal the advantage of the geometrical information in accurate endmember extraction/estimation. Finally, we should mention that SAD is scale-invariant and cannot solely be considered for the evaluation of unmixing models.
\end{itemize}

% \begin{figure*}[h]
%    \centering
%    \begin{tabular}{c} 
%    \includegraphics[width=.8\textwidth]{figs/Mixed100_30.pdf}\\
%    \includegraphics[width=.8\textwidth]{figs/Mixed85_30.pdf}
% \\ 
%    \includegraphics[width=.8\textwidth]{figs/Mixed70_30.pdf}
%  \end{tabular} 
%    \caption{Comparing abundance SRE in dB using different unmixing techniques applied to synthetic DC4 with different mixing ratio (from top to bottom) $\rho=$ 1, 0.85, 0.70.}
%    \label{fig:DC4_30dB_SRE}
% \end{figure*}

% \begin{figure*}[h]
%    \centering
%    \includegraphics[width=\textwidth]{figs/DC1_30.pdf}
%    \caption{TODO.}
%    \label{fig:DC1_30dB_SRE}
% \end{figure*}
% \begin{figure*}[h]
%    \centering
%    \includegraphics[width=\textwidth]{figs/DC2_30.pdf}
%    \caption{TODO.}
%    \label{fig:DC2_30dB_SRE}
% \end{figure*}
% \begin{figure*}[h]
%    \centering
%    \includegraphics[width=\textwidth]{figs/DC3_30.pdf}
%    \caption{TODO.}
%    \label{fig:DC3_30dB_SRE}
% \end{figure*}
% \begin{figure*}[h]
%    \centering
%    \includegraphics[width=\textwidth]{figs/Mixed100_30.pdf}
%    \caption{TODO.}
%    \label{fig:Mixed100_30dB_SRE}
% \end{figure*}
% \begin{figure*}[h]
%    \centering
%    \includegraphics[width=\textwidth]{figs/Mixed85_30.pdf}
%    \caption{TODO.}
%    \label{fig:Mixed85_30dB_SRE}
% \end{figure*}

% \begin{figure*}[h]
%    \centering
%    \includegraphics[width=\textwidth]{figs/Mixed70_30.pdf}
%    \caption{TODO.}
%    \label{fig:Mixed70_30dB_SRE}
% \end{figure*}

\begin{table*}[ht]
    \centering
    \caption{\textcolor{black}{Spectral Angle Distance (SAD, in degrees) results on the simulated datasets.}}
    \begin{adjustbox}{width=\linewidth}
    \begin{tabular}{c|c| c c c c c c c c | c c c}
    \toprule
    Dataset & SNR & ADMMNet & BayesianSMA & CNNAEU & EDAA & MSNet & MiSiCNet & NMFQMV & PGMSU & SISAL & SiVM & VCA \\
    \midrule
    DC1 & 30 & 0.84 $\pm$ 0.33 & 0.49 $\pm$ 0.02 & 8.12 $\pm$ 0.12 & 4.93 $\pm$ 1.15 & 1.77 $\pm$ 1.27 & \textbf{0.43} $\pm$ 0.01 & 0.99 $\pm$ 0.22 & 1.63 $\pm$ 0.32 & 1.85 $\pm$ 0.21 & \textbf{0.43} $\pm$ 0.02 & 0.45 $\pm$ 0.01\\
    DC2 & 30 & 2.20 $\pm$ 0.13 & 1.65 $\pm$ 0.13 & 4.63 $\pm$ 0.95 & 0.65 $\pm$ 0.51 & 0.76 $\pm$ 0.03 & 2.98 $\pm$ 0.40 & 1.07 $\pm$ 0.01 & 1.97 $\pm$ 0.09 & 9.26 $\pm$ 0.80 & \textbf{0.45} $\pm$ 0.05 & 0.53 $\pm$ 0.04 \\
    DC3 & 30 & 8.07 $\pm$ 3.61 & 9.26 $\pm$ 2.87 & 9.68 $\pm$ 0.50 & 4.82 $\pm$ 0.04 & 4.30 $\pm$ 0.22 & 0.87 $\pm$ 0.03 & \textbf{0.45} $\pm$ 0.08 & 6.73 $\pm$ 1.08 & 2.32 $\pm$ 0.17 & 7.09 $\pm$ 0.36 & 4.70 $\pm$ 0.75\\
    %Mixed70 & 30 & 8.92 $\pm$ 1.73 & 13.13 $\pm$ 2.53 & 14.80 $\pm$ 0.04 & 8.93 $\pm$ 0.20 & 7.99 $\pm$ 0.27 & 2.49 $\pm$ 0.42 & \textbf{0.75} $\pm$ 0.03 & 8.71 $\pm$ 1.54 & 2.27 $\pm$ 0.21 & 8.12 $\pm$ 1.04 & 8.42 $\pm$ 1.42 \\
    %Mixed85 & 30 & 2.28 $\pm$ 0.55 & 5.25 $\pm$ 0.08 & 14.73 $\pm$ 0.08 & 2.84 $\pm$ 0.05 & 2.93 $\pm$ 0.08 & 1.50 $\pm$ 0.04 & \textbf{0.84} $\pm$ 0.09 & 2.22 $\pm$ 0.27 & 2.72 $\pm$ 0.20 & 4.03 $\pm$ 0.55 & 3.99 $\pm$ 0.93 \\
    %Mixed100 & 30 & 1.14 $\pm$ 0.0 & 2.52 $\pm$ 0.02 & 14.79 $\pm$ 0.25 & 0.47 $\pm$ 0.04 & 1.16 $\pm$ 0.04 & 1.07 $\pm$ 0.04 & 0.84 $\pm$ 0.62 & 1.02 $\pm$ 0.15 & 2.96 $\pm$ 0.20 & \textbf{0.47} $\pm$ 0.03 & 0.70 $\pm$ 0.30 \\
    \bottomrule
    \end{tabular}
    \end{adjustbox}
    \label{SAD}
\end{table*}

\subsection{Experimental Results: Real Data}
%Will be added soon 
We selected three methods per category to conduct unmixing on the Cuprite dataset. 
Blind unmixing: MiSiCNet, MSNet and NMFQMV. 
Semi-supervised: SUnAA, MUA\_SLIC, S$^2$WSU.
Supervised: UnDIP combined with SISAL, SiVM and VCA.
We describe the hyperparameters that were fine-tuned for the following techniques.
The hyperparameters set as default are omitted.
\begin{itemize}
    \item MiSiCNet: $\lambda=100$, \texttt{projection=True}.
    \item MSNet: $\alpha=0.1$, $\beta=0.1$.
    \item MUA\_SLIC: $\beta=30$, $\lambda_1=0.001$, $\lambda_2=0.001$, $\text{slic\_size}=200$.
    \item S$^2$WSU: $\lambda=0.001$.
    \item SISAL: $\tau=1\text{e-}6$.
\end{itemize}

\begin{figure*}[htbp]
\centering
  \begin{tabular}{cc}
   \includegraphics[width=.14\textwidth]{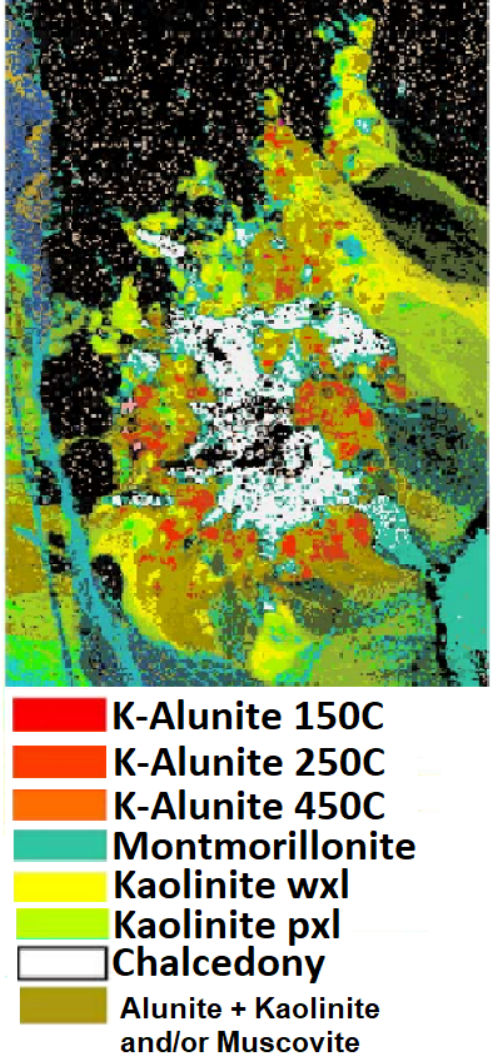}&   \includegraphics[width=.75\textwidth]{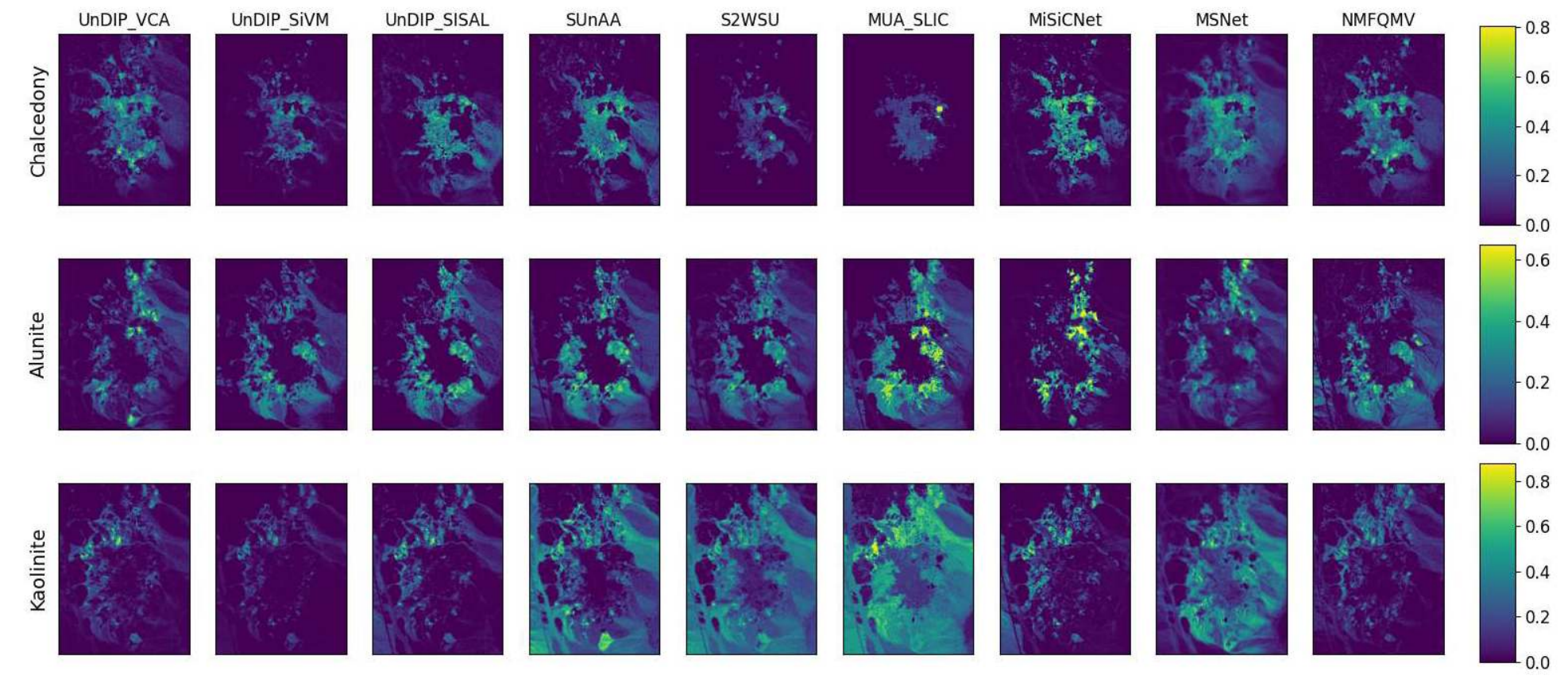}\\
 (a) Geological Ref. Map & (b) Estimate abundances
  \end{tabular}
  \caption{Estimate abundances obtained by applying different unmixing techniques to Cuprite compared with the geological reference map.}
\label{Real dataset}
\end{figure*}

Fig. \ref{Real dataset} (b) visually compares the estimated abundances for three dominant minerals, i.e., Chalcedony, Alunite, and Kaolinite. The comparison with the geological reference map Fig. \ref{Real dataset} (a) reveals that the estimated abundances obtained by semisupervised methods show more resemblance to the reference map for all three minerals. SUnAA visually outperforms the other techniques, particularly in the case of Chalcedony. The blind unmixing methods can better estimate Chalcedony compared to MUA\_SLIC and S$^2$WSU. This could be attributed to the mismatch of the endmember with the library's endmembers for this mineral. It is worth mentioning that SUnAA does not entirely rely on the library, and it estimates the endmember. Therefore, it can compensate for such a mismatch. More importantly, SUnAA is a parameter-free technique. We should note that selecting optimum parameters for the unmixing techniques is not a trivial task in real-world applications since the abundance RMSE cannot be computed.   

\subsection{Experimental Results: Processing Time}
The processing time of the unmixing methods from the HySUPP are given in Table \ref{tab:runtime} in seconds.  The processing times were obtained using a computer with an Intel(R) Xeon(R) Silver 4110 processor (2.10 GHz), 32 cores, 64GB of memory, and a NVIDIA
GeForce RTX (2080 Ti) graphical processing unit. In supervised unmixing, SISAL+FCLSU is the most efficient method. In semisupervised unmixing, SUnSAL is the most efficient one. We should mention that the efficiency of MUA\_SLIC considerably drops by increasing the number of pixels which highly affects the segmentation used in that method. For blind unmixing, CNNAEU and MSNet are more efficient than the other methods.
\begin{table}[ht]\addtolength{\tabcolsep}{-3pt}
    \centering
    \caption{Processing times (in seconds).}
 \begin{tabular}{c|c c c c c}
    \toprule
         Method & DC1 (30dB) &  DC2(30dB) & DC3 (30dB) & Cuprite \\
    \midrule
          VCA + FCLSU & \textbf{3} & 8 & \textbf{10}  & 92 \\
          SiVM + FCLSU & \textbf{3} & 9 & \textbf{10} & 58 \\
          SISAL + FCLSU & 6 & \textbf{7} & \textbf{10}  & \textbf{54} \\
          VCA + UnDIP & 31 & 23 & 29  & 492\\
          SiVM + UnDIP & 31 & 28 & 30 & 492 \\
          SISAL + UnDIP & 29 & 26 & 32 & 495 \\
    \midrule
          CLSUnSAL & \textbf{12} & 23 & 56  & 457 \\
          MUA\_SLIC & \textbf{12} & \textbf{15} & \textbf{36} &  310 \\
          S$^2$WSU & 35 & 60 & 116 &  767 \\
          SUnAA & 80 & 178 &  146 &  1201 \\
          SUnCNN & 75 & 65 & 97 &  492 \\
          SUnSAL & 13 & 17 & 44 & \textbf{176} \\
    \midrule
          ADMMNet & 33 & 57 & 61 & 969 \\
          BayesianSMA & 122 & 367 & 255 & 3593 \\
          CNNAEU & \textbf{9} & 22 & \textbf{24} & \textbf{20} \\
          EDAA & 274 & 51 & 59 &  114 \\
          MiSiCNet & 50 & 66 & 58 &   483 \\
          MSNet & 20 & \textbf{18} & 25 &  33 \\
          NMFQMV & 29 & 28 & 28 &  293 \\
          PGMSU & 43 & 41 & 49 &  135 \\
    \bottomrule
    \end{tabular}
    \label{tab:runtime}
\end{table}
\section{Conclusion, Discussion, Summary}
%\subsection{Method Selection}
Assuming that we have captured a spectral dataset and now have an unmixing problem in hand, we need to estimate the abundances of materials. The main question is which method to choose and which group of methods to select to tackle the problem. Indeed, the first step is to evaluate our problem and see if the linear mixing model or its variations are suitable for our problem. This decision needs prior knowledge of the physics of the problem. For instance, if you are dealing with intimate mixtures or close-range and microscopic scenarios, you should use nonlinear models. If you are dealing with macroscopic Earth observation problems, then linear models or their variations will be suitable. In some research, nonlinear models perform better than linear ones, however, one may pay attention to the selection of the model in a real-world application. Usually, linear models are more general. 

Here, we clarify the keys to the success of each category. The success of supervised (or sequential) unmixing lies in the confidence of the endmember measured, extracted, or estimated (pure/no pure scenario). Therefore, we should not use the supervised method if we are not confident about the endmembers. In other words, supervised methods perhaps are the best choice for endmembers with high confidence. When we have prior information on the material in the scene and a well-designed endmember library, semisupervised unmixing could be successful. Semisupervised are also suitable to capture the spectral variability. The success of semisupervised unmixing lies in the quality of the endmember library. Blind unmixing methods should be selected when there is no library, no pure pixels in the data set (including highly mixed scenarios), or the confidence of the measure, selected, or estimated endmembers is low. They should be used with caution, and the estimated endmembers should always go through physical interpretation.

% Question to answer? How do we know there are pure pixels in the data? In other words, when should we use pure pixel-based methods and no pure pixel methods? 

% \section{Future Challenges}
Despite the considerable advances, spectral unmixing can be considered one of the most challenging tasks in hyperspectral analysis. Here, we describe some of the main unmixing challenges. 
\begin{itemize}
    \item The linear models are much more general than the nonlinear models. However, from one dataset to another dataset, their performance may significantly decrease.
    \item Parameter selection considerably affects the performance of the unmixing methods. The selection of optimum parameters is very challenging when it comes to real-world datasets. 
    \item The performance of the linear unmixing methods often degrades by increasing the number of endmembers. The linear unmixing may fail for a dataset with a higher number of endmembers. 
    \item Spectral variability is still a big challenge and may considerably downgrade the performance of linear unmixing. 
    \item One of the main challenges is the absence of a real dataset with ground truth. 
    \item Multitemporal and multisource spectral unmixing are also challenging tasks. 
    \item Considering the large volume of hyperspectral data, scalable unmixing is the main key for global monitoring. 
\end{itemize}
Finally, we would like to further emphasize the importance of unmixing compared to the other signal and image processing approaches such as feature extraction/selection, clustering, and classification. The estimated features (i.e., abundances) in unmixing have physical meaning and can be attributed to materials. Therefore, estimated abundance maps can be used for quantitative analysis such as estimating vegetation cover, mineral composition, or pollutant concentrations providing quantitative information about the spatial distribution and relative quantities of materials present in the scene. The extracted/estimated endmembers can be used to identify and map different materials or surface types within the scene. By comparing the spectral signatures of unknown pixels with the reference endmembers, it is possible to classify and map land cover types, vegetation species, geological formations, or man-made structures. Temporal changes can be detected more accurately. The estimated abundance maps provide quantitative changes at a subpixel level. The extracted/estimated endmembers can be used for the identification of material changes. Intrinsic changes in the materials can be considered or ignored based on the endmembers assumptions. A future point is to investigate the performance of foundation models for spectral unmixing. 

\section*{Acknowledgments}
AZ, JM, and JC  were supported by ANR 3IA MIAI@Grenoble Alpes (ANR-19-P3IA-0003). AZ and JM were supported by the ERC grant number 101087696 (APHELAIA project). The authors would like to thank Dr. Michael Arbel (the author of the MLXP package) for the guidance on the MLXP package and for making it adaptable to the Windows operating system.

\appendix
\section*{Minimum Volume Simplex Using Total Variation}
\label{app: TV}
The total variation penalty enforces the data simplex to have minimum volume by pulling the endmembers towards each other. Assuming $\Bar{\bf e}=\frac{1}{r}{{\bf e}{\bf 1}_r}$ we have
\begin{align}\label{eqA: TV}\nonumber
 &TV({\bf E})=\sum_{i,j=1}^r \frac{1}{2}|| {\bf e}_i-{\bf e}_j||_2^{2}=\\& \nonumber \frac{1}{2}\sum_{i,j=1}^r|| {\bf e}_i-\Bar{\bf e}-{\bf e}_j+\Bar{\bf e}||_2^{2}=\\& \nonumber \frac{1}{2}\sum_{i=1}^r|| {\bf e}_i-\Bar{\bf e}||_2^{2}+\frac{1}{2}\sum_{j=1}^r|| {\bf e}_j-\Bar{\bf e}||_2^{2}- \\&
 \left(\sum_{i=1}^r\left({\bf e}_i-\Bar{\bf e}\right)\right)^T\sum_{j=1}^r\left( {\bf e}_j-\Bar{\bf e}\right)
\end{align}
where the third term is zero. Therefore, 
\begin{align}\label{eqA: TV1}\nonumber
 &TV({\bf E})=\sum_{i=1}^r|| {\bf e}_i-\Bar{\bf e}||_2^{2}=\\&\sum_{i=1}^r|| {\bf e}_i-\frac{1}{r}{{\bf e}{\bf 1}_r}||_2^{2}=||{\bf E}({\bf I}_r-\frac{1}{r}{\bf 1}_r{\bf 1}_r^T)||_F^2.  
\end{align}


\bibliographystyle{IEEEbib}
\bibliography{refs,new_ref,refs1,HyDe,refs_misicnet,Ref_non}

\begin{thebibliography}{100}

\bibitem{unmixing-review}
J.~M. {Bioucas-Dias}, A.~{Plaza}, N.~{Dobigeon}, M.~{Parente}, Q.~{Du},
  P.~{Gader}, and J.~{Chanussot},
\newblock ``Hyperspectral unmixing overview: Geometrical, statistical, and
  sparse regression-based approaches,''
\newblock {\em IEEE Journal of Selected Topics in Applied Earth Observations
  and Remote Sensing}, vol. 5, no. 2, pp. 354--379, April 2012.

\bibitem{Dobigeon_2014_NU}
N.~{Dobigeon}, J.~{Tourneret}, C.~{Richard}, J.~C.~M. {Bermudez},
  S.~{McLaughlin}, and A.~O. {Hero},
\newblock ``Nonlinear unmixing of hyperspectral images: Models and
  algorithms,''
\newblock {\em IEEE Signal Processing Magazine}, vol. 31, no. 1, pp. 82--94,
  2014.

\bibitem{Spec_var}
R.~A. Borsoi, T.~Imbiriba, J.~C.~Moreira Bermudez, C.~Richard, J.~Chanussot,
  L.~Drumetz, J.-Y. Tourneret, A.~Zare, and C.~Jutten,
\newblock ``Spectral variability in hyperspectral data unmixing: A
  comprehensive review,''
\newblock {\em IEEE Geoscience and Remote Sensing Magazine}, vol. 9, no. 4, pp.
  223--270, 2021.

\bibitem{HyDe}
D.~Coquelin, B.~Rasti, M.~Götz, P.~Ghamisi, R.~Gloaguen, and A.~Streit,
\newblock ``Hyde: The first open-source, python-based, gpu-accelerated
  hyperspectral denoising package,''
\newblock in {\em 2022 12th Workshop on Hyperspectral Imaging and Signal
  Processing: Evolution in Remote Sensing (WHISPERS)}, 2022, pp. 1--5.

\bibitem{BR_Rev1}
B.~Rasti, P.~Scheunders, P.~Ghamisi, G.~Licciardi, and J.~Chanussot,
\newblock ``Noise reduction in hyperspectral imagery: Overview and
  application,''
\newblock {\em Remote Sensing}, vol. 10, no. 3, pp. 482, 2018.

\bibitem{ImRes2021}
B.~Rasti, Y.~Chang, E.~Dalsasso, Loic Denis, and Pedram Ghamisi,
\newblock ``{Image Restoration for Remote Sensing: Overview and Toolbox},''
\newblock {\em IEEE Geoscience and Remote Sensing Magazine}, pp. 2--31, 2021.

\bibitem{hapke_2012}
B.~Hapke,
\newblock {\em Theory of Reflectance and Emittance Spectroscopy},
\newblock Cambridge University Press, 2 edition, 2012.

\bibitem{BHapke1981}
B.~Hapke,
\newblock ``Bidirectional reflectance spectroscopy: 1. theory,''
\newblock {\em Journal of Geophysical research}, vol. 86, pp. 3039--3054, 1981.

\bibitem{Keshava2003A}
N.~Keshava,
\newblock ``{A Survey of Spectral Unmixing Algorithms},''
\newblock {\em Lincoln Laboratory Journal}, vol. 14, no. 1, pp. 55--78, 2003.

\bibitem{Palsson_rev}
B.~Palsson, J.~R. Sveinsson, and M.~O. Ulfarsson,
\newblock ``Blind hyperspectral unmixing using autoencoders: A critical
  comparison,''
\newblock {\em IEEE Journal of Selected Topics in Applied Earth Observations
  and Remote Sensing}, vol. 15, pp. 1340--1372, 2022.

\bibitem{PB_DD_Un}
J.~Chen, M.~Zhao, X.~Wang, C.~Richard, and S.~Rahardja,
\newblock ``Integration of physics-based and data-driven models for
  hyperspectral image unmixing: A summary of current methods,''
\newblock {\em IEEE Signal Processing Magazine}, vol. 40, no. 2, pp. 61--74,
  2023.

\bibitem{End_Var_Rev}
Ben S., Gregory~P. A., Laurent T., and Pol C.,
\newblock ``Endmember variability in spectral mixture analysis: A review,''
\newblock {\em Remote Sensing of Environment}, vol. 115, no. 7, pp. 1603--1616,
  2011.

\bibitem{End_Var_Zare}
A.~Zare and K.C. Ho,
\newblock ``Endmember variability in hyperspectral analysis: Addressing
  spectral variability during spectral unmixing,''
\newblock {\em IEEE Signal Processing Magazine}, vol. 31, no. 1, pp. 95--104,
  2014.

\bibitem{End_extr_Plaza}
A.~Plaza, G.~Mart{\'i}n, J.~Plaza, M.~Zortea, and S.~S{\'a}nchez,
\newblock {\em Recent Developments in Endmember Extraction and Spectral
  Unmixing}, pp. 235--267,
\newblock Springer Berlin Heidelberg, Berlin, Heidelberg, 2011.

\bibitem{Dias_HS_Rev2013}
J.~M. {Bioucas-Dias}, A.~{Plaza}, G.~{Camps-Valls}, P.~{Scheunders},
  N.~{Nasrabadi}, and J.~{Chanussot},
\newblock ``Hyperspectral remote sensing data analysis and future challenges,''
\newblock {\em IEEE Geoscience and Remote Sensing Magazine}, vol. 1, no. 2, pp.
  6--36, 2013.

\bibitem{Ghamisi-review-2017}
P.~Ghamisi, N.~Yokoya, J.~Li, W.~Liao, S.~Liu, J.~Plaza, B.~Rasti, and
  A.~Plaza,
\newblock ``Advances in hyperspectral image and signal processing: A
  comprehensive overview of the state of the art,''
\newblock {\em IEEE Geoscience and Remote Sensing Magazine}, vol. 5, no. 4, pp.
  37--78, Dec 2017.

\bibitem{Rev_Tensor_HS}
M.~Wang, D.~Hong, Z.~Han, J.~Li, J.~Yao, L.~Gao, B.~Zhang, and J.~Chanussot,
\newblock ``Tensor decompositions for hyperspectral data processing in remote
  sensing: A comprehensive review,''
\newblock {\em IEEE Geoscience and Remote Sensing Magazine}, vol. 11, no. 1,
  pp. 26--72, 2023.

\bibitem{Rob_NonRev}
R.~Heylen, M.~Parente, and P.~Gader,
\newblock ``A review of nonlinear hyperspectral unmixing methods,''
\newblock {\em IEEE Journal of Selected Topics in Applied Earth Observations
  and Remote Sensing}, vol. 7, no. 6, pp. 1844--1868, 2014.

\bibitem{HyperMix}
L.~I. Jimenez and A.~Plaza,
\newblock ``Hypermix: An open source tool for hyperspectral imaging,''
\newblock in {\em 2015 IEEE International Geoscience and Remote Sensing
  Symposium (IGARSS)}, 2015, pp. 1749--1752.

\bibitem{Arbel2023MLXP}
M.~Arbel,
\newblock ``Mlxp: A framework for conducting machine learning experiments in
  python,'' Github, 2023.

\bibitem{SUnCNN}
B.~Rasti and B.~Koirala,
\newblock ``{SUnCNN}: Sparse unmixing using unsupervised convolutional neural
  network,''
\newblock {\em IEEE Geoscience and Remote Sensing Letters}, vol. 19, pp. 1--5,
  2022.

\bibitem{FCLSU}
D.~C. {Heinz} and {Chein-I-Chang},
\newblock ``Fully constrained least squares linear spectral mixture analysis
  method for material quantification in hyperspectral imagery,''
\newblock {\em IEEE Transactions on Geoscience and Remote Sensing}, vol. 39,
  no. 3, pp. 529--545, 2001.

\bibitem{RHeylen_2011}
R.~{Heylen}, D.~{Burazerovic}, and P.~{Scheunders},
\newblock ``Fully constrained least squares spectral unmixing by simplex
  projection,''
\newblock {\em IEEE Transactions on Geoscience and Remote Sensing}, vol. 49,
  no. 11, pp. 4112--4122, Nov 2011.

\bibitem{SISAL}
J.~M. {Bioucas-Dias},
\newblock ``A variable splitting augmented lagrangian approach to linear
  spectral unmixing,''
\newblock in {\em 2009 First Workshop on Hyperspectral Image and Signal
  Processing: Evolution in Remote Sensing}, 2009, pp. 1--4.

\bibitem{UnDIP}
B.~Rasti, B.~Koirala, P.~Scheunders, and P.~Ghamisi,
\newblock ``{UnDIP}: Hyperspectral unmixing using deep image prior,''
\newblock {\em IEEE Transactions on Geoscience and Remote Sensing}, pp. 1--15,
  2021.

\bibitem{VCA}
J.~Nascimento and J.~Bioucas-Dias,
\newblock ``Vertex component analysis: A{\textasciitilde}fast algorithm to
  extract endmembers spectra from hyperspectral data,''
\newblock in {\em Pattern Recognition and Image Analysis}, Francisco~Jos{\'e}
  Perales, Aur{\'e}lio J.~C. Campilho, Nicol{\'a}s~P{\'e}rez de~la Blanca, and
  Alberto Sanfeliu, Eds., Berlin, Heidelberg, 2003, pp. 626--635, Springer
  Berlin Heidelberg.

\bibitem{Collaborative}
M.~{Iordache}, J.~M. {Bioucas-Dias}, and A.~{Plaza},
\newblock ``Collaborative sparse regression for hyperspectral unmixing,''
\newblock {\em IEEE Transactions on Geoscience and Remote Sensing}, vol. 52,
  no. 1, pp. 341--354, 2014.

\bibitem{RABorsoi2019}
R.~A. {Borsoi}, T.~{Imbiriba}, J.~C.~M. {Bermudez}, and C.~{Richard},
\newblock ``A fast multiscale spatial regularization for sparse hyperspectral
  unmixing,''
\newblock {\em IEEE Geoscience and Remote Sensing Letters}, vol. 16, no. 4, pp.
  598--602, 2019.

\bibitem{SZhang2018}
S.~{Zhang}, J.~{Li}, H.~{Li}, C.~{Deng}, and A.~{Plaza},
\newblock ``Spectral–spatial weighted sparse regression for hyperspectral
  image unmixing,''
\newblock {\em IEEE Transactions on Geoscience and Remote Sensing}, vol. 56,
  no. 6, pp. 3265--3276, 2018.

\bibitem{SUnAA}
B.~Rasti, A.~Zouaoui, J.~Mairal, and J.~Chanussot,
\newblock ``Sunaa: Sparse unmixing using archetypal analysis,''
\newblock {\em IEEE Geoscience and Remote Sensing Letters}, pp. 1--1, 2023.

\bibitem{SUnSAL}
J.~M. {Bioucas-Dias} and M.~A.~T. {Figueiredo},
\newblock ``Alternating direction algorithms for constrained sparse regression:
  Application to hyperspectral unmixing,''
\newblock in {\em 2nd Workshop on Hyperspectral Image and Signal Processing:
  Evolution in Remote Sensing}, 2010, pp. 1--4.

\bibitem{ADMMNet}
C.~Zhou and M.~R. Rodrigues,
\newblock ``Admm-based hyperspectral unmixing networks for abundance and
  endmember estimation,''
\newblock {\em IEEE Transactions on Geoscience and Remote Sensing}, vol. 60,
  pp. 1--18, 2021.

\bibitem{BayesianSMA}
N.~Dobigeon, S.~Moussaoui, M.~Coulon, J.-Y. Tourneret, and A.~O. Hero,
\newblock ``Joint bayesian endmember extraction and linear unmixing for
  hyperspectral imagery,''
\newblock {\em IEEE Transactions on Signal Processing}, vol. 57, no. 11, pp.
  4355--4368, 2009.

\bibitem{CNNAEU}
B.~Palsson, M.~O. Ulfarsson, and J.~R. Sveinsson,
\newblock ``Convolutional autoencoder for spectral–spatial hyperspectral
  unmixing,''
\newblock {\em IEEE Transactions on Geoscience and Remote Sensing}, vol. 59,
  no. 1, pp. 535--549, 2021.

\bibitem{EDAA}
A.~Zouaoui, G.~Muhawenayo, B.~Rasti, J.~Chanussot, and J.~Mairal,
\newblock ``Entropic descent archetypal analysis for blind hyperspectral
  unmixing,''
\newblock {\em IEEE Transactions on Image Processing}, pp. 1--1, 2023.

\bibitem{MiSiCNet}
B.~Rasti, B.~Koirala, P.~Scheunders, and J.~Chanussot,
\newblock ``Misicnet: Minimum simplex convolutional network for deep
  hyperspectral unmixing,''
\newblock {\em IEEE Transactions on Geoscience and Remote Sensing}, pp. 1--1,
  2022.

\bibitem{MSNet}
Y.~Yu, Y.~Ma, X.~Mei, F.~Fan, J.~Huang, and H.~Li,
\newblock ``Multi-stage convolutional autoencoder network for hyperspectral
  unmixing,''
\newblock {\em International Journal of Applied Earth Observation and
  Geoinformation}, vol. 113, pp. 102981, 2022.

\bibitem{NMF_QMV}
L.~{Zhuang}, C.~{Lin}, M.~A.~T. {Figueiredo}, and J.~M. {Bioucas-Dias},
\newblock ``Regularization parameter selection in minimum volume hyperspectral
  unmixing,''
\newblock {\em IEEE Transactions on Geoscience and Remote Sensing}, vol. 57,
  no. 12, pp. 9858--9877, 2019.

\bibitem{PGMSU}
S.~Shi, M.~Zhao, L.~Zhang, Y.~Altmann, and J.~Chen,
\newblock ``Probabilistic generative model for hyperspectral unmixing
  accounting for endmember variability,''
\newblock {\em IEEE Transactions on Geoscience and Remote Sensing}, vol. 60,
  pp. 1--15, 2022.

\bibitem{BR_rev_2021}
B.~Rasti, Y.~Chang, E.~Dalsasso, L.~Denis, and P.~Ghamisi,
\newblock ``Image restoration for remote sensing: Overview and toolbox,''
\newblock {\em IEEE Geoscience and Remote Sensing Magazine}, vol. 10, no. 2,
  pp. 201--230, 2022.

\bibitem{BR_UnDN}
B.~Rasti, B.~Koirala, P.~Scheunders, and P.~Ghamisi,
\newblock ``How hyperspectral image unmixing and denoising can boost each
  other,''
\newblock {\em Remote Sensing}, vol. 12, no. 11, pp. 1728, May 2020.

\bibitem{TInce_2019}
T.~{Ince} and T.~{Dundar},
\newblock ``Simultaneous nonconvex denoising and unmixing for hyperspectral
  imaging,''
\newblock {\em IEEE Access}, vol. 7, pp. 124426--124440, 2019.

\bibitem{JYang_2016}
J.~{Yang}, Y.~{Zhao}, J.~C. {Chan}, and S.~G. {Kong},
\newblock ``Coupled sparse denoising and unmixing with low-rank constraint for
  hyperspectral image,''
\newblock {\em IEEE Transactions on Geoscience and Remote Sensing}, vol. 54,
  no. 3, pp. 1818--1833, March 2016.

\bibitem{WSRRR}
B.~Rasti, J.R. Sveinsson, and M.O. Ulfarsson,
\newblock ``{Wavelet-Based Sparse Reduced-Rank Regression for Hyperspectral
  Image Restoration},''
\newblock {\em IEEE Transactions on Geoscience and Remote Sensing}, vol. 52,
  no. 10, pp. 6688--6698, 2014.

\bibitem{OTVCA}
B.~Rasti, M.O. Ulfarsson, and J.~Sveinsson,
\newblock ``{Hyperspectral Feature Extraction Using Total Variation Component
  Analysis},''
\newblock {\em IEEE Transactions on Geoscience and Remote Sensing}, vol. 54, 08
  2016.

\bibitem{FE_Rev_BR20}
B.~{Rasti}, D.~{Hong}, R.~{Hang}, P.~{Ghamisi}, X.~{Kang}, J.~{Chanussot}, and
  J.~A. {Benediktsson},
\newblock ``Feature extraction for hyperspectral imagery: The evolution from
  shallow to deep (overview and toolbox),''
\newblock {\em IEEE Geoscience and Remote Sensing Magazine}, pp. 0--0, 2020.

\bibitem{HySime}
J.M. Bioucas-Dias and J.M.P. Nascimento,
\newblock ``Hyperspectral subspace identification,''
\newblock {\em IEEE Transactions on Geoscience and Remote Sensing}, vol. 46,
  no. 8, pp. 2435--2445, 2008.

\bibitem{HySURE}
B.~Rasti, M.O. Ulfarsson, and J.R. Sveinsson,
\newblock ``Hyperspectral subspace identification using {SURE},''
\newblock {\em IEEE Geoscience and Remote Sensing Letter}, vol. 12, no. 12, pp.
  2481--2485, {D}ec. 2015.

\bibitem{Sub_id_Chang_Du}
C.-I Chang and Q.~Du,
\newblock ``Estimation of number of spectrally distinct signal sources in
  hyperspectral imagery,''
\newblock {\em IEEE Transactions on Geoscience and Remote Sensing}, vol. 42,
  no. 3, pp. 608--619, 2004.

\bibitem{HFC}
J.~Harsanyi, W~Farrand, and C.I Chang,
\newblock ``Determining the number and identity of spectral endmembers: An
  integrated approach using neyman?pearson eigenthresholding and iterative
  constrained rms error minimization,''
\newblock in {\em 9th Thematic Conf. Geologic Remote Sens.}, 1993.

\bibitem{HFC_2}
B.~Luo, J.~Chanussot, S.~Doute, and L.~Zhang,
\newblock ``Empirical automatic estimation of the number of endmembers in
  hyperspectral images,''
\newblock {\em IEEE Geoscience and Remote Sensing Letters}, vol. 10, no. 1, pp.
  24--28, 2013.

\bibitem{ICE}
M.~Berman, H.~Kiiveri, R.~Lagerstrom, A.~Ernst, R.~Dunne, and J.F. Huntington,
\newblock ``Ice: a statistical approach to identifying endmembers in
  hyperspectral images,''
\newblock {\em IEEE Transactions on Geoscience and Remote Sensing}, vol. 42,
  no. 10, pp. 2085--2095, 2004.

\bibitem{Zare_ICE}
A.~Zare and P.~Gader,
\newblock ``Sparsity promoting iterated constrained endmember detection in
  hyperspectral imagery,''
\newblock {\em IEEE Geoscience and Remote Sensing Letters}, vol. 4, no. 3, pp.
  446--450, 2007.

\bibitem{GENE}
A~Ambikapathi, T.-H. Chan, C.-Y. Chi, and K.~Keizer,
\newblock ``Hyperspectral data geometry-based estimation of number of
  endmembers using p-norm-based pure pixel identification algorithm,''
\newblock {\em IEEE Transactions on Geoscience and Remote Sensing}, vol. 51,
  no. 5, pp. 2753--2769, {M}ay 2013.

\bibitem{PPI}
J.~Boardman, F.~A. Kruse, and R.~Green,
\newblock ``Mapping target signatures via partial unmixing of aviris data: in
  summaries,''
\newblock in {\em JPL Airborne Earth Sci. Workshop}, 1995, pp. 23--26.

\bibitem{OSP}
J.C. Harsanyi and C.-I. Chang,
\newblock ``Hyperspectral image classification and dimensionality reduction: an
  orthogonal subspace projection approach,''
\newblock {\em IEEE Transactions on Geoscience and Remote Sensing}, vol. 32,
  no. 4, pp. 779--785, 1994.

\bibitem{N-FINDR}
M.~E. Winter,
\newblock ``{N-FINDR: an algorithm for fast autonomous spectral end-member
  determination in hyperspectral data},''
\newblock in {\em Imaging Spectrometry V}, Michael~R. Descour and Sylvia~S.
  Shen, Eds. International Society for Optics and Photonics, 1999, vol. 3753,
  pp. 266 -- 275, SPIE.

\bibitem{MESM}
A.~Bateson and B.~Curtiss,
\newblock ``A method for manual endmember selection and spectral unmixing,''
\newblock {\em Remote Sensing of Environment}, vol. 55, no. 3, pp. 229--243,
  1996.

\bibitem{End_Bund1}
T.~Bajjouk, J.~Populus, and B.~Guillaumont,
\newblock ``Quantification of subpixel cover fractions using principal
  component analysis and a linear programming method: Application to the
  coastal zone of roscoff (france),''
\newblock {\em Remote Sensing of Environment}, vol. 64, no. 2, pp. 153--165,
  1998.

\bibitem{End_Bund2}
C.A. Bateson, G.P. Asner, and C.A. Wessman,
\newblock ``Endmember bundles: a new approach to incorporating endmember
  variability into spectral mixture analysis,''
\newblock {\em IEEE Transactions on Geoscience and Remote Sensing}, vol. 38,
  no. 2, pp. 1083--1094, 2000.

\bibitem{End_bundle}
B.~Somers, M.~Zortea, A.~Plaza, and G.~P. Asner,
\newblock ``Automated extraction of image-based endmember bundles for improved
  spectral unmixing,''
\newblock {\em IEEE Journal of Selected Topics in Applied Earth Observations
  and Remote Sensing}, vol. 5, no. 2, pp. 396--408, 2012.

\bibitem{AMEE}
A.~Plaza, P.~Martinez, R.~Perez, and J.~Plaza,
\newblock ``Spatial/spectral endmember extraction by multidimensional
  morphological operations,''
\newblock {\em IEEE Transactions on Geoscience and Remote Sensing}, vol. 40,
  no. 9, pp. 2025--2041, 2002.

\bibitem{SSEE}
D.M. Rogge, B.~Rivard, J.~Zhang, A.~Sanchez, J.~Harris, and J.~Feng,
\newblock ``Integration of spatial-spectral information for the improved
  extraction of endmembers,''
\newblock {\em Remote Sensing of Environment}, vol. 110, no. 3, pp. 287--303,
  2007.

\bibitem{SSEE_2}
M.~Xu, L.~Zhang, and B.~Du,
\newblock ``An image-based endmember bundle extraction algorithm using both
  spatial and spectral information,''
\newblock {\em IEEE Journal of Selected Topics in Applied Earth Observations
  and Remote Sensing}, vol. 8, no. 6, pp. 2607--2617, 2015.

\bibitem{SSEE_3}
X.~Shen, W.~Bao, and K.~Qu,
\newblock ``Spatial-spectral hyperspectral endmember extraction using a spatial
  energy prior constrained maximum simplex volume approach,''
\newblock {\em IEEE Journal of Selected Topics in Applied Earth Observations
  and Remote Sensing}, vol. 13, pp. 1347--1361, 2020.

\bibitem{SPP}
M.~Zortea and A.~Plaza,
\newblock ``Spatial preprocessing for endmember extraction,''
\newblock {\em IEEE Transactions on Geoscience and Remote Sensing}, vol. 47,
  no. 8, pp. 2679--2693, 2009.

\bibitem{MVES}
T.-H. Chan, C.-Y. Chi, Y.-M. Huang, and W.-K. Ma,
\newblock ``A convex analysis-based minimum-volume enclosing simplex algorithm
  for hyperspectral unmixing,''
\newblock {\em IEEE Transactions on Signal Processing}, vol. 57, no. 11, pp.
  4418--4432, 2009.

\bibitem{MVES2}
M.D. Craig,
\newblock ``Minimum-volume transforms for remotely sensed data,''
\newblock {\em IEEE Transactions on Geoscience and Remote Sensing}, vol. 32,
  no. 3, pp. 542--552, 1994.

\bibitem{MVIE}
C.-H. Lin, R.~Wu, W.-K. Ma, C.-Y. Chi, and Y.~Wang,
\newblock ``Maximum volume inscribed ellipsoid: A new simplex-structured matrix
  factorization framework via facet enumeration and convex optimization,''
\newblock {\em SIAM Journal on Imaging Sciences}, vol. 11, no. 2, pp.
  1651--1679, 2018.

\bibitem{MVSA}
J.~{Li} and J.~M. {Bioucas-Dias},
\newblock ``Minimum volume simplex analysis: A fast algorithm to unmix
  hyperspectral data,''
\newblock in {\em IGARSS 2008 - 2008 IEEE International Geoscience and Remote
  Sensing Symposium}, 2008, vol.~3, pp. III -- 250--III -- 253.

\bibitem{ADMM}
J.~Eckstein and D.~P. Bertsekas,
\newblock ``On the douglas-rachford splitting method and the proximal point
  algorithm for maximal monotone operators,''
\newblock {\em Mathematical Programming}, vol. 55, pp. 293--318, 1992.

\bibitem{ULS_CI}
C.-I. Chang, X.-L. Zhao, M.L.G. Althouse, and J.~J. Pan,
\newblock ``Least squares subspace projection approach to mixed pixel
  classification for hyperspectral images,''
\newblock {\em IEEE Transactions on Geoscience and Remote Sensing}, vol. 36,
  no. 3, pp. 898--912, 1998.

\bibitem{NCLS}
J.J. Settle and N.~Drake,
\newblock ``Linear mixing and the estimation of ground cover proportions,''
\newblock {\em International Journal of Remote Sensing - INT J REMOTE SENS},
  vol. 14, pp. 1159--1177, 04 1993.

\bibitem{NCLSU}
{Chein-I Chang} and D.~C. {Heinz},
\newblock ``Constrained subpixel target detection for remotely sensed
  imagery,''
\newblock {\em IEEE Transactions on Geoscience and Remote Sensing}, vol. 38,
  no. 3, pp. 1144--1159, 2000.

\bibitem{WLS}
Y.E. Shimabukuro and J.A. Smith,
\newblock ``The least-squares mixing models to generate fraction images derived
  from remote sensing multispectral data,''
\newblock {\em IEEE Transactions on Geoscience and Remote Sensing}, vol. 29,
  no. 1, pp. 16--20, 1991.

\bibitem{CLS}
Y.E. Shimabukuro,
\newblock {\em Shade Images Derived from Linear Mixing Models of Multispectral
  Measurements of Forested Areas},
\newblock Colorado State University, 1987.

\bibitem{FCLSU_Simplex}
R.~Heylen, D.~Burazerovic, and P.~Scheunders,
\newblock ``Fully constrained least squares spectral unmixing by simplex
  projection,''
\newblock {\em IEEE Transactions on Geoscience and Remote Sensing}, vol. 49,
  no. 11, pp. 4112--4122, 2011.

\bibitem{Ghassemian2020}
F.~{Khajehrayeni} and H.~{Ghassemian},
\newblock ``Hyperspectral unmixing using deep convolutional autoencoders in a
  supervised scenario,''
\newblock {\em IEEE Journal of Selected Topics in Applied Earth Observations
  and Remote Sensing}, vol. 13, pp. 567--576, 2020.

\bibitem{DIP}
D.~Ulyanov, A.~Vedaldi, and V.~Lempitsky,
\newblock ``Deep image prior,''
\newblock {\em International Journal of Computer Vision}, vol. 128, no. 7, pp.
  1867–1888, Mar 2020.

\bibitem{DIP_CVPR}
D.~Ulyanov, A.~Vedaldi, and V.~Lempitsky,
\newblock ``Deep image prior,''
\newblock in {\em Proceedings of the IEEE Conference on Computer Vision and
  Pattern Recognition (CVPR)}, June 2018.

\bibitem{EGU_Net}
Z.~Han, D.~Hong, L.~Gao, B.~Zhang, and J.~Chanussot,
\newblock ``Deep half-siamese networks for hyperspectral unmixing,''
\newblock {\em IEEE Geoscience and Remote Sensing Letters}, vol. 18, no. 11,
  pp. 1996--2000, 2021.

\bibitem{EGU_Net2}
D.~Hong, L.~Gao, J.~Yao, N.~Yokoya, J.~Chanussot, U.~Heiden, and B.~Zhang,
\newblock ``Endmember-guided unmixing network (egu-net): A general deep
  learning framework for self-supervised hyperspectral unmixing,''
\newblock {\em IEEE Transactions on Neural Networks and Learning Systems}, vol.
  33, no. 11, pp. 6518--6531, 2022.

\bibitem{TV_End}
M.~Berman, H.~Kiiveri, R.~Lagerstrom, A.~Ernst, R.~Dunne, and J.F. Huntington,
\newblock ``Ice: a statistical approach to identifying endmembers in
  hyperspectral images,''
\newblock {\em IEEE Transactions on Geoscience and Remote Sensing}, vol. 42,
  no. 10, pp. 2085--2095, 2004.

\bibitem{MVC}
L.~{Miao} and H.~{Qi},
\newblock ``Endmember extraction from highly mixed data using minimum volume
  constrained nonnegative matrix factorization,''
\newblock {\em IEEE Transactions on Geoscience and Remote Sensing}, vol. 45,
  no. 3, pp. 765--777, 2007.

\bibitem{CoNMF}
J.~{Li}, J.~M. {Bioucas-Dias}, and A.~{Plaza},
\newblock ``Collaborative nonnegative matrix factorization for remotely sensed
  hyperspectral unmixing,''
\newblock in {\em IEEE International Geoscience and Remote Sensing Symposium
  (IGARSS)}, July 2012, pp. 3078--3081.

\bibitem{R-CoNMF}
J.~Li, J.~M. Bioucas-Dias, A.~Plaza, and L.~Liu,
\newblock ``Robust collaborative nonnegative matrix factorization for
  hyperspectral unmixing,''
\newblock {\em IEEE Transactions on Geoscience and Remote Sensing}, vol. 54,
  no. 10, pp. 6076--6090, 2016.

\bibitem{L_0.5}
Y.~Qian, S.~Jia, J.~Zhou, and A.~Robles-Kelly,
\newblock ``Hyperspectral unmixing via $l_{1/2}$ sparsity-constrained
  nonnegative matrix factorization,''
\newblock {\em IEEE Transactions on Geoscience and Remote Sensing}, vol. 49,
  no. 11, pp. 4282--4297, 2011.

\bibitem{NMF_multiplic}
D.~Lee and H.~S. Seung,
\newblock ``Algorithms for non-negative matrix factorization,''
\newblock in {\em Advances in Neural Information Processing Systems}, T.~Leen,
  T.~Dietterich, and V.~Tresp, Eds. 2000, vol.~13, p. 556–562, MIT Press.

\bibitem{NMF_L1_multiplic}
P.O. Hoyer,
\newblock ``Non-negative sparse coding,''
\newblock in {\em Proceedings of the 12th IEEE Workshop on Neural Networks for
  Signal Processing}, 2002, pp. 557--565.

\bibitem{JakubLq}
J.~{Sigurdsson}, M.~O. {Ulfarsson}, and J.~R. {Sveinsson},
\newblock ``Hyperspectral unmixing with $l_{q}$ regularization,''
\newblock {\em IEEE Transactions on Geoscience and Remote Sensing}, vol. 52,
  no. 11, pp. 6793--6806, 2014.

\bibitem{JackobTV}
J.~Sigurdsson, M.O. Ulfarsson, J.R. Sveinsson, and J.A. Benediktsson,
\newblock ``Smooth spectral unmixing using total variation regularization and a
  first order roughness penalty,''
\newblock in {\em IEEE International Geoscience and Remote Sensing Symposium
  (IGARSS)}, July 2013, pp. 2160--2163.

\bibitem{NMF_TV_RS}
W.~He, H.~Zhang, and L.~Zhang,
\newblock ``Total variation regularized reweighted sparse nonnegative matrix
  factorization for hyperspectral unmixing,''
\newblock {\em IEEE Transactions on Geoscience and Remote Sensing}, vol. 55,
  no. 7, pp. 3909--3921, 2017.

\bibitem{NMF_1}
J.~Yao, D.~Meng, Q.~Zhao, W.~Cao, and Z.~Xu,
\newblock ``Nonconvex-sparsity and nonlocal-smoothness-based blind
  hyperspectral unmixing,''
\newblock {\em IEEE Transactions on Image Processing}, vol. 28, no. 6, pp.
  2991--3006, 2019.

\bibitem{NMF_2}
C.~G. Tsinos, A.~A. Rontogiannis, and K.~Berberidis,
\newblock ``Distributed blind hyperspectral unmixing via joint sparsity and
  low-rank constrained non-negative matrix factorization,''
\newblock {\em IEEE Transactions on Computational Imaging}, vol. 3, no. 2, pp.
  160--174, 2017.

\bibitem{NMF_3}
Z.~Yang, G.~Zhou, S.~Xie, S.~Ding, J.-M. Yang, and J.~Zhang,
\newblock ``Blind spectral unmixing based on sparse nonnegative matrix
  factorization,''
\newblock {\em IEEE Transactions on Image Processing}, vol. 20, no. 4, pp.
  1112--1125, 2011.

\bibitem{NMF_4}
S.~Zhang, G.~Zhang, F.~Li, C.~Deng, S.~Wang, A.~Plaza, and J.~Li,
\newblock ``Spectral-spatial hyperspectral unmixing using nonnegative matrix
  factorization,''
\newblock {\em IEEE Transactions on Geoscience and Remote Sensing}, vol. 60,
  pp. 1--13, 2022.

\bibitem{PnP}
M.~Zhao, X.~Wang, J.~Chen, and W.~Chen,
\newblock ``A plug-and-play priors framework for hyperspectral unmixing,''
\newblock {\em IEEE Transactions on Geoscience and Remote Sensing}, vol. 60,
  pp. 1--13, 2022.

\bibitem{cutler1994archetypal}
A.~Cutler and L.~Breiman,
\newblock ``Archetypal analysis,''
\newblock {\em Technometrics}, vol. 36, no. 4, pp. 338--347, 1994.

\bibitem{ME_AA}
G.~Zhao, X.~Jia, and C.~Zhao,
\newblock ``Multiple endmembers based unmixing using archetypal analysis,''
\newblock in {\em 2015 IEEE International Geoscience and Remote Sensing
  Symposium (IGARSS)}, 2015, pp. 5039--5042.

\bibitem{FKAA}
C.~Zhao, G.~Zhao, and X.~Jia,
\newblock ``Hyperspectral image unmixing based on fast kernel archetypal
  analysis,''
\newblock {\em IEEE Journal of Selected Topics in Applied Earth Observations
  and Remote Sensing}, vol. 10, no. 1, pp. 331--346, 2017.

\bibitem{L1AA}
M.~Xu, Z.~Yang, G.~Ren, H.~Sheng, S.~Liu, W.~Liu, and C.~Ye,
\newblock ``L1 sparsity-constrained archetypal analysis algorithm for
  hyperspectral unmixing,''
\newblock {\em IEEE Geoscience and Remote Sensing Letters}, vol. 19, pp. 1--5,
  2022.

\bibitem{NTF}
Q.~Zhang, H.~Wang, R.~Plemmons, and V.~P. Pauca,
\newblock ``Spectral unmixing using nonnegative tensor factorization,''
\newblock in {\em Proceedings of the 45th Annual Southeast Regional
  Conference}, New York, NY, USA, 2007, ACM-SE 45, p. 531–532, Association
  for Computing Machinery.

\bibitem{TD}
A.~Huck and M.~Guillaume,
\newblock ``Estimation of the hyperspectral tucker ranks,''
\newblock in {\em 2009 IEEE International Conference on Acoustics, Speech and
  Signal Processing}, 2009, pp. 1281--1284.

\bibitem{BTD}
L.~De~Lathauwer,
\newblock ``Decompositions of a higher-order tensor in block terms—part ii:
  Definitions and uniqueness,''
\newblock {\em SIAM Journal on Matrix Analysis and Applications}, vol. 30, no.
  3, pp. 1033--1066, 2008.

\bibitem{MVNTF}
Y.~Qian, F.~Xiong, S.~Zeng, J.~Zhou, and Y.~Y. Tang,
\newblock ``Matrix-vector nonnegative tensor factorization for blind unmixing
  of hyperspectral imagery,''
\newblock {\em IEEE Transactions on Geoscience and Remote Sensing}, vol. 55,
  no. 3, pp. 1776--1792, 2017.

\bibitem{SLNTF}
Pan Zheng, Hongjun Su, and Qian Du,
\newblock ``Sparse and low-rank constrained tensor factorization for
  hyperspectral image unmixing,''
\newblock {\em IEEE Journal of Selected Topics in Applied Earth Observations
  and Remote Sensing}, vol. 14, pp. 1754--1767, 2021.

\bibitem{WSLNTF}
P.~Yang, Ti.-Z. Huang, J.~Huang, and J.-J. Wang,
\newblock ``Efficient weighted-adaptive sparse constrained nonnegative tensor
  factorization for hyperspectral unmixing,''
\newblock {\em IEEE Journal of Selected Topics in Applied Earth Observations
  and Remote Sensing}, vol. 15, pp. 10113--10130, 2022.

\bibitem{TV_NTF}
F.~Xiong, Y.~Qian, J.~Zhou, and Y.~Y. Tang,
\newblock ``Hyperspectral unmixing via total variation regularized nonnegative
  tensor factorization,''
\newblock {\em IEEE Transactions on Geoscience and Remote Sensing}, vol. 57,
  no. 4, pp. 2341--2357, 2019.

\bibitem{MultiHU_TD}
M.~Jouni, M.~D. Mura, L.~Drumetz, and P.~Comon,
\newblock ``Multihu-td: Multifeature hyperspectral unmixing based on tensor
  decomposition,''
\newblock {\em IEEE Transactions on Geoscience and Remote Sensing}, vol. 61,
  pp. 1--21, 2023.

\bibitem{ELMM1}
M.A. Veganzones, L.~Drumetz, G.~Tochon, M.~Dalla~Mura, A.~Plaza,
  J.~Bioucas-Dias, and J.~Chanussot,
\newblock ``A new extended linear mixing model to address spectral
  variability,''
\newblock in {\em 2014 6th Workshop on Hyperspectral Image and Signal
  Processing: Evolution in Remote Sensing (WHISPERS)}, 2014, pp. 1--4.

\bibitem{ELMM2}
L.~Drumetz, M.-A. Veganzones, S.~Henrot, R.~Phlypo, J.~Chanussot, and
  C.~Jutten,
\newblock ``Blind hyperspectral unmixing using an extended linear mixing model
  to address spectral variability,''
\newblock {\em IEEE Transactions on Image Processing}, vol. 25, no. 8, pp.
  3890--3905, 2016.

\bibitem{GLMM}
T.~Imbiriba, R.~A. Borsoi, and J.~C. Moreira~Bermudez,
\newblock ``Generalized linear mixing model accounting for endmember
  variability,''
\newblock in {\em 2018 IEEE International Conference on Acoustics, Speech and
  Signal Processing (ICASSP)}, 2018, pp. 1862--1866.

\bibitem{GLMM2}
R.~A. Borsoi, T.~Imbiriba, and J.~C. Moreira~Bermudez,
\newblock ``Improved hyperspectral unmixing with endmember variability
  parametrized using an interpolated scaling tensor,''
\newblock in {\em ICASSP 2019 - 2019 IEEE International Conference on
  Acoustics, Speech and Signal Processing (ICASSP)}, 2019, pp. 2177--2181.

\bibitem{PLMM}
P.-A. Thouvenin, N.~Dobigeon, and J.-Y. Tourneret,
\newblock ``Hyperspectral unmixing with spectral variability using a perturbed
  linear mixing model,''
\newblock {\em IEEE Transactions on Signal Processing}, vol. 64, no. 2, pp.
  525--538, 2016.

\bibitem{PLMM_MT}
P.-A. Thouvenin, N.~Dobigeon, and J.-Y. Tourneret,
\newblock ``Online unmixing of multitemporal hyperspectral images accounting
  for spectral variability,''
\newblock {\em IEEE Transactions on Image Processing}, vol. 25, no. 9, pp.
  3979--3990, 2016.

\bibitem{GLMM_MT}
R.~A. Borsoi, T.~Imbiriba, P.~Closas, J.~C.~M. Bermudez, and C.~Richard,
\newblock ``Kalman filtering and expectation maximization for multitemporal
  spectral unmixing,''
\newblock {\em IEEE Geoscience and Remote Sensing Letters}, vol. 19, pp. 1--5,
  2022.

\bibitem{yao_sparsity-enhanced_2021}
J.~Yao, D.~Hong, L.~Xu, D.~Meng, J.~Chanussot, and Z.~Xu,
\newblock ``Sparsity-enhanced convolutional decomposition: {A} novel
  tensor-based paradigm for blind hyperspectral unmixing,''
\newblock {\em IEEE Transactions on Geoscience and Remote Sensing}, vol. 60,
  pp. 1--14, 2021.

\bibitem{hong2018augmented}
D.~Hong, N.~Yokoya, J.~Chanussot, and X.~Zhu,
\newblock ``An augmented linear mixing model to address spectral variability
  for hyperspectral unmixing,''
\newblock {\em IEEE Transactions on Image Processing}, vol. 28, no. 4, pp.
  1923--1938, 2018.

\bibitem{SULoRA}
D.~Hong and X.~X. Zhu,
\newblock ``Sulora: Subspace unmixing with low-rank attribute embedding for
  hyperspectral data analysis,''
\newblock {\em IEEE Journal of Selected Topics in Signal Processing}, vol. 12,
  no. 6, pp. 1351--1363, 2018.

\bibitem{EndNet}
S.~{Ozkan}, B.~{Kaya}, and G.~B. {Akar},
\newblock ``Endnet: Sparse autoencoder network for endmember extraction and
  hyperspectral unmixing,''
\newblock {\em IEEE Transactions on Geoscience and Remote Sensing}, vol. 57,
  no. 1, pp. 482--496, 2019.

\bibitem{YSU2018}
Y.~{Su}, A.~{Marinoni}, J.~{Li}, J.~{Plaza}, and P.~{Gamba},
\newblock ``Stacked nonnegative sparse autoencoders for robust hyperspectral
  unmixing,''
\newblock {\em IEEE Geoscience and Remote Sensing Letters}, vol. 15, no. 9, pp.
  1427--1431, 2018.

\bibitem{DAEN}
Y.~{Su}, J.~{Li}, A.~{Plaza}, A.~{Marinoni}, P.~{Gamba}, and S.~{Chakravortty},
\newblock ``Daen: Deep autoencoder networks for hyperspectral unmixing,''
\newblock {\em IEEE Transactions on Geoscience and Remote Sensing}, vol. 57,
  no. 7, pp. 4309--4321, 2019.

\bibitem{uDAS}
Y.~{Qu} and H.~{Qi},
\newblock ``udas: An untied denoising autoencoder with sparsity for spectral
  unmixing,''
\newblock {\em IEEE Transactions on Geoscience and Remote Sensing}, vol. 57,
  no. 3, pp. 1698--1712, 2019.

\bibitem{Hua2021}
Z.~Hua, X.~Li, Q.~Qiu, and L.~Zhao,
\newblock ``Autoencoder network for hyperspectral unmixing with adaptive
  abundance smoothing,''
\newblock {\em IEEE Geoscience and Remote Sensing Letters}, vol. 18, no. 9, pp.
  1640--1644, 2021.

\bibitem{CNN_unmixing}
X.~{Zhang}, Y.~{Sun}, J.~{Zhang}, P.~{Wu}, and L.~{Jiao},
\newblock ``Hyperspectral unmixing via deep convolutional neural networks,''
\newblock {\em IEEE Geoscience and Remote Sensing Letters}, vol. 15, no. 11,
  pp. 1755--1759, 2018.

\bibitem{Burkni20SSAE}
B.~{Palsson}, J.~R. {Sveinsson}, and M.~O. {Ulfarsson},
\newblock ``Spectral-spatial hyperspectral unmixing using multitask learning,''
\newblock {\em IEEE Access}, vol. 7, pp. 148861--148872, 2019.

\bibitem{Burkni20CAE}
B.~{Palsson}, M.~O. {Ulfarsson}, and J.~R. {Sveinsson},
\newblock ``Convolutional autoencoder for spectral-spatial hyperspectral
  unmixing,''
\newblock {\em IEEE Transactions on Geoscience and Remote Sensing}, pp. 1--15,
  2020.

\bibitem{CyCUNet}
L.~Gao, Z.~Han, D.~Hong, B.~Zhang, and J.~Chanussot,
\newblock ``Cycu-net: Cycle-consistency unmixing network by learning cascaded
  autoencoders,''
\newblock {\em IEEE Transactions on Geoscience and Remote Sensing}, pp. 1--14,
  2021.

\bibitem{Trans_SU}
P.~Ghosh, S.~K. Roy, B.~Koirala, B.~Rasti, and P.~Scheunders,
\newblock ``Hyperspectral unmixing using transformer network,''
\newblock {\em IEEE Transactions on Geoscience and Remote Sensing}, vol. 60,
  pp. 1--16, 2022.

\bibitem{AAE}
Qiwen J., Yong M., F.~Fan, J.~Huang, X.~Mei, and J.~Ma,
\newblock ``Adversarial autoencoder network for hyperspectral unmixing,''
\newblock {\em IEEE Transactions on Neural Networks and Learning Systems}, pp.
  1--15, 2021.

\bibitem{DeepGUn}
R.~A. {Borsoi}, T.~{Imbiriba}, and J.~C.~M. {Bermudez},
\newblock ``Deep generative endmember modeling: An application to unsupervised
  spectral unmixing,''
\newblock {\em IEEE Transactions on Computational Imaging}, vol. 6, pp.
  374--384, 2020.

\bibitem{GAN_VAE}
S.~Shi, L.~Zhang, Y.~Altmann, and J.~Chen,
\newblock ``Deep generative model for spatial–spectral unmixing with multiple
  endmember priors,''
\newblock {\em IEEE Transactions on Geoscience and Remote Sensing}, vol. 60,
  pp. 1--14, 2022.

\bibitem{MV_proof}
W.-K. Lin, Ch.-H.and~Ma, W.-Ch. Li, Ch.-Y. Chi, and A.~Ambikapathi,
\newblock ``Identifiability of the simplex volume minimization criterion for
  blind hyperspectral unmixing: The no-pure-pixel case,''
\newblock {\em IEEE Transactions on Geoscience and Remote Sensing}, vol. 53,
  no. 10, pp. 5530--5546, 2015.

\bibitem{SUNSAL_Unrol}
C.~Zhou and M.~R.~D. Rodrigues,
\newblock ``Admm-based hyperspectral unmixing networks for abundance and
  endmember estimation,''
\newblock {\em IEEE Transactions on Geoscience and Remote Sensing}, vol. 60,
  pp. 1--18, 2022.

\bibitem{ISTA_Unrol}
Y.~Qian, F.~Xiong, Q.~Qian, and J.~Zhou,
\newblock ``Spectral mixture model inspired network architectures for
  hyperspectral unmixing,''
\newblock {\em IEEE Transactions on Geoscience and Remote Sensing}, vol. 58,
  no. 10, pp. 7418--7434, 2020.

\bibitem{ISTA}
I.~Daubechies, M.~Defrise, and C.~De~Mol,
\newblock ``An iterative thresholding algorithm for linear inverse problems
  with a sparsity constraint,''
\newblock {\em Communications on Pure and Applied Mathematics}, vol. 57, no.
  11, pp. 1413--1457, 2004.

\bibitem{NMF_Lp_Unrol}
F.~Xiong, J.~Zhou, S.~Tao, J.~Lu, and Y.~Qian,
\newblock ``Snmf-net: Learning a deep alternating neural network for
  hyperspectral unmixing,''
\newblock {\em IEEE Transactions on Geoscience and Remote Sensing}, vol. 60,
  pp. 1--16, 2022.

\bibitem{GST}
W.~Zuo, D.~Meng, L.~Zhang, X.~Feng, and D.~Zhang,
\newblock ``A generalized iterated shrinkage algorithm for non-convex sparse
  coding,''
\newblock in {\em 2013 IEEE International Conference on Computer Vision}, 2013,
  pp. 217--224.

\bibitem{Un_MAP}
L.~C. Parra, C.~Spence, P.~Sajda, A.~Ziehe, and K.-R. M\"{u}ller,
\newblock ``Unmixing hyperspectral data,''
\newblock in {\em Advances in Neural Information Processing Systems 12}. 2000,
  pp. 942--948, MIT Press.

\bibitem{Un_MMSE}
N.~{Dobigeon}, S.~{Moussaoui}, M.~{Coulon}, J.~{Tourneret}, and A.~O. {Hero},
\newblock ``Joint bayesian endmember extraction and linear unmixing for
  hyperspectral imagery,''
\newblock {\em IEEE Transactions on Signal Processing}, vol. 57, no. 11, pp.
  4355--4368, 2009.

\bibitem{Elad_AvS}
M.~Elad, P.~Milanfar, and Ron Rubinstein,
\newblock ``Analysis versus synthesis in signal priors,''
\newblock {\em Inverse Problems}, vol. 23, no. 3, pp. 947--968, apr 2007.

\bibitem{ICA}
C.~Jutten and J.~Herault,
\newblock ``Blind separation of sources, part 1: An adaptive algorithm based on
  neuromimetic architecture,''
\newblock {\em Sig. Process.}, vol. 24, no. 1, pp. 1--10, Aug. 1991.

\bibitem{ICA_BU}
J.~D. Bayliss, J.~A. Gualtieri, and R.~F. Cromp,
\newblock ``{Analyzing hyperspectral data with independent component
  analysis},''
\newblock in {\em 26th AIPR Workshop: Exploiting New Image Sources and
  Sensors}, J.~Michael Selander, Ed. International Society for Optics and
  Photonics, 1998, vol. 3240, pp. 133 -- 143, SPIE.

\bibitem{ICA_BU2}
C.~H. Chen and X.~Zhang,
\newblock ``{Independent component analysis for remote sensing study},''
\newblock in {\em Image and Signal Processing for Remote Sensing V},
  Sebastiano~Bruno Serpico, Ed. International Society for Optics and Photonics,
  1999, vol. 3871, pp. 150 -- 158, SPIE.

\bibitem{ICA_No}
J.M.P. Nascimento and J.M.B. Dias,
\newblock ``Does independent component analysis play a role in unmixing
  hyperspectral data?,''
\newblock {\em IEEE Transactions on Geoscience and Remote Sensing}, vol. 43,
  no. 1, pp. 175--187, 2005.

\bibitem{DCA_1}
J.~M.~P Nascimento and J.~M. Bioucas-Dias,
\newblock ``Hyperspectral unmixing algorithm via dependent component
  analysis,''
\newblock in {\em 2007 IEEE International Geoscience and Remote Sensing
  Symposium}, 2007, pp. 4033--4036.

\bibitem{DECA}
J.~M.~P. Nascimento and J.~M. Bioucas-Dias,
\newblock ``Hyperspectral unmixing based on mixtures of dirichlet components,''
\newblock {\em IEEE Transactions on Geoscience and Remote Sensing}, vol. 50,
  no. 3, pp. 863--878, 2012.

\bibitem{Bayes_Gibbs}
N.~Dobigeon, S.~Moussaoui, J.-Y. Tourneret, and C.~Carteret,
\newblock ``Bayesian separation of spectral sources under non-negativity and
  full additivity constraints,''
\newblock {\em Signal Processing}, vol. 89, no. 12, pp. 2657--2669, 2009,
\newblock Special Section: Visual Information Analysis for Security.

\bibitem{Lib_Chall}
B.~Somers, L.~Tits, D.~Roberts, and E.~Wetherley,
\newblock ``Chapter 17 - endmember library approaches to resolve spectral
  mixing problems in remotely sensed data: Potential, challenges, and
  applications,''
\newblock in {\em Resolving Spectral Mixtures}, Cyril Ruckebusch, Ed., vol.~30
  of {\em Data Handling in Science and Technology}, pp. 551--577. Elsevier,
  2016.

\bibitem{MESMA}
D.~Roberts, M.~Gardner, R.~Church, S.~Ustin, G.~Scheer, and R.O. Green,
\newblock ``Mapping chaparral in the santa monica mountains using multiple
  endmember spectral mixture models,''
\newblock {\em Remote Sensing of Environment}, vol. 65, 09 1998.

\bibitem{MisMat_Lib}
S.~Tompkins, J.~F. Mustard, C.~M. Pieters, and D.~W. Forsyth,
\newblock ``Optimization of endmembers for spectral mixture analysis,''
\newblock {\em Remote Sensing of Environment}, vol. 59, no. 3, pp. 472--489,
  1997.

\bibitem{SUn}
M.~{Iordache}, J.~M. {Bioucas-Dias}, and A.~{Plaza},
\newblock ``Sparse unmixing of hyperspectral data,''
\newblock {\em IEEE Transactions on Geoscience and Remote Sensing}, vol. 49,
  no. 6, pp. 2014--2039, 2011.

\bibitem{MESMA_Lib_EE}
C.~Quintano, A.~Fernandez-Manso, and D.~A. Roberts,
\newblock ``Multiple endmember spectral mixture analysis (mesma) to map burn
  severity levels from landsat images in mediterranean countries,''
\newblock {\em Remote Sensing of Environment}, vol. 136, pp. 76--88, 2013.

\bibitem{GSIMN}
L.~{Drumetz}, T.~R. {Meyer}, J.~{Chanussot}, A.~L. {Bertozzi}, and C.~{Jutten},
\newblock ``Hyperspectral image unmixing with endmember bundles and group
  sparsity inducing mixed norms,''
\newblock {\em IEEE Transactions on Image Processing}, vol. 28, no. 7, pp.
  3435--3450, 2019.

\bibitem{Rad_Tra_Lib}
L.~Tits, B.~Somers, and P.~Coppin,
\newblock ``The potential and limitations of a clustering approach for the
  improved efficiency of multiple endmember spectral mixture analysis in plant
  production system monitoring,''
\newblock {\em IEEE Transactions on Geoscience and Remote Sensing}, vol. 50,
  no. 6, pp. 2273--2286, 2012.

\bibitem{PROSPECT}
S.~Jacquemoud and F.~Baret,
\newblock ``Prospect: A model of leaf optical properties spectra,''
\newblock {\em Remote Sensing of Environment}, vol. 34, no. 2, pp. 75--91,
  1990.

\bibitem{MESMA_VAE}
R.~A. Borsoi, T.~Imbiriba, J.~C.~M. Bermudez, and C.~Richard,
\newblock ``Deep generative models for library augmentation in multiple
  endmember spectral mixture analysis,''
\newblock {\em IEEE Geoscience and Remote Sensing Letters}, vol. 18, no. 10,
  pp. 1831--1835, 2021.

\bibitem{End_Rob}
R.~Heylen, A.~Zare, P.~Gader, and P.~Scheunders,
\newblock ``Hyperspectral unmixing with endmember variability via alternating
  angle minimization,''
\newblock {\em IEEE Transactions on Geoscience and Remote Sensing}, vol. 54,
  no. 8, pp. 4983--4993, 2016.

\bibitem{Comp_MESMA}
Philip~E. D., Kerry~Q. H., and Dar~A. R.,
\newblock ``A comparison of error metrics and constraints for multiple
  endmember spectral mixture analysis and spectral angle mapper,''
\newblock {\em Remote Sensing of Environment}, vol. 93, no. 3, pp. 359--367,
  2004.

\bibitem{MELSUM}
J.-Ph. Combe, S.~{Le MouÃ©lic}, C.~Sotin, A.~Gendrin, J.F. Mustard, L.~{Le
  Deit}, P.~Launeau, J.-P. Bibring, B.~Gondet, Y.~Langevin, and P.~Pinet,
\newblock ``Analysis of omega/mars express data hyperspectral data using a
  multiple-endmember linear spectral unmixing model (melsum): Methodology and
  first results,''
\newblock {\em Planetary and Space Science}, vol. 56, no. 7, pp. 951--975,
  2008.

\bibitem{MESMA_LibR1}
D.A. Roberts, P.E. Dennison, M.E. Gardner, Y.~Hetzel, S.L. Ustin, and C.T. Lee,
\newblock ``Evaluation of the potential of hyperion for fire danger assessment
  by comparison to the airborne visible/infrared imaging spectrometer,''
\newblock {\em IEEE Transactions on Geoscience and Remote Sensing}, vol. 41,
  no. 6, pp. 1297--1310, 2003.

\bibitem{MESMA_LibR2}
Carmen Q., Alfonso F.-M., and Dar~A. R.,
\newblock ``Multiple endmember spectral mixture analysis (mesma) to map burn
  severity levels from landsat images in mediterranean countries,''
\newblock {\em Remote Sensing of Environment}, vol. 136, pp. 76--88, 2013.

\bibitem{MESMA_LibR3}
Philip~E. D. and Dar~A. R.,
\newblock ``Endmember selection for multiple endmember spectral mixture
  analysis using endmember average rmse,''
\newblock {\em Remote Sensing of Environment}, vol. 87, no. 2, pp. 123--135,
  2003.

\bibitem{MESMA_Rob}
R.~Heylen, A.~Zare, P.~Gader, and P.~Scheunders,
\newblock ``Hyperspectral unmixing with endmember variability via alternating
  angle minimization,''
\newblock {\em IEEE Transactions on Geoscience and Remote Sensing}, vol. 54,
  no. 8, pp. 4983--4993, 2016.

\bibitem{MESMA_Rob2}
L.~Tits, R.~Heylen, B.~Somers, P.~Scheunders, and P.~Coppin,
\newblock ``A geometric unmixing concept for the selection of optimal binary
  endmember combinations,''
\newblock {\em IEEE Geoscience and Remote Sensing Letters}, vol. 12, no. 1, pp.
  82--86, 2015.

\bibitem{SUNSAL_arX}
J.~M. Bioucas-Dias and M.~A.~T. Figueiredo,
\newblock ``Alternating direction algorithms for constrained sparse regression:
  Application to hyperspectral unmixing,''
\newblock {\em arXiv}, 2012.

\bibitem{SU_3Jose}
J.~M.~P. Nascimento, J.~M. Bioucas-Dias, J.~M. Rodríguez~Alves, V.~Silva, and
  A.~Plaza,
\newblock ``Parallel hyperspectral unmixing on gpus,''
\newblock {\em IEEE Geoscience and Remote Sensing Letters}, vol. 11, no. 3, pp.
  666--670, 2014.

\bibitem{OMP}
Y.~C. Pati, R.~Rezaiifar, and P.~Krishnaprasad,
\newblock ``Orthogonal matching pursuit: Recursive function approximation with
  applications to wavelet decomposition,''
\newblock in {\em Proc. 27th Annu. Asilomar Conf. Signals, Syst., Comput., Los
  Alamitos, CA,}, 1993, p. 40–44.

\bibitem{MC}
D.~L. Donoho and M.~Elad,
\newblock ``Optimally sparse representation in general (nonorthogonal)
  dictionaries via $\ell_1$ minimization,''
\newblock {\em Proceedings of the National Academy of Sciences of the United
  States of America}, vol. 100, no. 5, pp. 2197--2202, 2003.

\bibitem{RICs}
E.~J. Candes, J.~K. Romberg, and T.~Tao,
\newblock ``Stable signal recovery from incomplete and inaccurate
  measurements,''
\newblock {\em Communications on Pure and Applied Mathematics}, vol. 59, no. 8,
  pp. 1207--1223, 2006.

\bibitem{SUnSAL-TV}
M.~{Iordache}, J.~M. {Bioucas-Dias}, and A.~{Plaza},
\newblock ``Total variation spatial regularization for sparse hyperspectral
  unmixing,''
\newblock {\em IEEE Transactions on Geoscience and Remote Sensing}, vol. 50,
  no. 11, pp. 4484--4502, 2012.

\bibitem{GSUNSAL}
M.-D. Iordache, J.~M. Bioucas-Dias, and A.~Plaza,
\newblock ``Hyperspectral unmixing with sparse group lasso,''
\newblock in {\em 2011 IEEE International Geoscience and Remote Sensing
  Symposium}, 2011, pp. 3586--3589.

\bibitem{GLASSO}
L.~Meier, S.~V.~D. Geer, and P.~Bühlmann,
\newblock ``The group lasso for logistic regression,''
\newblock {\em Journal of the Royal Statistical Society. Series B}, vol. 70,
  no. 1, pp. 53--71, 2008.

\bibitem{SU_lq}
F.~Chen and Y.~Zhang,
\newblock ``Sparse hyperspectral unmixing based on constrained lp - l2
  optimization,''
\newblock {\em IEEE Geoscience and Remote Sensing Letters}, vol. 10, no. 5, pp.
  1142--1146, 2013.

\bibitem{collab_l0}
Z.~Shi, T.~Shi, M.~Zhou, and X.~Xu,
\newblock ``Collaborative sparse hyperspectral unmixing using $l_{0}$ norm,''
\newblock {\em IEEE Transactions on Geoscience and Remote Sensing}, vol. 56,
  no. 9, pp. 5495--5508, 2018.

\bibitem{DRSU_TV}
R.~Wang, H.-C. Li, A.~Pizurica, J.~Li, A.~Plaza, and W.~J. Emery,
\newblock ``Hyperspectral unmixing using double reweighted sparse regression
  and total variation,''
\newblock {\em IEEE Geoscience and Remote Sensing Letters}, vol. 14, no. 7, pp.
  1146--1150, 2017.

\bibitem{SZhang2016}
S.~{Zhang}, J.~{Li}, K.~{Liu}, C.~{Deng}, L.~{Liu}, and A.~{Plaza},
\newblock ``Hyperspectral unmixing based on local collaborative sparse
  regression,''
\newblock {\em IEEE Geoscience and Remote Sensing Letters}, vol. 13, no. 5, pp.
  631--635, 2016.

\bibitem{SU_LR}
P.~V. Giampouras, K.~E. Themelis, A.~A. Rontogiannis, and K.~D. Koutroumbas,
\newblock ``Simultaneously sparse and low-rank abundance matrix estimation for
  hyperspectral image unmixing,''
\newblock {\em IEEE Transactions on Geoscience and Remote Sensing}, vol. 54,
  no. 8, pp. 4775--4789, 2016.

\bibitem{SU_LR_SSW}
F.~Li, S.~Zhang, B.~Liang, C.~Deng, C.~Xu, and S.~Wang,
\newblock ``Hyperspectral sparse unmixing with spectral-spatial low-rank
  constraint,''
\newblock {\em IEEE Journal of Selected Topics in Applied Earth Observations
  and Remote Sensing}, vol. 14, pp. 6119--6130, 2021.

\bibitem{SU_SpNorm}
H.~Han, G.~Wang, M.~Wang, J.~Miao, S.~Guo, L.~Chen, M.~Zhang, and K.~Guo,
\newblock ``Hyperspectral unmixing via nonconvex sparse and low-rank
  constraint,''
\newblock {\em IEEE Journal of Selected Topics in Applied Earth Observations
  and Remote Sensing}, vol. 13, pp. 5704--5718, 2020.

\bibitem{SpNorm}
Y.~Xie, S.~Gu, Y.~Liu, W.~Zuo, W.~Zhang, and L.~Zhang,
\newblock ``Weighted schatten $p$ -norm minimization for image denoising and
  background subtraction,''
\newblock {\em IEEE Transactions on Image Processing}, vol. 25, no. 10, pp.
  4842--4857, 2016.

\bibitem{TD_SU1}
L.~Sun, F.~Wu, T.~Zhan, W.~Liu, J.~Wang, and B.~Jeon,
\newblock ``Weighted nonlocal low-rank tensor decomposition method for sparse
  unmixing of hyperspectral images,''
\newblock {\em IEEE Journal of Selected Topics in Applied Earth Observations
  and Remote Sensing}, vol. 13, pp. 1174--1188, 2020.

\bibitem{TD_SU2}
J.~Huang, T.-Z. Huang, X.-L. Zhao, and L.-J. Deng,
\newblock ``Nonlocal tensor-based sparse hyperspectral unmixing,''
\newblock {\em IEEE Transactions on Geoscience and Remote Sensing}, vol. 59,
  no. 8, pp. 6854--6868, 2021.

\bibitem{TanerInce2020}
T.~Ince,
\newblock ``Superpixel-based graph laplacian regularization for sparse
  hyperspectral unmixing,''
\newblock {\em IEEE Geoscience and Remote Sensing Letters}, pp. 1--5, 2020.

\bibitem{SUn_SV}
G.~Zhang, S.~Mei, B.~Xie, M.~Ma, Y.~Zhang, Y.~Feng, and Q.~Du,
\newblock ``Spectral variability augmented sparse unmixing of hyperspectral
  images,''
\newblock {\em IEEE Transactions on Geoscience and Remote Sensing}, vol. 60,
  pp. 1--13, 2022.

\bibitem{SPEE}
S.~Mei, M.~He, Z.~Wang, and D.~Feng,
\newblock ``Spatial purity based endmember extraction for spectral mixture
  analysis,''
\newblock {\em IEEE Transactions on Geoscience and Remote Sensing}, vol. 48,
  no. 9, pp. 3434--3445, 2010.

\bibitem{SU_LibP_BS}
M.-D. Iordache, J.~M. Bioucas-Dias, and A.~Plaza,
\newblock ``Potential and limitations of band selection and library pruning in
  sparse hyperspectral unmixing,''
\newblock in {\em 2015 7th Workshop on Hyperspectral Image and Signal
  Processing: Evolution in Remote Sensing (WHISPERS)}, 2015, pp. 1--4.

\bibitem{Lib_Prun}
M.-D. Iordache, J.~M. Bioucas-Dias, and A.~Plaza,
\newblock ``Dictionary pruning in sparse unmixing of hyperspectral data,''
\newblock in {\em 2012 4th Workshop on Hyperspectral Image and Signal
  Processing: Evolution in Remote Sensing (WHISPERS)}, 2012, pp. 1--4.

\bibitem{MUSIC_CSR}
M.-D. Iordache, J.~M. Bioucas-Dias, A.~Plaza, and B.~Somers,
\newblock ``Music-csr: Hyperspectral unmixing via multiple signal
  classification and collaborative sparse regression,''
\newblock {\em IEEE Transactions on Geoscience and Remote Sensing}, vol. 52,
  no. 7, pp. 4364--4382, 2014.

\bibitem{MUSIC}
R.~Schmidt,
\newblock ``Multiple emitter location and signal parameter estimation,''
\newblock {\em IEEE Transactions on Antennas and Propagation}, vol. 34, no. 3,
  pp. 276--280, 1986.

\bibitem{SU_LibP_MO}
X.~Xu, B.~Pan, Z.~Chen, Z.~Shi, and T.~Li,
\newblock ``Simultaneously multiobjective sparse unmixing and library pruning
  for hyperspectral imagery,''
\newblock {\em IEEE Transactions on Geoscience and Remote Sensing}, vol. 59,
  no. 4, pp. 3383--3395, 2021.

\bibitem{SUnSAL_Leg}
M.~Parente and M.-D. Iordache,
\newblock ``Sparse unmixing of hyperspectral data: The legacy of sunsal,''
\newblock in {\em 2021 IEEE International Geoscience and Remote Sensing
  Symposium IGARSS}, 2021, pp. 21--24.

\bibitem{SU_LibP_Mis}
X.~Fu, W.-K. Ma, J.~M. Bioucas-Dias, and T.-H. Chan,
\newblock ``Semiblind hyperspectral unmixing in the presence of spectral
  library mismatches,''
\newblock {\em IEEE Transactions on Geoscience and Remote Sensing}, vol. 54,
  no. 9, pp. 5171--5184, 2016.

\bibitem{RAA}
Y.~Chen, J.~Mairal, and Z.~Harchaoui,
\newblock ``Fast and robust archetypal analysis for representation learning,''
\newblock {\em 2014 IEEE Conference on Computer Vision and Pattern
  Recognition}, pp. 1478--1485, 2014.

\bibitem{SU_GM1}
H.~Su, C.~Jia, P.~Zheng, and Q.~Du,
\newblock ``Superpixel-based weighted collaborative sparse regression and
  reweighted low-rank representation for hyperspectral image unmixing,''
\newblock {\em IEEE Journal of Selected Topics in Applied Earth Observations
  and Remote Sensing}, vol. 15, pp. 393--408, 2022.

\bibitem{SU_GM2}
L.~Sun, F.~Wu, C.~He, T.~Zhan, W.~Liu, and D.~Zhang,
\newblock ``Weighted collaborative sparse and l1/2 low-rank regularizations
  with superpixel segmentation for hyperspectral unmixing,''
\newblock {\em IEEE Geoscience and Remote Sensing Letters}, vol. 19, pp. 1--5,
  2022.

\bibitem{SU_GM3}
T.~Chen, Y.~Liu, Y.~Zhang, B.~Du, and A.~Plaza,
\newblock ``Superpixel-based collaborative and low-rank regularization for
  sparse hyperspectral unmixing,''
\newblock {\em IEEE Transactions on Geoscience and Remote Sensing}, vol. 60,
  pp. 1--16, 2022.

\bibitem{SU_GM4}
L.~Qi, J.~Li, Y.~Wang, Y.~Huang, and X.~Gao,
\newblock ``Spectral–spatial-weighted multiview collaborative sparse unmixing
  for hyperspectral images,''
\newblock {\em IEEE Transactions on Geoscience and Remote Sensing}, vol. 58,
  no. 12, pp. 8766--8779, 2020.

\bibitem{SU_GM5}
X.~Shen, H.~Liu, X.~Zhang, K.~Qin, and X.~Zhou,
\newblock ``Superpixel-guided local sparsity prior for hyperspectral sparse
  regression unmixing,''
\newblock {\em IEEE Geoscience and Remote Sensing Letters}, vol. 19, pp. 1--5,
  2022.

\bibitem{SU_GM6}
H.~Li, R.~Feng, L.~Wang, Y.~Zhong, and L.~Zhang,
\newblock ``Superpixel-based reweighted low-rank and total variation sparse
  unmixing for hyperspectral remote sensing imagery,''
\newblock {\em IEEE Transactions on Geoscience and Remote Sensing}, vol. 59,
  no. 1, pp. 629--647, 2021.

\bibitem{SU_GM7}
L.~Ren, Z.~Ma, F.~Bovolo, and L.~Bruzzone,
\newblock ``A nonconvex framework for sparse unmixing incorporating the group
  structure of the spectral library,''
\newblock {\em IEEE Transactions on Geoscience and Remote Sensing}, vol. 60,
  pp. 1--19, 2022.

\bibitem{SU_Mix}
D.~Zhang, T.~Wang, S.~Yang, Y.~Jia, and F.~Li,
\newblock ``Spectral reweighting and spectral similarity weighting for sparse
  hyperspectral unmixing,''
\newblock {\em IEEE Geoscience and Remote Sensing Letters}, vol. 19, pp. 1--5,
  2022.

\bibitem{SMALU}
Y.~Lin and P.~Gader,
\newblock ``Addressing spectral variability in hyperspectral unmixing with
  unsupervised neural networks,''
\newblock in {\em 2022 12th Workshop on Hyperspectral Imaging and Signal
  Processing: Evolution in Remote Sensing (WHISPERS)}, 2022, pp. 1--5.

\bibitem{SVANT}
G.~Zhang, S.~Mei, B.~Xie, Y.~Feng, and Q.~Du,
\newblock ``Spectral variability augmented two-stream network for hyperspectral
  sparse unmixing,''
\newblock {\em IEEE Geoscience and Remote Sensing Letters}, vol. 19, pp. 1--5,
  2022.

\bibitem{SS_Net}
F.~Kong, M.~Chen, Y.~Li, and D.~Li,
\newblock ``A global spectral–spatial feature learning network for
  semisupervised hyperspectral unmixing,''
\newblock {\em IEEE Journal of Selected Topics in Applied Earth Observations
  and Remote Sensing}, vol. 15, pp. 3190--3203, 2022.

\bibitem{ISTA_Unrol_SUn}
Q.~Qian, F.~Xiong, and J.~Zhou,
\newblock ``Deep unfolded iterative shrinkage-thresholding model for
  hyperspectral unmixing,''
\newblock in {\em IGARSS 2019 - 2019 IEEE International Geoscience and Remote
  Sensing Symposium}, 2019, pp. 2151--2154.

\bibitem{SU_Unroll}
Y.~Shao, Q.~Liu, and L.~Xiao,
\newblock ``Iviu-net: Implicit variable iterative unrolling network for
  hyperspectral sparse unmixing,''
\newblock {\em IEEE Journal of Selected Topics in Applied Earth Observations
  and Remote Sensing}, vol. 16, pp. 1756--1770, 2023.

\bibitem{Fan}
W.~Fan, B.~Hu, J.~Miller, and M.~Li,
\newblock ``Comparative study between a new nonlinear model and common linear
  model for analysing laboratory simulated‐forest hyperspectral data,''
\newblock {\em International Journal of Remote Sensing}, vol. 30, no. 11, pp.
  2951--2962, 2009.

\bibitem{PPNM}
Y.~Altmann, A.~Halimi, N.~Dobigeon, and J.Y. Tourneret,
\newblock ``Supervised nonlinear spectral unmixing using a postnonlinear mixing
  model for hyperspectral imagery,''
\newblock {\em IEEE Transactions on Image Processing}, vol. 21, no. 6, pp.
  3017--3025, 2012.

\bibitem{AHalimi2011}
A.~{Halimi}, Y.~{Altmann}, N.~{Dobigeon}, and J.~{Tourneret},
\newblock ``Nonlinear unmixing of hyperspectral images using a generalized
  bilinear model,''
\newblock {\em IEEE Transactions on Geoscience and Remote Sensing}, vol. 49,
  no. 11, pp. 4153--4162, 2011.

\bibitem{LianruGAO2022}
L.~Gao, Z.~Wang, L.~Zhuang, H.~Yu, B.~Zhang, and J.~Chanussot,
\newblock ``Using low-rank representation of abundance maps and nonnegative
  tensor factorization for hyperspectral nonlinear unmixing,''
\newblock {\em IEEE Transactions on Geoscience and Remote Sensing}, vol. 60,
  pp. 1--17, 2022.

\bibitem{RHeylenMLM}
R.~Heylen and P.~Scheunders,
\newblock ``A multilinear mixing model for nonlinear spectral unmixing,''
\newblock {\em IEEE Transactions on Geoscience and Remote Sensing}, vol. 54,
  no. 1, pp. 240--251, Jan 2016.

\bibitem{AMarinoni20159}
A.~{Marinoni} and P.~{Gamba},
\newblock ``A novel approach for efficient $p$-linear hyperspectral unmixing,''
\newblock {\em IEEE Journal of Selected Topics in Signal Processing}, vol. 9,
  no. 6, pp. 1156--1168, Sep. 2015.

\bibitem{AMarinoni20158}
A.~{Marinoni}, J.~{Plaza}, A.~{Plaza}, and P.~{Gamba},
\newblock ``Nonlinear hyperspectral unmixing using nonlinearity order
  estimation and polytope decomposition,''
\newblock {\em IEEE Journal of Selected Topics in Applied Earth Observations
  and Remote Sensing}, vol. 8, no. 6, pp. 2644--2654, June 2015.

\bibitem{AMarinoni20169}
A.~{Marinoni}, A.~{Plaza}, and P.~{Gamba},
\newblock ``Harmonic mixture modeling for efficient nonlinear hyperspectral
  unmixing,''
\newblock {\em IEEE Journal of Selected Topics in Applied Earth Observations
  and Remote Sensing}, vol. 9, no. 9, pp. 4247--4256, Sep. 2016.

\bibitem{KFCLSU}
J.~Broadwater, R.~Chellappa, A.~Banerjee, and P.~Burlina,
\newblock ``Kernel fully constrained least squares abundance estimates,''
\newblock in {\em 2007 IEEE International Geoscience and Remote Sensing
  Symposium}, 2007, pp. 4041--4044.

\bibitem{KNMF}
X.~Wu, X.~Li, and L.~Zhao,
\newblock ``A kernel spatial complexity-based nonlinear unmixing method of
  hyperspectral imagery,''
\newblock in {\em International Conference on Intelligent Computing for
  Sustainable Energy and Environment}, Berlin, Heidelberg, 2010, LSMS/ICSEE'10,
  p. 451–458, Springer-Verlag.

\bibitem{SVM_Un}
M.~Brown, H.G. Lewis, and S.R. Gunn,
\newblock ``Linear spectral mixture models and support vector machines for
  remote sensing,''
\newblock {\em IEEE Transactions on Geoscience and Remote Sensing}, vol. 38,
  no. 5, pp. 2346--2360, 2000.

\bibitem{BHapke1998}
B.~Hapke, R.~Nelson, and W.~Smythe,
\newblock ``The opposition effect of the moon: Coherent backscatter and shadow
  hiding,''
\newblock {\em Icarus}, vol. 133, no. 1, pp. 89 -- 97, 1998.

\bibitem{NU_DAE}
M.~Wang, M.~Zhao, J.~Chen, and S.~Rahardja,
\newblock ``Nonlinear unmixing of hyperspectral data via deep autoencoder
  networks,''
\newblock {\em IEEE Geoscience and Remote Sensing Letters}, vol. 16, no. 9, pp.
  1467--1471, 2019.

\bibitem{BU_DAE_MTL}
Y.~Su, X.~Xu, J.~Li, H.~Qi, P.~Gamba, and A.~Plaza,
\newblock ``Deep autoencoders with multitask learning for bilinear
  hyperspectral unmixing,''
\newblock {\em IEEE Transactions on Geoscience and Remote Sensing}, vol. 59,
  no. 10, pp. 8615--8629, 2021.

\bibitem{DL_PPNM}
M.~Zhao, L.~Yan, and J.~Chen,
\newblock ``Lstm-dnn based autoencoder network for nonlinear hyperspectral
  image unmixing,''
\newblock {\em IEEE Journal of Selected Topics in Signal Processing}, vol. 15,
  no. 2, pp. 295--309, 2021.

\bibitem{3DCNN_PPNM}
M.~Zhao, M.~Wang, J.~Chen, and S.~Rahardja,
\newblock ``Hyperspectral unmixing for additive nonlinear models with a 3-d-cnn
  autoencoder network,''
\newblock {\em IEEE Transactions on Geoscience and Remote Sensing}, vol. 60,
  pp. 1--15, 2022.

\bibitem{AE_RBF_PPNM}
K.~T. Shahid and I.~D. Schizas,
\newblock ``Spatial-aware hyperspectral nonlinear unmixing autoencoder with
  endmember number estimation,''
\newblock {\em IEEE Journal of Selected Topics in Applied Earth Observations
  and Remote Sensing}, vol. 15, pp. 20--41, 2022.

\bibitem{MB_DAE_PPNM}
H.~Li, R.~A. Borsoi, T.~Imbiriba, P.~Closas, J.~C.~M. Bermudez, and
  D.~Erdoğmuş,
\newblock ``Model-based deep autoencoder networks for nonlinear hyperspectral
  unmixing,''
\newblock {\em IEEE Geoscience and Remote Sensing Letters}, vol. 19, pp. 1--5,
  2022.

\bibitem{AE_RBF_PPNM2}
K.~T. Shahid and I.~D. Schizas,
\newblock ``Unsupervised hyperspectral unmixing via nonlinear autoencoders,''
\newblock {\em IEEE Transactions on Geoscience and Remote Sensing}, vol. 60,
  pp. 1--13, 2022.

\bibitem{Simplex_AE_PPNM}
Q.~Lyu and X.~Fu,
\newblock ``Identifiability-guaranteed simplex-structured post-nonlinear
  mixture learning via autoencoder,''
\newblock {\em IEEE Transactions on Signal Processing}, vol. 69, pp.
  4921--4936, 2021.

\bibitem{NU_GAN}
M.~Tang, Y.~Qu, and H.~Qi,
\newblock ``Hyperspectral nonlinear unmixing via generative adversarial
  network,''
\newblock in {\em IGARSS 2020 - 2020 IEEE International Geoscience and Remote
  Sensing Symposium}, 2020, pp. 2404--2407.

\bibitem{HapkeCNN}
B.~Rasti, B.~Koirala, and P.~Scheunders,
\newblock ``Hapkecnn: Blind nonlinear unmixing for intimate mixtures using
  hapke model and convolutional neural network,''
\newblock {\em IEEE Transactions on Geoscience and Remote Sensing}, vol. 60,
  pp. 1--15, 2022.

\end{thebibliography}
\vspace{11pt}

\vfill

\begin{IEEEbiography}[{\includegraphics[width=1in,height=1.25in,clip,keepaspectratio]{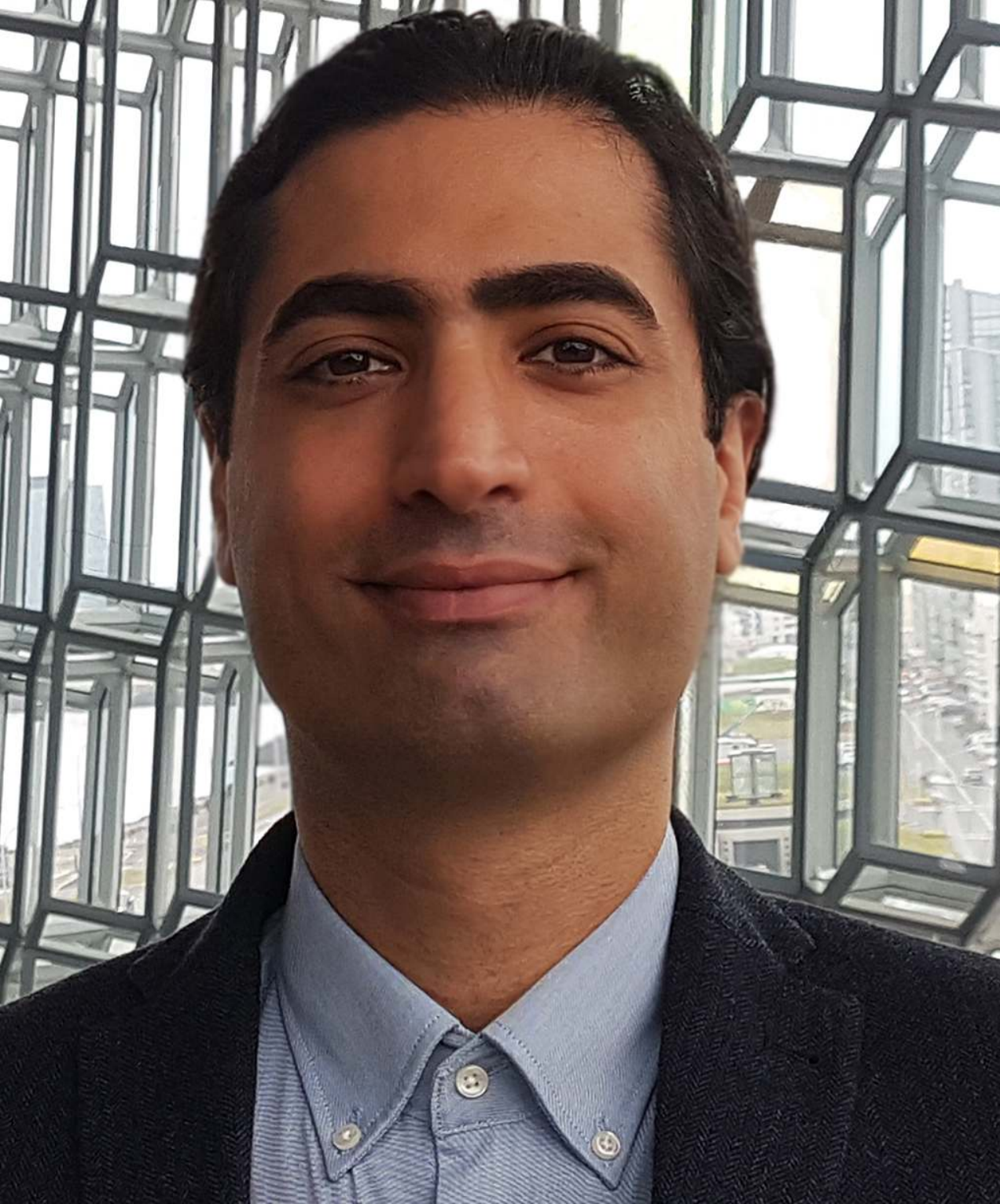}}]{Behnood Rasti (M’12–SM’19)} received the B.Sc. and M.Sc. degrees in electronics and electrical engineering from the Electrical Engineering Department, University of Guilan, Rasht, Iran, in 2006 and 2009, respectively, and the Ph.D. degree in electrical and computer engineering from the University of Iceland, Reykjavik, Iceland, in 2014. He was a Valedictorian as an M.Sc. Student in 2009. In 2015 and 2016, he worked as a Post-Doctoral Researcher with the Electrical and Computer Engineering Department, University of Iceland. From 2016 to 2019, he was a Lecturer with the Center of Engineering Technology and Applied Sciences, Department of Electrical and Computer Engineering, University of Iceland. Dr. Rasti was a Humboldt Research Fellow in 2020 and 2021. From 2022 to 2023, he was a Principal Research Associate with Helmholtz-Zentrum Dresden-Rossendorf (HZDR), Dresden, Germany. He is currently a Senior Research Scientist with the Faculty of Electrical Engineering and Computer Science, Technische Universität Berlin. His research interests include signal and image processing, machine/deep learning, remote sensing, and artificial intelligence. 

Dr. Rasti won the Doctoral Grant of the University of Iceland Research Fund “The Eimskip University Fund” and the “Alexander von Humboldt Research Fellowship Grant” in 2013 and 2019, respectively. He serves as an Associate Editor for the IEEE GEOSCIENCE AND REMOTE SENSING LETTERS (GRSL). 
\end{IEEEbiography}

\begin{IEEEbiography}
[{\includegraphics[width=1in,height=1.25in,clip,keepaspectratio]{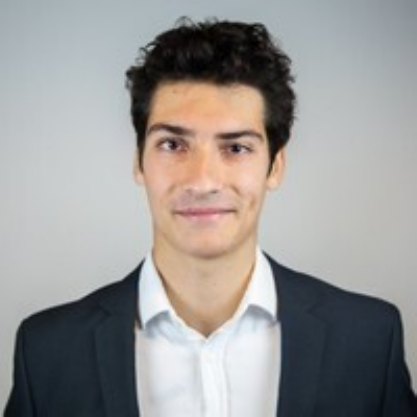}}]{Alexandre Zouaoui} received his M.Sc. degree in computer science from Télécom Paris, Paris, France, in 2019. He received his Ph.D. in applied mathematics at Université Grenoble Alpes, Grenoble, France, in 2024 on Sparse and Archetypal Decomposition Algorithms for Hyperspectral Images Restoration and Spectral Unmixing. His research interests include computer vision, interpretable machine learning and remote sensing image processing. In 2024, he joined Data Science Experts as a research engineer focusing on applications related to hyperspectral imagery.
\end{IEEEbiography}

\begin{IEEEbiography}[{\includegraphics[width=1in,height=1.25in,clip,keepaspectratio]{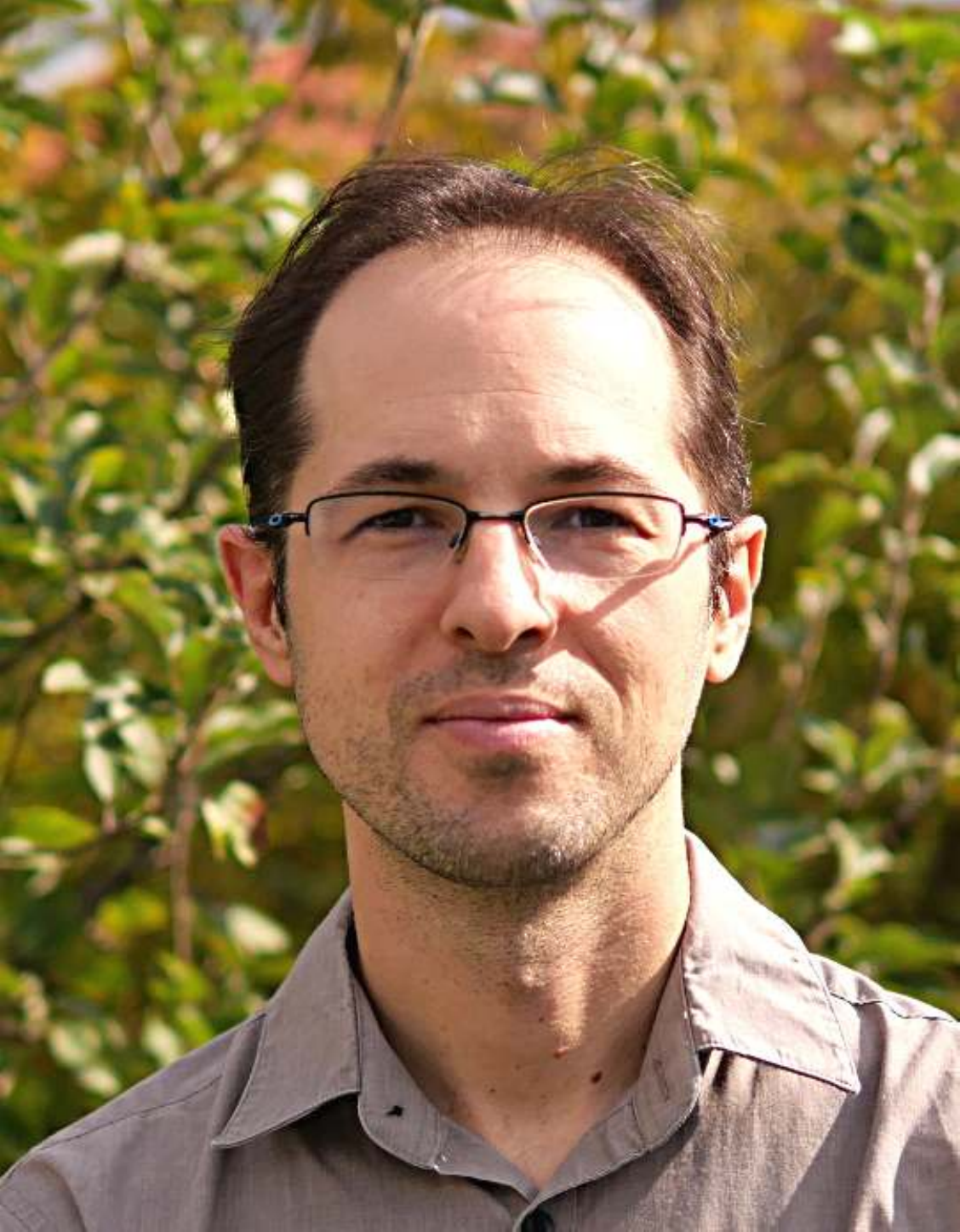}}]{Julien Mairal} received the graduate degree from Ecole Polytechnique, Palaiseau, France, in 2005, and the Ph.D. degree from Ecole Normale Superieure, Cachan, France, in 2010. After that, he joined the statistics department at UC Berkeley as a post-doctoral researcher. In 2012, he joined Inria, Grenoble, France, where he is currently a research director and head of the Thoth team. His research interests include machine learning, computer vision, mathematical optimization, and statistical image and signal processing. He received a starting grant and a consolidator grant from the European Research Council, respectively in 2016 and 2022. He was awarded the Cor Baayen prize in 2013, the IEEE PAMI young research award in 2017 and the test-of-time award at ICML 2019.
\end{IEEEbiography}

\begin{IEEEbiography}[{\includegraphics[width=1in,height=1.25in,clip,keepaspectratio]{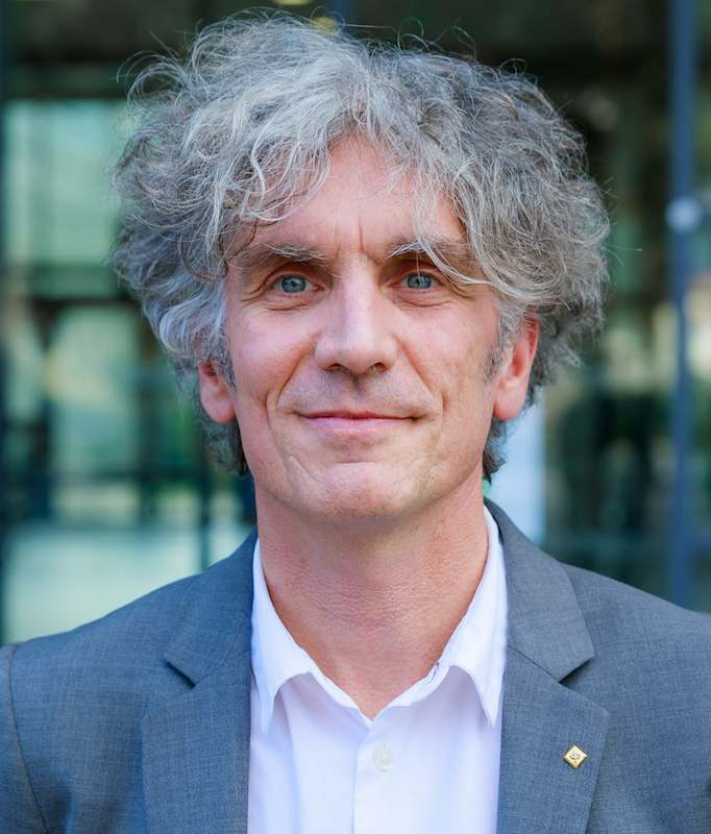}}]{Jocelyn Chanussot (M’04–SM’04–F’12)} received the M.Sc. degree in electrical engineering from the Grenoble Institute of Technology (Grenoble INP), Grenoble, France, in 1995, and the Ph.D. degree from the Université de Savoie, Annecy, France, in 1998. From 1999 to 2023, he has been with Grenoble INP, where he was a Professor of signal and image processing. He is currently a Research Director with INRIA, Grenoble. His research interests include image analysis, hyperspectral remote sensing, data fusion, machine learning and artificial intelligence. He has been a visiting scholar at Stanford University (USA), KTH (Sweden) and NUS (Singapore). Since 2013, he is an Adjunct Professor of the University of Iceland. In 2015-2017, he was a visiting professor at the University of California, Los Angeles (UCLA).  He holds the AXA chair in remote sensing and is an Adjunct professor at the Chinese Academy of Sciences, Aerospace Information research Institute, Beijing.
Dr. Chanussot is the founding President of IEEE Geoscience and Remote Sensing French chapter (2007-2010) which received the 2010 IEEE GRS-S Chapter Excellence Award. He has received multiple outstanding paper awards. He was the Vice-President of the IEEE Geoscience and Remote Sensing Society, in charge of meetings and symposia (2017-2019). He was the General Chair of the first IEEE GRSS Workshop on Hyperspectral Image and Signal Processing, Evolution in Remote sensing (WHISPERS). He was the Chair (2009-2011) and  Cochair of the GRS Data Fusion Technical Committee (2005-2008). He was a member of the Machine Learning for Signal Processing Technical Committee of the IEEE Signal Processing Society (2006-2008) and the Program Chair of the IEEE International Workshop on Machine Learning for Signal Processing (2009). He is an Associate Editor for the IEEE Transactions on Geoscience and Remote Sensing, the IEEE Transactions on Image Processing and the Proceedings of the IEEE. He was the Editor-in-Chief of the IEEE Journal of Selected Topics in Applied Earth Observations and Remote Sensing (2011-2015). In 2014 he served as a Guest Editor for the IEEE Signal Processing Magazine. He is a Fellow of the IEEE, an ELLIS Fellow, a Fellow of the Asia-Pacific Artificial Intelligence Association, a member of the Institut Universitaire de France (2012-2017) and a Highly Cited Researcher (Clarivate Analytics/Thomson Reuters, since 2018).
\end{IEEEbiography}

\end{document}